\documentstyle[12pt]{article}
\newcommand{\h}{\hspace*{5 ex}}

\begin{document}
\vspace{5cm}
\section*{{\centerline{Ettore Majorana: Vita e Opere}\\
\centerline{(L'opera scientifica edita e inedita) $^{(\dag)}$}}}
\footnotetext{$^{(\dag)}$ Work partially supported by CNR, MURST and INFN. \ No
part of this article (in particular, no letters, documents, testimonies contained
in it) may be further reproduced ---or transmitted or broadcast--- in any form or
by any means (electronic, mechanical, or other) without the writtem prmission
of the right holders: \copyright Erasmo Recami and Maria Majorana.}

\addcontentsline{toc}{chapter}{Ricordo di Ettore Majorana a
sessant'anni dalla sua scomparsa: L'opera scientifica edita e inedita}

\ \\[2mm]

\centerline{{\large{Erasmo RECAMI}}$^{(*)}$ \footnotetext{$^{(*)}$
Dip.to di Ingegneria, Univ. statale di Bergamo; \ I.N.F.N. -
Sezione di Milano, Italia.  \ Fax (39)-035-2052310; e-mail \
recami@mi.infn.it \ . }}

\vspace*{0.5 cm}

{\sc Riassunto} --- Ettore Majorana, probabilmente il pi\`u
brillante fisico teorico italiano di questo secolo (fu paragonato,
da Enrico Fermi, a Galileo e Newton), scomparve misteriosamente da
Napoli sessant'anni or sono, nel 1938, all'et\`a di trentun anni.
Nella prima parte di questo lavoro se ne tratteggiano la
personalit\`a scientifica (sulla base di lettere, documenti,
testimonianze da noi raccolti in circa vent'anni) e il significato
di alcune parti delle sue pubblicazioni.  Nella seconda porzione
di quest'articolo si presenta il preliminare Catalogo (redatto da
M.Baldo, R.Mignani e E.Recami) dei manoscritti scientifici inediti
lasciatici da E.Majorana e attualmente quasi tutti depositati
presso la ``Domus Galilaeana" di Pisa. \ Tutto il materiale che
segue \`e tratto dal nostro libro ``Il Caso Majorana: Epistolario,
Documenti, Testimonianze" (Mondadori, Milano, 1987,1991; Di Renzo,
Roma, 2000-2008), e si rimandano a tale libro i lettori
interessati a maggiori e pi\`u profonde informazioni; nonch\'e,
per questioni pi\`u tecniche, ai successivi volumi riproducenti ad
e-sempio parte dei manoscritti scientifici lasciati inediti da
Ettore Majorana: si veda, p.es., l'e-print
arXiv:0709.1183v1[physics.hist-ph]. \ [{\em Oss.\/}: Tutto il
materiale che segue \`e coperto da copyright, e non pu\`o essere
riprodotto senza il permesso scritto dell'autore, in solido con
gli altri detentori dei relativi diritti.]

\vspace*{1.6 cm}
%\markboth{\footnotesize \sl Erasmo Recami}
%{\sl \footnotesize Ricordo di Ettore Majorana a Sessant'Anni dalla sua
%Scomparsa $\ldots$}
% (L'Opera Scientifica Edita e Inedita)}
\thispagestyle{empty}
\noindent

\newpage

\section*{CAPITOLO I}

\

\section*{{\large {ETTORE MAJORANA: LO SCIENZIATO E L'UOMO}}}\setcounter{footnote}{0}

%\begin{quotation}
%\noindent
\hspace*{4cm}\hfill \begin{minipage}[t]{13cm}
{\em Chiss\`a quanti sono come me, nelle \\
mie stesse condizioni, fratelli miei. \\
Si lascia il cappello e la giacca, \\
con una lettera in tasca, sul parapetto \\
d'uno ponte, su un fiume; e poi, \\
invece di buttarsi gi\'u, si va via \\
tranquillamente: in America o altrove.} \\
L. Pirandello (Il fu Mattia Pascal, 1904).
\end{minipage}
%\end{quotation}

\vspace{1.5cm}

\section*{1. La Fama}

\subsection*{1.1. Genialit\`a}

\paragraph*{}

La fama di Ettore Majorana, ovvia per gli specialisti, pu\`o solidamente appoggiarsi
anche a  testimonianze come la seguente, dovuta alla memore penna di Giuseppe
Cocconi. Invitato da Edoardo Amaldi\footnote{Il primo storico di Ettore Majorana. Si vedano
di E. Amaldi: {\em ``La Vita e l'Opera di E. Majorana} (Accad. Naz. dei Lincei:
Roma (1966); ``Ricordo di Ettore Majorana'', {\em Giornale di
Fisica} {\bf 9} (1968) 300;
``Ettore Majorana: Man and scientist'', in {\em Strong and Weak Interactions},
a cura di A. Zichichi (New York, 1966); ``From the discovery of the neutron to the discovery
of nuclear fission'', {\em Phys. Reports} {\bf 111} (14) 1-322; ``I miei giorni
con Fermi'', in {\em La Repubblica, Supplemento al n. 285} del 31.12.86 (Roma).},
dal CERN gli scrive:

\begin{quotation}
\footnotesize{$<<$Ginevra, 1965 Luglio 18 --- Caro Amaldi, In una discussione che si ebbe
tempo fa sul libro {\em [poi edito dall'Accademia dei Lincei]} che stai scrivendo
su Ettore Majorana, ti dissi come io pure ebbi un tenue contatto con Majorana
poco prima della sua fine. Tu esprimesti allora il desiderio che ti descrivessi con
maggiore dettaglio il mio ricordo, e qui cerco di accontentarti.

Nel gennaio 1938, appena laureato, mi fu offerto, essenzialmente da te, di venire
a Roma per sei mesi nell'Istituto di Fisica dell'Universit\`a come assistente
incaricato, ed una volta l\'{\i} ebbi la fortuna di unirmi a Fermi, Bernardini
(che aveva avuto una Cattedra a Camerino pochi mesi prima) ed Ageno (lui
pure giovane laureato), nella ricerca dei prodotti di disintegrazione dei
``mesoni'' {\em mu} (allora chiamati mesotroni ed anche yukoni) prodotti
dai raggi cosmici. L'esistenza dei ``mesoni'' {\em mu} era stata proposta
circa un anno prima, ed il problema del loro decadimento era gi\`a molto attuale.

Fu proprio mentre mi trovavo con Fermi nella piccola officina del secondo piano,
intenti lui a lavorare al tornio un pezzo della camera di Wilson che doveva
servire a rivelare i mesoni in fine {\em range}, io a costruire un trabiccolo per
l'illuminazione della camera, utilizzante il flash prodotto dall'esplosione di
una fettuccia di alluminio cortocircuitata su una batteria, che Ettore Majorana
venne in cerca di Fermi. Gli fui presentato e scambiammo poche parole. Una faccia
scura. E fu tutto l\'{\i}. Un episodio dimenticabile se dopo poche settimane,
mentre ero ancora con Fermi nella medesima officinetta, non fosse arrivata la
notizia della scomparsa da Napoli del Majorana. Mi ricordo che Fermi si dette
da fare telefonando da varie parti sinch\'e, dopo alcuni giorni, si ebbe l'impressione
che non lo si sarebbe ritrovato pi\'u.

Fu allora che Fermi, cercando di farmi capire che cosa significasse tale perdita, si espresse
in modo alquanto insolito, lui che era cos\'{\i} serenamente severo quando si trattava
di giudicare il prossimo. Ed a questo punto vorrei ripetere le sue parole, cos\'{\i}
come da allora me le sento risuonare nella memoria: {\em ``Perch\'e, vede, al mondo
ci sono varie categorie di scienziati; gente di secondo e terzo rango, che fan
del loro meglio ma non vanno molto lontano. C'\`e anche gente di primo rango, che
arriva a scoperte di grande importanza, fondamentali per lo sviluppo della scienza}
(e qui ho netta l'impressione che in quella categoria volesse mettere se stesso).
{\em ``Ma poi ci sono i geni, come Galileo e Newton. Ebbene, Ettore era uno di quelli.
Majorana aveva quel che nessun altro mondo ha; sfortunatamente gli mancava quel che
invece \`e comune trovare negli altri uomini, il semplice buon senso''}.

Spero che queste mie righe ti dicano quanto desideravi. Cordiali saluti, \\}
\ \hfill {Giuseppe Cocconi$>>$}.
\end{quotation}

``Il semplice buon senso''; noi preferiremmo dire il {\em senso comune};
il quale non \`e detto sia sempre buono, o il migliore.

Enrico Fermi\footnote{Uno dei maggiori fisici della nostra epoca. Per quello
che ha fatto nel 1942 a Chicago (con la costruzione della prima ``pila atomica'')
il suo nome diverr\`a forse leggendario come quello di Prometeo.}  si espresse
in maniera insolita anche in un'altra occasione, il 27 luglio 1938
(dopo la scomparsa di Majorana, avvenuta il sabato 26 marzo 1938), scrivendo
da Roma al primo ministro Mussolini onde chiedere una intensificazione delle
ricerche di Ettore:
\begin{quotation}
\footnotesize{
$<<$ Io non esito a dichiararVi, e non lo dico quale espressione
iperbolica, che fra tutti gli studiosi italiani e stranieri che ho avuto occasione
di avvicinare il Majorana \`e fra tutti quello che per profondit\`a di ingegno
mi ha maggiormente colpito$>>$.}
\end{quotation}

E un testimone diretto, Bruno Pontecorvo, aggiunge: $<<$Qualche tempo dopo
l'ingresso nel gruppo di Fermi, Majorana possedeva gi\`a una erudizione
tale ed aveva raggiunto un tale livello di comprensione della fisica da
potere parlare con Fermi di problemi scientifici da pari a pari. Lo stesso
Fermi lo riteneva il pi\'u grande fisico teorico dei nostri tempi. Spesso
ne rimaneva stupito [...]. Ricordo esattamente queste parole di Fermi:
``Se un problema \`e gi\`a posto, nessuno al mondo lo pu\`o
risolvere meglio di Majorana".$>>$

Il mito della ``scomparsa'' ha contribuito a dare a Majorana, quindi,
null'altro che la
notoriet\`a  che gli spettava, per essere egli davvero un genio: e di una genialit\`a precorritrice
dei tempi. Anzi, cos\'{\i} come avviene quando \`e vera, la sua fama \`e cresciuta e cresce col
tempo, anche tra i colleghi. Da una decina d'anni \`e esplosa: e una elevata
percentuale di pubblicazioni scientifiche nel mondo (in alcuni settori della fisica
delle particelle elementari) contiene ora il suo nome {\em nel titolo}.

Enrico Fermi \`e stato forse uno degli ultimi ---e straordinari--- esempi di
grande teorico e contemporaneamente di grande
sperimentale. Majorana era invece un teorico puro, anzi (per dirla
con le stesse parole di Fermi, nel prosieguo del suo scritto a
Mussolini) Ettore aveva al massimo grado quel raro complesso di attitudini
che formano il fisico teorico {\em di gran classe}. Ettore
``portava'' la scienza, come ha detto Leonardo Sciascia: portava,
anzi, la fisica teorica. Non era inferiore a un Wigner [premio Nobel
1963] o a un Weyl: i quali, per il loro rigore fisico-matematico,
erano forse gli unici per i quali Ettore nutrisse ammirazione senza riserve.

Da un lato, quindi, non aveva alcuna propensione per le attivit\`a
sperimentali (neanche costretto, per intenderci, avrebbe mai potuto
recare contributi concreti a progetti come quello della costruzione
tecnologica della bomba atomica). Dall'altro lato, per\`o, sapeva
calarsi a profondit\`a insuperate nella sostanza dei fenomeni fisici,
leggendovi eleganti simmetrie e nuove potenti strutture matematiche,
o scoprendovi raffinate leggi. La sua acutezza lo portava a vedere al
di l\`a dei colleghi: ad essere cio\`e un pioniere. Perfino i suoi
appunti di studio ---redatti ciascuno in circa un anno, a partire dagli
inizi del 1927, cominciati cio\`e prima ancora del
passaggio dagli studi di ingegneria a quelli di
fisica--- sono un modello non solo di ordine, divisi come sono in
argomenti e persino muniti di indici, ma anche di originalit\`a,
scelta dell'essenziale, e sinteticit\`a. Tanto       che questi
quaderni, noti come {\em i Volumetti}, potrebbero essere riprodotti
fotograficamente e pubblicati cos\'{\i} come si trovano, in modo analogo
a quanto fece la Chicago University Press per gli appunti di meccanica
quantistica di Enrico   Fermi: e costituirebbero un ottimo testo
{\em moderno} (dopo oltre cinquant'anni!) di Istituzione di fisica teorica.

Ricordiamo che Majorana, passato a fisica agli inizi del '28, si
laure\`o\footnote{Tra i membri della commissione di laurea ricordiamo
Corbino, Fermi, Volterra, Levi-Civita, Lo Surdo, Armellini e
Trabacchi. Le votazioni riportate de Ettore durante gli studi erano
state: Geometria  analitica e proiettiva, Meccanica razionale, Fisica
superiore, Fisica matematica, Fisica terrestre: trenta su trenta e
lode; Analisi algebrica, Geometria descrittiva, Fisica sperimentale,
Esercizi di fisica: trenta; Chimica applicata: ventotto e mezzo;
Chimica generale, Analisi infinitesimale: ventisette.} con Fermi il 6
luglio 1929; e consegu\'{\i} la libera docenza in fisica teorica il 12
novembre 1932.

\subsection*{1.2. Da Galileo a Fermi}

\paragraph*{}
Per dare un'idea di cosa abbia significato per la cultura e la
scienza italiane l'attivit\`a romana di Fermi e del suo gruppo (con
questo senza dimenticare la contemporanea attivit\`a di altri gruppi,
{\em in primis} quello di Firenze), ricordiamo che la fisica italiana gi\`a
una volta aveva  conquistato una posizione di eccellenza a livello
internazionale: con Galileo. Ma la condanna da parte della Chiesa (22
luglio 1633) ---che, considerati i tempi, non ebbe in fondo
conseguenze molto gravi per Galileo--- risult\`o disastrosa per la
``scuola'' galileana, la quale avrebbe potuto continuare ad essere la
prima del mondo. Il vasto e promettente movimento scientifico creato
dal Galilei venne colpito alla radice dalla condanna del maestro;
cos\'{\i} che la scienza si trasfer\'{\i} al di l\`a delle Alpi. John
Milton, ricordando una visita fatta, a suo dire, al ``celebre
Galileo, oramai vegliardo e prigioniero dell'Inquisizione'' (Galileo
mor\'{\i} nel 1642), riassunse magistralmente la situazione''
annotando nel 1644 che
\begin{quotation}
\footnotesize{$<<$lo stato di servit\'u, in cui la
scienza era stata ridotta nella loro patria, era la cagione per cui lo
spirito italiano, cos\'{\i} vivo un giorno, si era ormai spento; e
da molti anni tutto ci\`o che si scriveva non era che adulazione e
banalit\`a$>>$.}
\end{quotation}

Devono poi passare quasi due secoli prima che si riveli un altro
{\em grande} fisico, Volta. Alessandro Volta genera un filone di
ricerche le quali portano alle applicazioni prevalentemente
tecnologiche di Antonio Pacinotti, Galileo Ferraris e Augusto Righi
e, pi\'u tardi, a quelle di Guglielmo Marconi. Ma non ne deriva una vera
``scuola'', tanto che alla fine del 1926, quando Fermi ottiene la cattedra di fisica
teorica a Roma, l'Italia certo non emerge nel mondo per la fisica.

\`E solo Fermi che, ben tre secoli dopo Galileo, riesce a generare di nuovo
un esteso e moderno movimento in seno alle scienze fisiche italiane. Ad esempio
l'articolo di Fermi che d\`a avvio alla teoria delle interazioni deboli (coronata
dopo cinquant'anni, nel 1983, dalle scoperte di Carlo Rubbia [premio Nobel
1984] e collaboratori) esce nel 1933: esattamente {\em trecento anni} dopo la condanna
definitiva della teoria galileana.

\subsection*{1.3. Il gruppo di Roma}

\paragraph*{}
La rinascita della fisica italiana non avrebbe avuto luogo, forse, senza
l'intervento di Orso Mario Corbino, siciliano, laureato in fisica a Palermo,
professore prima a Messina e poi a Roma, dal 1918 direttore dell'Istituto di fisica
di via Panisperna. Nel 1920 era stato nominato senatore e l'anno seguente Ministro
della pubblica istruzione. Ricorda Amaldi:

\begin{quotation}
\footnotesize{
$<<$Quando Fermi, nell'autunno 1926, iniziava a 25 anni la sua
attivit\`a di professore di Fisica Teorica a Roma, aveva gi\`a un nome internazionale
anche se non mancavano nelle diverse universit\`a italiane alcuni professori di vecchio
stile, che stentavano o addirittura non volevano riconoscere il suo valore.

L'intesa con O.M. Corbino era invece perfetta. Bisognava approfittare della
situazione e far nascere una scuola di fisica moderna. Come prima cosa Franco
Rasetti fu chiamato da Firenze a Roma, dove prese il posto di aiuto di Corbino,
lasciato libero da Persico che si era trasferito a Firenze.

L'abilit\`a sperimentale di Rasetti completava notevolmente quella di Fermi, che a quell'epoca rivolgeva
quasi tutte le sue energie alla fisica teorica. Circa due anni dopo Corbino, coadiuvato
da Fermi e da vari illustri e lungimiranti matematici [G. Castelnuovo, F. Enriques,
T. Levi-Civita,$\ldots$ ], riusc\'{\i} a creare  presso la Facolt\`a di Scienze di Roma
una cattedra di spettroscopia a cui fu chiamato Rasetti.

Creato cos\'{\i} un gruppo veramente notevole di docenti giovani, attivi in fisica
moderna e di grande valore, Corbino, Fermi e Rasetti si adoperarono per raccogliere alcuni
allievi. Corbino, per esempio, rivolse un appello agli studenti durante una lezione
dicendo che con la chiamata di Fermi a Roma e nella situazione di fermento di idee,
che esisteva ormai in tutta Europa nel campo della Fisica, si apriva, a suo
giudizio, un periodo del tutto eccezionale per i giovani che avevano gi\`a cominciato
a dare prova di essere sufficientemente dotati e che si sentivano disposti
ad intraprendere uno sforzo non comune di studio e di lavoro teorico e sperimentale. Fu
cos\'{\i} che negli anni successivi all'Istituto di Fisica, a via Panisperna 89A, si form\`o
una vera e propria scuola di fisica moderna.

Fra gli allievi teorici qui ricorder\`o, in ordine di ingresso in Istituto:
Ettore Majorana, Gian Carlo Wick, Giulio Racah, Giovanni Gentile Jr., Ugo Fano, Bruno
Ferretti e Piero Caldirola, l'ultimo dei quali giunse a Roma da Pavia poco prima
che Fermi lasciasse l'Italia nel dicembre del 1938.

Gli allievi nel campo sperimentale furono: Emilio Segr\'e, Edoardo Amaldi, Bruno
Pontecorvo, Eugenio Fubini, Mario Ageno e Giuseppe Cocconi, il quale giunse a Roma da Milano
circa un anno prima della partenza di Fermi$>>$.}
\end{quotation}

Nel 1923 il diciassettenne Majorana (che viveva a Roma dall'et\`a di otto o nove
anni in collegio, insieme con fratelli e cugini, al ``Convitto {\em Massimo} alle
Terme'' dei padri gesuiti: finch\'e nel 1921 l'intera famiglia non si trasfer\'{\i} a Roma
ed Ettore pass\`o da ``interno'' ad ``esterno'') si era iscritto al corso di laurea
in Ingegneria dell'universit\`a di Roma. Aveva come compagni il fratello Luciano,
Emilio Segr\'e, Gastone Piqu\'e, Enrico Volterra, Giovannino Gentile e
Giovanni Enriques. Gli ultimi tre erano figli, rispettivamente, del matematico Vito
Volterra (uno dei pochi professori universitari ---tra l'altro--- a rifiutare
il giuramento al regime fascista), del filosofo Giovanni Gentile, e di Federigo
Enriques, matematico ed epistemologo. Nel giugno del 1927 Corbino rivolge il suo
appello agli studenti; ed Amaldi, allora al termine del secondo anno, ne raccoglie
l'invito. Quasi simultaneamente Segr\'e conosce Rasetti, e quindi Fermi: e lui pure decide
di passare a Fisica. Qui egli inizia a parlare delle doti straordinarie di Ettore;
ed un giorno convince Majorana ad incontrare Fermi. Il passaggio di Ettore a Fisica ha luogo all'inizio
del suo quinto anno di studi universitari dopo un colloquio con Fermi narrato
nei particolari da Segr\'e, e di cui diremo.

Amaldi racconta:
\begin{quotation}
\footnotesize{$<<$Fu in quell'occasione che vidi Majorana per la prima volta.
Di lontano appariva smilzo, con un'andatura timida e quasi incerta; da vicino si
notavano i capelli nerissimi, la carnagione scura, le gote lievemente scavate, gli
occhi vivacissimi e scintillanti: nell'insieme l'aspetto di un saraceno$>>$.}
\end{quotation}

Pi\'u tardi si un\'{\i} al gruppo Bruno Pontecorvo; cos\'{\i} che questo risult\`o
formato essenzialmente da Fermi, Rasetti, Majorana, Amaldi, Segr\'e e
Pontecorvo, oltre al chimico Oscar D'Agostino.

Fermi ricever\`a il premio Nobel nel 1938. Del gruppo romano, solo Amaldi
rester\`a in patria, sobbarcandosi al peso maggiore per mantenere viva in
Italia la scuola creata da Fermi; e sar\`a uno dei
padri dei laboratori {\em europei} di Ginevra. Gli altri emigreranno:
Fermi e Segr\'e negli Stati Uniti (ove Segr\'e ricever\`a
il premio Nobel nel 1959 per la scoperta dell'antiprotone); Pontecorvo ---nel
settembre 1950, dopo un lungo periodo in USA, Canada, Inghilterra--- in Unione Sovietica
(ove, noto universalmente come Bruno Maksimovich,  diverr\`a presto Accademico
delle Scienze dell'URSS, nonch\'e uno dei direttori dei grandi laboratori di Dubna).
Rasetti, invece, dopo avere svolto con tanto successo il ruolo di principale fisico
sperimentale del gruppo romano, abbandona in seguito la fisica, divenendo  un
paleontologo di fama, e pi\'u tardi un botanico di gran valore: la sua personalit\`a
\`e cos\'{\i} straordinaria, che vogliamo riportare qui la lettera scrittaci da Waremme (Belgio) il
22.6.1984, all'et\`a di quasi ottantatre anni, in risposta alla nostra consueta
richiesta di ``testimonianza'':
\begin{quotation}
\footnotesize {$<<$Caro Professore, --- La prego di scusarmi se rispondo soltanto oggi alla Sua
 lettera del 30/4 scorso. Questa \'e arrivata durante la mia assenza per
un viaggio a scopo botanico durata dal 6/5 al 17/6; quindi ne ho preso visione
 solo da qualche giorno. Per il momento non ho il tempo di fare tutti i commenti
 che merita, quindi Le comunico soltanto le prime impressioni, riservandomi di scriverLe pi\'u a lungo
 pi\'u tardi.

Non potei partecipare al convegno ``fermiano''\footnote{Il convegno tenutosi il 26 aprile 1984
a Bologna in occasione del cinquantesimo anniversario della formulazione della
teoria di Fermi sul decadimento beta {\em (``Cinquant'anni di Fisica delle
 Interazioni Deboli'')}.} perch\'e soffrii di un malessere per un mese e mezzo, ci\`o
 che mi fece rinunciare a un viaggio a Palermo, per il quale avevo gi\`a acquistato il
 biglietto dell'aereo, viaggio che aveva come scopo di fotografare le orchidacee
 caratteristiche della Provincia di Palermo. Ne ho fatto un altro, come ho detto sopra,
 nell'Italia centro-meridionale. Come Lei sapr\`a, da molti anni ho perso
 ogni interesse per la fisica, dedicandomi con molta soddisfazione e successo prima
 alla paleontologia del Cambriano di Canada, Stati Uniti e Sardegna, poi alla
 botanica delle Alpi, e per ultimo alle Orchidacee italiane.

 Anche da una prima e sommaria visione dei Suoi scritti, comprendo che Lei
 ha  informazioni e conoscenze sulla vita di Ettore ben pi\'u complete
e profonde di quanto sia stato scritto da altri con ben pi\'u  immaginazione che seriet\`a.

$\ldots$Voglio segnalare anche un incredibile ``granchio'' preso dal Corriere della
 Sera {\em [del 13.12.83, pag. 16]} nella didascalia sotto alla figura che riproduce la famosa
 fotografia detta dei ``tre preti'' (perch\'e una amica di Laura Fermi, vedendone una
 copia appesa alla parete del salotto di questa, esclam\`o ``Chi sono quei tre preti?'').
 Dalla didascalia del Corriere sembrerebbe che la fotografia
rappresentasse Fermi, Rasetti e Segr\'e come laureandi! Invece Fermi
ed
io ci laureammo nel 1922, quando ancora non esisteva il regime fascista e i laureandi
non portavano la camicia nera! Si tratta invece di una fotografia fatta verso la
met\`a degli anni trenta, e tutti e tre eravamo membri di una commissione
di laurea, non laureandi. Se simili mostruosit\`a si trovano in un giornale che dovrebbe essere
serio, che cosa dobbiamo credere di quanto scrivono giornali meno rispettabili?

Con questo chiudo le mie osservazioni per il momento, e La prego di gradire
distinti e cordiali saluti. \\}
\hfill {Franco Rasetti}

\

\footnotesize{
P.S.: Mi dimenticavo di dirle che sarebbe vano aspettarsi da me informazioni su Majorana
che non siano gi\`a ben note. Tra i fisici di Via Panisperna sono certamente la persona
che lo ha meno conosciuto. Intanto io ero alquanto pi\'u anziano di lui, mentre Segr\'e
e Amaldi erano press'a poco coetanei. Inoltre con questi ero legato dalle attivit\`a
sportive, mentre Majorana era del tutto estraneo a qualsiasi sport. Con Segr\'e
e Amaldi giuocavamo a tennis, ma soprattutto eravamo compagni di alpinismo (per esempio,
abbiamo fatto varie ascensioni difficili, sempre senza guida, tra cui
la traversata del Cervino).$>>$}
\end{quotation}

Il medesimo concorso che assegn\`o a Fermi la cattedra romana ne attibu\'{\i}
una anche a Enrico Persico (a Firenze) e ad Aldo Pontremoli (a Milano), quest'ultimo
perito nella spedizione Nobile col dirigibile {\em Italia} al Polo Nord.
Persico ebbe a Firenze una influenza forte e benefica. Lasciamo parlare
Segr\'e\footnote{E. Segr\'e: {\em Enrico Fermi, Fisico} (Zanichelli; Bologna, 1971).}:
\begin{quotation}
\footnotesize{
$<<$Persico a Firenze era l'unico professore di ruolo della nuova generazione
e insegnava con grande successo. I giovani fisici che si trovavano a Firenze
e affrontavano problemi moderni erano: Bruno Rossi, famoso per gli
studi sui raggi cosmici; Giuseppe Occhialini, che collabor\`o in modo decisivo
alla scoperta sia degli sciami\footnote{Segr\'e si riferisce al fatto
che Occhialini collabor\`o attivamente
(insieme con l'inglese P. Blackett) alla scoperta della prima antiparticella,
il {\em positone}, e dei fenomeni di creazione e annichilazione di coppie
elettrone-positone.
Ma il premio Nobel fu attribuito (nel 1936) solo all'americano Anderson, che aveva rivelato
il positone con lieve anticipo. Con Blackett si rimedi\`o assegnandogli il Nobel in un'altra
occasione (nel 1948). In quanto alla scoperta del primo mesone, il {\em pione} (la particella
prevista teoricamente nel 1934 da H. Yukawa), essa fu effettuata nel 1947 da C. Lattes, G. Occhialini e
C. Powell: ma ancora una volta il premio Nobel fu dato solo all'inglese Powell (nel 1950).},
sia del pione; Gilberto Bernardini, pi\'u tardi direttore scientifico del CERN di Ginevra
e direttore della Scuola Normale Superiore di Pisa; e Giulio Racah, inventore
dei coefficienti che da lui prendono il nome $\ldots$ , che divenne pi\'u tardi Rettore dell'Universit\`a
di Gerusalemme. Anche questi fisici fiorentini, avevano da poco passati i vent'anni,
$\ldots$ erano pieni di entusiasmo, energia e idee originali. I gruppi di Firenze e Roma
 erano legati da sincera amicizia e spesso si scambiavano visite o seminari$>>$.}
\end{quotation}

\section*{2. Il Concorso a Cattedre del 1937}

\subsection*{2.1. Dall'Archivio Centrale dello Stato}

\paragraph*{}
 Dopo il concorso del 1926, in cui ottennero la cattedra Fermi, Persico e
Pontremoli, passarono
 altri dieci anni prima che si aprisse, nel 1937, un nuovo concorso per la fisica
 teorica, richiesto dall'universit\`a di Palermo per opera di Emilio Segr\'e.
 Le vicende di questo Concorso, e specialmente i suoi antecedenti,
hanno dato luogo nel 1975 ad una vivace polemica tra Leonardo Sciascia, Edoardo Amaldi,
 e altri (Segr\'e, Zichichi, e chi scrive).

 Qui ci limiteremo, secondo la nostra propensione, a riprodurre i documenti certi, esistenti
 presso l'Archivio Centrale dello Stato ({\em Serie} Direz. Gen. Istruzione Superiore;
 {\em Busta} Personali --- $II^{\underline{a}}$ serie;  {\em Fascicolo} Ettore Majorana):
 in nostro possesso, questa volta, grazie ad una collaborazione coi ticinesi fratelli Dubini,
 residenti a Colonia. I concorrenti furono numerosi, e molti di essi di elevato valore;
 soprattutto quattro: Ettore Majorana, Giulio Racah (ebreo, che successivamente
 passer\`a da Firenze in Israele fondandovi la fisica teorica), GianCarlo Wick
 (di madre torinese e nota antifascista), e Giovanni Gentile Jr. (come sappiamo
 figliolo dell'omonimo filosofo, gi\`a ministro ---come si direbbe
ora--- della Pubblica
 Istruzione), ideatore delle ``parastatistiche'' in meccanica quantica. La
 commissione giudicatrice era costituita da: Enrico Fermi (presidente), Antonio
 Carrelli, Orazio Lazzarino, Enrico Persico e Giovanni Polvani.

 Il verbale n.1 recita:

 \begin{quotation}
\footnotesize{$<<$La commissione nominata da S.E. il Ministro
dell'Educazione Nazionale, e
 formata dai Professori Carrelli Antonio, Fermi S.E. Enrico, Lazzarino Orazio, Persico Enrico, Polvani Giovanni si
 \`e riunita alle ore 16 del giorno 25 ottobre 1937-XV in un'aula dell'Istituto
 Fisico della R. Universit\`a di Roma. La commissione si \`e costituita nominando
 come Presidente S.E. Fermi, e come Segretario Carrelli.

 Dopo esauriente scambio di idee, la Commissione si trova unanime nel riconoscere
 la posizione scientifica assolutamente eccezionale del Prof. Majorana Ettore
 che \`e uno dei concorrenti. E pertanto la Commissione decide di inviare una lettera
 e una relazione a S.E. il Ministro per prospettargli l'opportunit\`a di nominare
 il Majorana professore di Fisica Teorica per alta e meritata fama in una
 Universit\`a del Regno, indipendentemente dal concorso chiesto dalla Universit\`a
 di Palermo. La Commissione, in attesa di ricevere istruzioni da S.E. il Ministro,
 si aggiorna fino a nuova convocazione.

 La seduta \`e tolta alle ore 19. Letto approvato e sottoscritto seduta
 stante.\hfill\break

\hfill E. Fermi, O. Lazzarino,

\hfill E. Persico, G. Polvani, A. Carrelli.$>>$}
\end{quotation}

\vspace{0.5cm}
\noindent

La lettera inviata lo stesso giorno a S.E. il Ministro, sulla quale il ministro Giuseppe Bottai
verg\`o a mano la parola ``Urgente'', ripete il contenuto del verbale, dichiarando il Prof.
Majorana Ettore avere tra i concorrenti una posizione scientifica nazionale e
internazionale di tale risonanza che
\begin{quotation}
\footnotesize{
$<<$la Commissione esita ad applicare a lui la
procedura  normale dei concorsi universitari$>>$.}
\end{quotation}
Tale lettera ha un allegato,
{\em Relazione sulla attivit\`a scientifica del Prof. Ettore Majorana};
firmata, come sempre, nell'ordine: Fermi, Lazzarino, Persico, Polvani e
Carrelli. Vediamola:
\begin{quotation}
\footnotesize{$<<$Prof. Majorana Ettore si \`e laureato in Fisica a Roma nel 1929. Fin dall'inizio
della sua carriera scientifica ha dimostrato una profondit\`a di pensiero ed
una genialit\`a di concezione da attirare su di lui la attenzione degli studiosi
di Fisica Teorica di tutto il mondo. Senza elencarne i lavori, tutti notevolissimi
per l'originalit\`a dei metodi impiegati e per l'importanza dei risultati raggiunti,
ci si limita qui alle seguenti segnalazioni:

Nelle teorie nucleari moderne il contributo portato da questo ricercatore con la introduzione
delle forze dette ``Forze di Majorana'' \`e universalmente riconosciuto, tra i
 pi\'u fondamentali, come quello che permette di comprendere teoricamente le ragioni
 della stabilit\`a dei nuclei. I lavori del Majorana servono oggi di base alle pi\'u
 importanti ricerche in questo campo.

 Nell'atomistica spetta al Majorana il merito di aver risolto, con semplici ed eleganti
 considerazioni  di simmetria, alcune tra le pi\'u intricate questioni sulla
 struttura degli spettri.

 In un recente lavoro infine ha escogitato un brillante metodo che permette di
 trattare in modo simmetrico l'elettrone positivo e negativo, eliminando finalmente
la necessit\`a di ricorrere all'ipotesi estremamente artificiosa ed insoddisfacente di una
carica elettrica infinitamente grande diffusa in tutto lo spazio,
questione che era stata invano affrontata da molti
altri studiosi$>>$}.
\end{quotation}

Uno dei lavori pi\'u importanti di Ettore, quello in cui introduce la sua
``equazione a infinite
componenti'' (di cui diciamo nel Par. 5.4), non \`e menzionato:
ancora non era stato capito.
\`E interessante notare, per\`o, che viene dato giusto rilievo alla sua teoria
simmetrica per l'elettrone e l'anti-elettrone (oggi in auge, per la sua
applicazione a neutrini e anti-neutrini); e a causa
della capacit\`a di eliminare l'ipotesi cosiddetta ``del {\em    mare} di Dirac''
[P.A.M. Dirac, premio Nobel 1933]: ipotesi che viene definita ``estremamente artificiosa
e insoddisfacente'', nonostante che essa dai pi\'u sia sempre stata accettata
in maniera acritica. E questo tocco di originalit\`a in un documento burocratico
\`e rallegrante; e l'argomento ci trova del tutto consenzienti.

Una volta attribuita la cattedra a Ettore ``fuori concorso'' ---applicando
una legge che era stata usata per dare una cattedra universitaria, appunto fuori concorso, a
Guglielmo Marconi [premio Nobel 1909]---, la commissione riprendeva i suoi lavori
giungendo all'unanimit\`a alla formazione della terna vincente: $1^{\underline{0}}$) Gian Carlo Wick;
$2^{\underline{0}}$) Giulio Racah; $3^{\underline{0}}$) Giovannino Gentile.
Wick and\`o a Palermo, Racah a Pisa,
e Gentile Jr. a Milano. Giovannino Gentile, grande amico di Ettore\footnote{Due
belle e interessanti lettere di Ettore a Giovannino Gentile ci sono recentemente
pervenute grazie al cortese interessamento di L. Sciascia, L. Canfora
e F. Valentini.}, scomparir\`a prematuramente nel 1942.

\subsection*{2.2. Gli atti ufficiali}

\paragraph*{}
Il 2 novembre 1937 il ministro Bottai emette il decreto di nomina a professore ordinario
di Ettore Majorana; il 4 dicembre il decreto \`e gi\`a registrato alla Corte dei Conti;
e quindi la nomina viene partecipata dal Ministero a Ettore, presso la sua abitazione
di viale Regina Margherita 37, in Roma. Lo vedremo pi\'u avanti.

\section*{3. La Famiglia}

\subsection*{3.1. Il nonno, capostipite della famiglia}

\paragraph*{}
L'{\em antenato} della famiglia di Ettore \`e il nonno, Salvatore Majorana
Calatabiano, nato a Militello Val di Catania la vigilia di Natale del 1825. Nato pressoch\'e
dal nulla, nel 1865 \`e professore ordinario all'universit\`a di Messina (e poco
dopo a Catania), e  l'anno successivo viene eletto deputato al parlamento. Nel primo
ministero della    Sinistra, Depretis gli affida il portafoglio di Agricoltura,
Industria e Commercio; e, dopo la crisi della fine del 1877, torna al suo posto
di ministro nel terzo governo Depretis. Ci pu\`o interessare qui una citazione;
convinto che le leggi economiche siano leggi naturali di indole matematica,
scrive:
\begin{quotation}
\footnotesize{$<<$ \`e lo sprezzo dei dettami scientifici ---che in
conclusione dovrebbero essere
nel campo delle cose legislative quello che nelle applicazioni
tecniche sono i teoremi della
fisica e del calcolo,--- codesto sprezzo, codesto divorzio
tra il pensiero e la pratica, tra la scienza e l'arte sociale, la causa potente del disagio
in cui la cosa pubblica si trova$>>$}.
\end{quotation}

\subsection*{3.2 Deputati, rettori, scienziati: gli zii paterni}

\paragraph*{}
Sposatosi in seconde nozze con Rosa Campisi, Salvatore ne aveva avuto sette
figli: Giuseppe, Angelo, Quirino, Dante, Fabio Massimo (il padre di Ettore), Elvira
ed Emilia: si veda l'albero genealogico della famiglia (Tabella I). Tre
dei figli arrivano ad essere deputati, nonch\'e rettori dell'universit\`a di Catania (s\'{\i}:
tre fratelli rettori della stessa universit\`a: rispettivamente negli
anni 1895-1898, 1911-1919
e 1944-1947). Dovendo scegliere, accenneremo pi\'u avanti soprattutto ad Angelo, il
quale sar\`a ministro delle Finanze, poi del Tesoro.

Quirino si laurea a 19 anni in Ingegneria e a 21 in Scienze fisiche e matematiche,
e diviene poi presidente della  Societ\`a Italiana di Fisica. Lasciamo parlare
il Dizionario enciclopedico Treccani (che, tra parentesi, ricorda vari Majorana:
{\em tutti} discendenti di Salvatore); di lui, ancora in vita, diceva:

\begin{quotation}
\footnotesize{$<<$Fisico
(n. Catania 1871), fratello di Angelo; direttore dell'Istituto superiore dei
telefoni e telegrafi dello stato (1904-1914), poi professore di fisica sperimentale
al Politecnico di Torino, e (dal 1921) a Bologna, dove succede ad A. Righi come direttore
dell'istituto fisico. \`E socio dell'Accademia dei Lincei$\ldots$ Ha compiuto importanti
ricerche di fisica sperimentale (sui raggi catodici, sull'effetto Volta, sui fenomeni fotoelettrici, sulla
costanza della velocit\`a della luce emessa da una sorgente in moto,
ecc.). Consegu\'{\i}
notevoli risultati nel campo delle telecomunicazioni, eseguendo numerose esperienze
di radiotelegrafia a grande distanza$\ldots$ e di telefonia ottica
con luce ordinaria, ultravioletta e infrarossa;
ide\`o, nel corso di queste esperienze, un  microfono$\ldots$ {\em (microfono    idraulico di M.)}.
Si \`e altres\'{\i} occupato attivamente
di varie questioni connesse con la teoria della Relativit\`a, e in particolar
modo dell'assorbimento della gravitazione nella materia$>>$.}
\end{quotation}

Aggiungiamo che fu
membro della Commissione che diede la cattedra a Fermi; scopr\'{\i} la birifrangenza
magnetica; scrisse il suo primo libro, sui Raggi X, a 25 anni (Roma, 1897),
e dedic\`o molte energie a mostrare falsa la teoria della Relativit\`a Speciale:
ma, essendo uno sperimentale provetto e rigoroso, non pot\'e che confermare la teoria
einsteiniana.  Muore nel 1957. In conclusione un validissimo scienziato;
ma ad Ettore non paragonabile.

Elvira ed Emilia compiono la loro educazione ed istruzione a Roma, e quivi si sposano (la
prima col Consigliere di Stato Savini-Nicci, la seconda con l'Avv. Domined\`o).

\subsection*{3.3. Il padre}

\paragraph*{}
Anche Fabio, il padre di Ettore (n. a Catania nel 1875, m. a Roma nel 1934), si
laurea giovanissimo ---a 19 anni--- in Ingegneria, e poi in Scienze fisiche e
matematiche. Sar\`a lui a educare culturalmente e scolasticamente il piccolo Ettore
(che fece le prime classi elementari in casa), fino agli otto o nove anni,
quando ---come sappiamo--- Ettore passa all'Istituto Parificato M. Massimo dei
padri gesuiti in Roma, onde terminarvi le scuole elementari e frequentare poi le
medie superiori. Ettore rester\`a sempre molto attaccato al padre; e
senz'altro ne sentir\`a profondamente la dipartita, nel 1934.

L'ing. Fabio Majorana fonda a Catania la prima impresa telefonica,
tanto che in citt\`a il suo nome diviene sinonimo di ``societ\`a
telefonica''.  Trasferitosi a Roma, nel 1928
\`e nominato Capo divisione e, qualche anno dopo, Ispettore generale del Ministero
delle comunicazioni. Si dedica pure all'ingegneria ed\'{\i}le: un recente testo
sul Liberty a Catania riproduce la casa di famiglia da lui costruita in Via Sei
Aprile.

Scrive Edoardo Amaldi:
\begin{quotation}
\footnotesize{
$<<$Dal matrimonio dell'Ing. Fabio con la Sig.ra Dorina
Corso (n. a Catania nel 1876, m. a Roma nel 1966), anch'essa di famiglia catanese,
nacquero cinque figli: Rosina, sposata pi\'u tardi con Werner Schultze; Salvatore, dottore in legge
e studioso di filosofia; Luciano, ingegnere civile, specializzato in costruzioni
aeronautiche ma che poi si dedic\`o alla progettazione e costruzione
di strumenti
per l'astronomia ottica; Ettore (n. a Catania il 5 agosto 1906); e, quinta e ultima,
Maria, musicista$\ldots >>$.}
\end{quotation}

Ma torniamo un momento agli zii Rettori e deputati di Ettore. Il pi\'u anziano,
Giuseppe (n.~a Catania nel 1863), di vasta cultura, dal punto di
vista scientifico
si occupa di economia statistica, pubblicando molti libri con la casa editrice
{\em Loescher} di Roma, tra cui ``Teoria del valore'' (1887), ``La Statistica e l'Economia
di Stato'', ``La legge del grande numero e l'Assicurazione'', ``Teoria della Statistica''
(tutti del 1889), ``Le leggi naturali dell'economia politica'' (1890), ``Principio
della popolazione'' (1891) e ``I dati statistici nella questione bancaria''
(1894): gli ultimi due, in  particolare, venendo lodati anche in Germania, Francia,
Croazia, Spagna. Pure successo all'estero ottiene un volume pubblicato dapprima
a Firenze (Barbera Ed., 1889) e quindi, in forma pi\'u completa, a Catania
col titolo ``Programma di Statistica Teorica e Applicata'' (1893). Autore pure di
alcune opere letterarie, pi\'u tardi le ripudia, adoperandosi a distruggere
tutte le copie che gli capitano per mano.

Il pi\'u giovane dei tre, Dante, avvocato, si dedica sotrattutto al diritto. Ricordiamo
per esempio i suoi volumi, pubblicati sempre per i tipi della {\em E. Loescher \& C.} di
Roma, ``La caccia e la sua legislazione'' (1898), ``La frazione di Comune nel
diritto amministrativo italiano'' (1899), ``La concezione giuridica delle scienze
dello stato'' (1899), ecc. Nel 1924 verr\`a eletto deputato.

Ma, di questi tre zii, quello che lascia pi\'u sconcertati \`e senza dubbio Angelo,
giurista e sociologo. Egli brilla precocissimo, ma presto si spegne. E la
parabola della sua vita lascia pensosi, quando la si paragoni ---per quanto
ne sappiamo--- a quella di Ettore.

\subsection*{3.4 Lo zio ministro}

\paragraph*{}
Angelo nasce a Catania, secondo dei sette figli di Salvatore, nel dicembre del
1865. Maturo a 12 anni, all'et\`a di 16 \`e dottore in Legge a Roma. Tra i 18
e i 20 anni d\`a alle stampe le sue prime opere. Ma ---conseguita la libera
docenza a 17 anni--- gi\`a \`e ``professore pareggiato'' all'universit\`a di Catania.
Nel 1886 si presenta a tre concorsi per le cattedre di Diritto costituzionale di Catania,
Messina e Pavia: e, non ancora maggiorenne, li vince tutti e tre. Passa cos\'{\i} titolare
a Catania, ove diverr\`a, a 29 anni, magnifico rettore. A 28 anni si affaccia
alla politica attiva. Di coloritura liberale, Giolitti gli affida nel 1904 il dicastero delle
Finanze. Due anni dopo \`e di nuovo ministro con Giolitti: questa volta del Tesoro.
Ma presto il suo organismo si consuma, esausto per l'imponente attivit\`a sostenuta. Si
spegne a Catania a soli 44 anni.

Abbiamo un poco indugiato sul precoce, geniale accendersi  e sul rapido morire dell'attivit\`a
dello ``zio Angelo''; e ne avevamo motivo. Ma non resistiamo alla tentazione
---questa volta senza pi\'u scuse (se non quelle che possono derivare dai nostri
rapporti con l'universit\`a italica)--- di riportare un brano di uno dei
tanti discorsi tenuti da Angelo Majorana alla Camera dei deputati: sui problemi, appunto,
dell'Universit\`a. Il titolo \`e, senza mezzi termini, {\em La questione degli
``spostati'', e la riforma dell'istruzione pubblica}. La data la diciamo dopo.
\begin{quotation}
\footnotesize{$<<$Permetta la Camera che io riferisca alcune delle cifre calcolate dal collega Ferraris in un suo
pregevole studio statistico; e che ci\`o faccia io, appartenente a quelle provincie
del Mezzogiorno per le quali le cifre medesime danno argomento a men liete considerazioni$\ldots$
Abbiamo gi\`a visto che noi riversiamo dalle nostre Universit\`a, ogni anno, sul
pubblico mercato della vita nazionale (mi si perdoni questa frase di stile
economico) ben 1070 laureati in giurisprudenza: ossia, pi\'u del doppio del necessario!
Parimenti, mentre il fabbisogno di medici e di chirurghi sarebbe appena di 500
all'anno, i laureati in medicina e chirurgia, secondo una media che d'altronde
va sempre crescendo, sono 928 all'anno!$\ldots$

Orbene: il numero degli iscritti nelle Universit\`a e Istituti superiori oscilla, fra
le diverse regioni e le diverse Facolt\`a, nel seguente modo. Nell'Italia settentrionale abbiamo
per la Facolt\`a di giurisprudenza una quota di iscritti che \`e valutata a 13,85
sovra 100.000 abitanti; nella meridionale continentale, una quota di 24,14;
nella Sicilia di 22,82. Per la Facolt\`a di medicina, poi, abbiamo, sempre
una quota di 17,73 nell'Alta Italia, di fronte ad una di 19,06 in Sicilia e di ben
27,26 nel Mezzogiorno continentale. {\em (Commenti)}$\ldots$

Aggiungo che, guardando agl'iscritti nelle Facolt\`a di scienze, e quindi anche nelle
Scuole di ingegneria, si manifesta un fenomeno formalmente inverso, ma identico nella
sostanza. Nelle provincie settentrionali, per ogni centomila abitanti abbiamo una quota
di 9,08, mentre in quelle meridionali la quota \`e di 5,39, ed in Sicilia \`e di 7,72$\ldots$

Noi ci chiediamo: ma tutti questi laureati al di l\`a del bisogno, tutti questi avvocati che non
trovano clienti, tutti questi medici che non trovano ammalati, tutti questi
professori che non trovano scuole n\'e salari, che cosa fanno in seno alla societ\`a
che pur li contiene? E dovr\`o dirlo io? Ma chi di noi, interrogando se stesso,
non potr\`a darvi una risposta? N\'e io vi dir\`o, onorevoli colleghi, che, se non
vi sono medici i quali inventano le malattie per poterle curare, vi sono purtroppo
avvocati, specie in materia civile, che inventano le cause per poterle difendere
{\em (Ilarit\`a)}$\ldots$ N\'e ripeter\`o come, per fatale conseguenza, sia moltiplicato il numero
dei concorrenti ai pubblici impieghi; e come, precisamente per ci\`o, e lo Stato
e i Comuni e le Provincie e tutte le pubbliche Amministrazioni sieno costrette
ad accrescere le loro competenze ed uffici e mansioni, con relativi        impiegati,
enormemente aggravando i loro bilanci e quindi offendendo, di rimbalzo, il bilancio
della nazione$\ldots$

Questo male non \`e esclusivo all'Italia, ma si trova anche in altre regioni. In
Germania son celebri il detto di Virchow, che difin\'{\i} le Universit\`a ``semenzaio
di spostati'', e l'altro dell'imperatore Guglielmo, che defin\'{\i} i frequentatori
delle scuole classiche ``candidati alla fame''.$>>$

{\em E pi\'u avanti, parlando contro la burocratizzazione centralistica degli
studi superiori, conclude:} $<<$Trovo completamente sbagliato il sistema di sottoporre
giovani che sono all'apogeo della loro forza, che ordinariamente hanno pi\'u di
vent'anni di et\`a, che sono cio\`e maggiorenni con pienezza di responsabilit\`a
e di capacit\`a giuridica, cos\'{\i} pei diritti civili come pei politici: trovo
sbagliato, dico, il fatto di volerli ancora sottoporre ad un regime cos\'{\i}
{\em vincolato} di esami, che tende a sopprimere in essi il senso della iniziativa, senza
eccitarli alle gare degli studi sani e fecondi$\ldots$ Dianzi vi ricordavo che
gran parte della nostra popolazione universitaria ha un miraggio: l'ufficio
pubblico; ha un ideale: il 27 del mese. Orbene, notate adesso quale singolare
riscontro psicologico corra tra il 27 del mese, che \`e l'ideale  futuro, e l'approvazione all'esame
che \`e l'ideale pi\'u prossimo! Dichiaro francamente che sono contrario agli esami
annuali; mentre sono favorevole a un ``esame di stato'' finale$\ldots$ \`E
necessario togliere l'umiliante tutela, inutile al bene, efficace all'incoraggiamento dell'ignavia,
che sui nostri giovani \`e stabilita dal presente regime di accentramento
universitario$\ldots >>$}
\end{quotation}

Pur senza entrare nel merito delle idee espressevi, riesce difficile credere che questo
discorso fu tenuto oltre cento anni or sono, nella seduta dell'11 marzo 1899.

\section*{4. L'uomo: Il Ricordo dei Colleghi}

\paragraph*{}
Dalla Scuola Normale di Pisa il 9.5.1984 Gilberto Bernardini scrive:
\begin{quotation}
\footnotesize{
$<<$Caro Recami:$\ldots$ Ti sono grato
di aver riesumato, con la copia della lettera\footnote{Lettera da Lipsia del 22.1.1933} di Ettore a la mamma,
un sepolto ricordo dei miei sporadici incontri con lui durante il periodo che trascorsi a Berlino
con Lise Meitner$\ldots$ A Bologna\footnote{Ved. nota $^4$.} ho parlato con Bruno
Pontecorvo e penso che lui pi\'u e meglio di me potrebbe scrivere come mi hai amichevolmente
chiesto ``una lettera di testimonianze intorno a Ettore Majorana''. Nel parlarne
con Bruno si sono ravvivate alcune reminiscenze e, fra queste, che io con Ettore
evitavo di parlare di fisica perch\'e quello che avrei potuto dirgli sarebbe
stato per lui insignificante. Come mi \`e poi successo con Pauli, credo che considerassi
pi\'u agevole per me e meno banale per lui comunicare per esempio come fosse da
rallegrarsi di essere nati dopo Michelangelo e Beethoven. --- Non ho ancora ricevuto
quanto hai pubblicato nel '75. Non so se con la sorella {\em [di Ettore]} Maria tu ne
stia predisponendo una nuova e diversa pubblicazione. Se cos\'{\i} fosse, mi permetterei
il suggerimento di prescindere dall'eccezionale ingegno di Ettore come fisico per
accentuare quanto di lui potesse rievocare la complessa spiritualit\`a umana, tanto pi\'u estesa
ed illuminata di quella sulla quale hanno fantasticato dei romanzieri. Questa, spiritualit\`a anche
emotivamente, \`e evidente quando si legga ``La vita e l'opera di Ettore Majorana''
scritta da Edoardo Amaldi e pubblicata dall'Accademia dei Lincei nel
1966$\ldots$ --- Gilberto Bernardini$>>$.}
\end{quotation}

Ettore era persona sensibilissima e introversa, ma profondamente buona.
La sua ritrosia e timidezza\footnote{Anche Hans Bethe [premio Nobel 1967], a Erice,
il 27.4.81 ci ha confermato di essere riuscito a scambiare qualche parola con Ettore
(durante il proprio soggiorno a Roma) solo due o tre  volte, data la sua grande
timidezza. Bethe ricorda, inoltre, di avere sentito Segr\'e parlare molto bene
di Majorana.}, e la sua difficolt\`a di contatto umano ---reso ancor pi\'u difficile
dalla sua stessa intelligenza---, non gli impedivano di essere sinceramente affettuoso. E la sua critica
severa si addolciva quando il giudizio riguardava gli amici.

Nella prima lettera a Carrelli (qui riportata in
nota\footnote{$<<$Napoli, 25 marzo 1938-XVI. --- Caro
Carrelli, Ho preso una decisione che era ormai inevitabile. Non vi \`e in essa un solo
granello di egoismo, ma mi rendo conto delle noie che la mia improvvisa
scomparsa potr\`a procurare a te e agli studenti. Anche per questo ti prego di
perdonarmi, ma sopra tutto per avere deluso tutta la fiducia, la sincera
 amicizia e la simpatia che mi hai dimostrato in questi mesi. Ti prego anche
 di ricordarmi a coloro che ho  imparato a conoscere e ad apprezzare nel tuo
 Istituto, particolarmente a Sciuti, dei quali tutti conserver\`o un caro
 ricordo almeno fino alle undici di questa sera, e possibilmente
anche dopo. ---
 E. Majorana$>>$.}), il giorno prima della scomparsa, Ettore sente perfino il
bisogno di ricordare
{\em particolarmente Sciuti}: Sciuti (ora professore alla facolt\`a
di Ingegneria di Roma) era stato uno dei suoi pochi studenti a Napoli.  Di
questi allievi Ettore si interessava, e ne era soddisfatto; il 2 marzo 1938,
nella sua ultima
lettera a Gentile, scrive: $<<$Sono
contento degli studenti, alcuni dei quali sembrano risoluti a prendere la
fisica sul serio$>>$. E quando prendeva in mano il gesso, la timidezza
scompariva ed Ettore si trasfigurava, mentre dalla sua mano uscivano con
facilit\`a intere, eleganti lavagne di simboli fisici e matematici. Ci\`o
\`e stato ricordato di fronte alla discreta telecamera di Bruno Russo, e
recentemente al nuovo Dipartimento di Fisica dell'Universit\`a di Napoli, da
un'altra sua allieva  [Gilda Senatore ved. Cennamo: allora giovane e valorosa
(e bella) studentessa di fisica]: ma era facile immaginarselo.

L'intera serie degli appunti autografi di lezione redatti da
Majorana\footnote{Probabilmente Ettore
stava pensando di scrivere un libro per gli studenti, cos\'{\i} \ come
aveva pensato ad un libro nello stendere i suoi {\em Volumetti} inediti
di fisica teorica.} sembra fu consegnata all'allieva Gilda Senatore e (essendo
intermediari Cennamo, Carrelli e Gilberto Bernardini) fin\'{\i} nelle mani
di Edoardo Amaldi, probabilmente soltanto in parte, e quindi negli archivi
della ``Domus Galilaeana". La parte cos\'{\i} sopravvissuta \`e stata
finalmente pubblicata per interessamento di Bruno Preziosi e della S.I.F.
(Bibliopolis; Napoli, 1987). Gli appunti per la prolusione al corso
---la lezione inaugurale---
sono stati invece rinvenuti da noi, e li trascriviamo pi\'u sotto.

Tutti sanno dell'eccezionale spirito critico, e autocritico, di Ettore.
 Ma pochi sanno che, almeno fino al 1933 (anno in cui Ettore trascorse vari mesi
 a Lipsia, presso Werner Heisenberg) Ettore era di carattere allegro. La sorella
 Maria ne ricordava
 soprattutto le barzellette, le risate, il gioco alla palla fatto nel corridoio
 di casa. Da ragazzo, con amici, si mise al volante della macchina
di famiglia ---una delle prime auto---, ovviamente senza saper guidare, e altrettanto ovviamente
 fin\'{\i} contro un muro, serbandone ricordo con una lunga cicatrice su una mano. Anche
 l'Ing. Enrico Volterra (gi\`a professore presso l'Universit\`a del Texas ad
 Austin; ora scomparso) ci ha  ricordato il gran tempo trascorso con Ettore al
 bar ``Il Faraglino''$^4$ di Roma, o le chiacchierate e discussioni
culturali alla ``Casina delle Rose'' di Villa Borghese.

Ettore era poi ricchissimo di {\em humor}, cosa abbondantemente confermata da
tanti episodi anedd\`otici e dal suo epistolario, che abbiamo pubblicato  altrove
per i tipi della Di Renzo (Roma; genn.2001: terza ediz.).
Leggiamo ad esempio un brano della
sua lettera del 22.1.33 dall'Insititut f\"{u}r theoretische Physik di Lipsia:
\begin{quotation}
\footnotesize{
$<<$ Cara mamma: $\ldots$Io mi trovo benissimo$\ldots$ Nevica spesso dolcemente$\ldots$
L'Istituto di Fisica con molti altri affini \`e posto in posizione {\em ridente}, un po' fuori
di mano tra il cimitero e il manicomio$\ldots$ La situazione politica interna
appare permanentemente catastrofica, ma non mi sembra che interessi molto
la gente. Ho notato in treno la rigidit\`a di un ufficiale della Reichswehr solo
nello scompartimento con me, a cui non riusciva di deporre un oggetto sulla rete e quasi
di fare il minimo movimento senza sbattere insieme con forza i talloni.  Tale rigidit\`a era evidentemente determinata
dalla mia presenza, e in realt\`a sembra che la cortesia squisita ma sostenuta verso
gli stranieri faccia parte dello spirito soldatesco prussiano, poich\'e, mentre egli si sarebbe
sentito disonorato se non si fosse precipitato per accendermi una
sigaretta, d'altra parte il suo contegno mi ha impedito di scambiare
con lui una sola parola all'infuori dei cortesissimi saluti di rigore$>>$.}
\end{quotation}

L'esperienza in Germania modifica le opinioni di Ettore circa il
fascismo e l'incipiente nazismo; probabilmente anche per l'effetto
che gli fa il riuscire a vivere da solo (e forse non tanto male)
nella bene organizzata e accogliente ---``cortesissima e simpatica''---
citt\`a tedesca di {\em Leipzig}. Tanto che, da un lato, \`e
stato detto che Ettore nutrir\`a poi delle simpatie per il nazismo.
Dall'altro lato, per\`o, abbiamo la testimonianza di un altro grande
fisico, Rudolf Peierls, il quale dichiara che verso la fine del 1932
(cio\`e prima di partire per la Germania, per lo meno) era
profondamente antifascista. Peierls, infatti, il 2.7.84 scrive da
Oxford a Donatello Dubini.

\begin{quotation}
\footnotesize{
$<<$Sehr geehrter Herr Dubini, --- Ihr Brief vom 4.VI, der nach
Cambridge addresiert war, hat mich jetzt erreicht.

Ich war mit Ettore Majorana in Fermis Institut in Rom zusammen, im
Winter 1932-3. {\em Er erschien mir als ein ausserordentlich
begabter Physiker, etwas sch\"{u}chtern, {\em und dem Faschismus sehr
entgegengesetzt}.} Das was bevor er seine ber\"{u}hmten Arbeiten
\"{u}ber Kernkr\"{a}fte und \"{u}ber Neutrinos schrieb. Ich habe
diese Arbeiten nat\"{u}rlich mit grossem Interesse verfolgt, aber ich
habe ihn nicht mehr gesehen, soweit ich mich erinnere. \"{U}ber sein
Verschwinden habe ich wohl schon in 1938 geh\"{o}rt, aber ich weiss
nicht mehr wann oder von wem. Nat\"{u}rlich hat dass allen Physikern
sehr leidgetan, aber wir wussten zu wenige Einzelheiten, um \"{u}ber die
Ursachen zu spekulieren.[...]

Mit besten Gr\"{u}ssen --- Ihr --- Rudolf Peierls$>>$,}
\end{quotation}
dove la frase
da noi messa in corsivo significa ``Mi apparve come un fisico
straordinariamente dotato, un poco timido, e veramente contrario al
fascismo''. Oltre alla presente testimonianza, al riguardo si possono
ricordare le assennate, non sospette ---e per noi ``conclusive''---
osservazioni di Sciascia\footnote{L. Sciascia: {\em La scomparsa di Majorana}
(Einaudi; Torino, 1975), pp.42-43.}.

\section*{5. Lo Scienziato}

\subsection*{5.1. Introduzione}

\paragraph*{}
Nella lettera scritta il 27 luglio 1938 a Mussolini, e che gi\`a
abbiamo incontrato, Fermi ---dopo avere dichiarato che fra tutti gli
scienziati il Majorana era quello che pi\'u al mondo l'avesse colpito---
aggiunge:
\begin{quotation}
\footnotesize{
$<<$Capace nello stesso tempo di svolgere ardite
ipotesi e di criticare l'opera sua e degli altri; calcolatore
espertissimo e matematico profondo che mai per altro perde di vista
sotto il velo delle cifre e degli algoritmi l'essenza reale del
problema fisico, Ettore Majorana ha al massimo grado quel raro
complesso di attitudini che formano il fisico teorico di gran classe.
Ed invero, nei pochi anni in cui si \`e svolta fino ad ora la sua
attivit\`a, egli ha saputo imporsi all'attenzione degli studiosi
di tutto il mondo, che hanno riconosciuto in lui uno dei forti ingegni
del nostro tempo. E le successive notizie della sua scomparsa hanno
costernato quanti vedono in lui chi potr\`a ancora molto aggiungere
al prestigio della Scienza Italiana$>>$.}
\end{quotation}
Altre volte, come gi\`a
sappiamo, Fermi aveva paragonato Ettore a Galileo e Newton;
considerandolo a tutti, e a se stesso, superiore. Tanto che (come ci
ha raccontato Piero Caldirola) una volta Bruno Pontecorvo
rimprover\`o Fermi di ``umiliarsi'' troppo di fronte ad Ettore.

I dettagli del primo incontro di Majorana con Fermi ci illuminano
circa alcuni aspetti, scientifici e no, di Ettore. Essi sono noti da
quando li ha narrati Segr\'e; ma vale la pena di rileggerli con
attenzione\footnote{Ved. nota $^{(5)}$.}:
\begin{quotation}
\footnotesize{
$<<$Il primo lavoro
importante scritto da Fermi a Roma {\em [su alcune propriet\`a
statistiche dell'atomo]}$\ldots$ \`e oggi noto come metodo di
Thomas-Fermi $\ldots$Quando Fermi trov\`o che per procedere gli
occorreva la soluzione di un'equazione differenziale non lineare,
caratterizzata da condizioni al contorno insolite, con la sua
abituale energia in una settimana di assiduo lavoro calcol\`o la
soluzione con una piccola calcolatrice a mano. Majorana, che era
entrato da poco in Istituto e che era sempre molto scettico, decise
che probabilmente la soluzione numerica di Fermi era sbagliata e che
sarebbe stato meglio verificarla. And\`o a casa, trasform\`o
l'equazione originale di Fermi in una equazione del tipo di Riccati e
la risolse senza l'aiuto di nessuna calcolatrice, servendosi della
sua straordinaria attitudine al calcolo numerico (egli avrebbe potuto
facilmente diventare un numero da teatro di variet\`a esibendo
la facilit\`a con cui eseguiva a memoria le operazioni aritmetiche
pi\'u complicate). Quando torn\`o in Istituto confront\`o con aria
scettica il pezzetto di carta, su cui aveva riportato i dati
ottenuti, col quaderno di Fermi, e quando trov\`o che i risultati
coincidevano esattamente non pot\'e nascondere la sua meraviglia$>>$.}
\end{quotation}

Ettore, invece, invitato a redigere un curriculum\footnote{Scritto su
insistenza di Fermi, che gli fece avere dal CNR una sovvenzione onde
si recasse per circa sei mesi (a partire dal gennaio 1933) a Lipsia e
Copenaghen.}, scriver\`a di s\'e con l'usuale modestia (e ci\`o nel
maggio 1932, quando gi\`a aveva completato o in corso i suoi lavori
pi\'u importanti):

\begin{quotation}
\footnotesize{
$<<${\em Notizie sulla carriera didattica} --- Sono
nato a Catania il 5 agosto 1906. Ho seguito gli studi classici
conseguendo la licenza liceale nel 1923; ho poi atteso regolarmente
agli studi di ingegneria in Roma fino alla soglia dell'ultimo anno.
--- Nel 1928, desiderando occuparmi di scienza pura, ho chiesto e
ottenuto il passaggio alla Facolt\`a di Fisica e nel 1929 mi sono
laureato in Fisica Teorica sotto la direzione di S.E. Enrico Fermi
svolgendo la tesi ``La teoria quantistica dei nuclei
radioattivi''\footnote{Le tesine orali avevano per titolo: 1) Su un
effetto fotoelettrico constatato negli {\em audion}; 2) Sulle configurazioni
di equilibrio di un fluido rotante; 3) Sulle correlazioni
statistiche.} e ottenendo i pieni voti e la lode. --- Negli anni
successivi ho frequentato liberamente l'Istituto di Fisica di Roma
seguendo\footnote{Scrive {\em seguendo}, non {\em seguendone}. Si
noti inoltre che il titolo di S.E. spettava a Fermi quale membro
dell'Accademia d'Italia.} il movimento scientifico e attendendo a
ricerche teoriche di varia indole. Ininterrottamente mi sono giovato
della guida sapiente e animatrice di S.E. il professor Enrico
Fermi$>>$.}
\end{quotation}
 ``Liberamente'' significa {\em volontariamente}: cio\`e
senza ricevere una lira di stipendio (e cosi sar\`a fino al Novembre
1937).

\subsection*{5.2. L'opera scientifica}

``En science, nous devons nous int\'eresser aux choses, non aux
personnes'', ebbe a {\em dover dire} la polacca Marya Sklodowska in
Curie [Madame Curie: premio Nobel 1903-Fisica e 1911-Chimica].

Ettore scrisse pochi articoli scientifici: nove; oltre allo scritto
semi-divulgativo ``Il valore delle leggi statistiche nella fisica e
nelle scienze sociali'', pubblicato postumo su {\em Scientia}
[{\bf 36} (1942) 55-66] a cura di G. Gentile. Si ricordi che
Majorana pass\`o da ingegneria a fisica nel 1928 (anno in cui
pubblic\`o gi\`a un articolo, il primo: scritto insieme con l'amico
Gentile), e poi si dedic\`o alla pubblicistica in fisica teorica solo per
pochissimi anni, in pratica fino al 1933.

Ma Ettore ci ha lasciato anche vari manoscritti scientifici
inediti, pure depositati presso la ``Domus Galilaeana''; di cui
abbiamo redatto un catalogo in collaborazione
con M. Baldo e R. Mignani. L'analisi di questi manoscritti permette
di rilevare: \ 1) come Ettore fosse estremamente diligente e preciso
nel lavoro. Tutte le sue scoperte risultano precedute da una
indefessa serie di calcoli, fatti e rifatti: anche per i pi\'u
dotati, naturalmente, la scienza non pu\`o essere solo un semplice
gioco di intuizioni, come invece la leggenda aveva voluto farci
credere; \ 2) che fra il materiale inedito molti spunti hanno ancora
interesse scientifico {\em attuale} (insieme coi colleghi citati,
abbiamo operato una selezione: alcune centinaia di pagine [trasmesse
in copia anche al {\em Center for History of Physics} dell'A.I.P.,
New York, e relativa ``Niels Bohr Library''] possono ancora essere
utili per la ricerca contemporanea; ma solo poche pagine sono state
da noi interpretate e pubblicate\footnote{M. Baldo, R. Mignani \& E.
Recami: ``About a Dirac--like equation for the photon, according to
Ettore Majorana'', {\em Lett. Nuovo Cimento} {\bf 11} (1974) 568,
interessante pure ai fini di una possibile interpretazione fisica
della funzione d'onda del fotone. Ved. anche E.Giannetto, {\em Lett.
Nuovo Cimento}  {\bf 44} (1985) 140 e 145; e S.Esposito, {\em Found.
Phys.} {\bf 28} (1998) 231}; \ 3) che tutto il materiale
esistente {\em sembra} scritto entro il 1933 (anche la  bozza
dell'ultimo articolo, sulla ``Teoria simmetrica dell'elettrone e del
positrone'', che Ettore pubblicher\`a alle soglie del concorso a
cattedra nel 1937, pare fosse gi\`a pronta dal 1933, anno in cui si
ha la conferma della scoperta ---appunto--- del positone); \ 4) che quasi
nulla ci \`e noto di ci\`o che egli fece negli anni a seguire (1934--1938).
A parte una lunga serie di 34 lettere di risposta, scritte da Ettore in
quegli anni (precisamente dal 17.3.31
fino al 16.11.37) allo zio Quirino, il quale lo sollecitava a fornire una
spiegazione teorica dei risultati dei propri esperimenti.  Queste lettere,
pervenute a suo tempo a Franco Bassani e a noi per concessione di Silvia
Quirino Toniolo, sono di carattere essenzialmente tecnico: tanto che
ne abbiamo pubblicato altrove solo una piccola parte; ma esse
mostrano in tal modo che anche negli ultimi anni Ettore, almeno per
amore dello zio,  ben sapeva tornare
alla fisica, mostrando di possedere sempre le sue doti di eccelso
teorico.

Invero la sorella Maria ricordava che anche in quegli anni Ettore ---il
quale aveva diradato sempre pi\'u le sue visite all'Istituto, a cominciare dalla
fine del 1933, cio\`e dal suo rientro da Lipsia--- continu\`o a studiare e
lavorare a casa parecchie ore al giorno; e la notte. Si diede Ettore solo a
studi di letteratura e filosofia (amava particolarmente Pirandello,
Schopenhauer e Shakespeare), o di ``teoria dei giochi" e strategia navale
(sua passione fin dall'infanzia), nonch\'e di economia, di politica e infine
di medicina; oppure continu\`o a dedicarsi anche alla Fisica?  Dalla lettera
a Quirino  del 16.1.1936 ci viene ora una risposta; perch\'e veniamo a
sapere che Ettore si occupava
``da qualche tempo di elettrodinamica quantistica".  Conoscendo la modestia
di Ettore nell'esprimersi, ci\`o significa che durante l'anno 1935 Majorana
si era dedicato a fondo a ricerche originali nel settore  ---per lo meno---
della elettrodinamica quantistica. \ E ancora nel 1938, a Napoli, Carrelli
avr\`a l'impressione che Ettore stesse lavorando a qualcosa di importante,
di cui non voleva parlare. \ Altri lumi ci giungono, indirettamente, dalle
importanti lettere scritte al C.N.R. da Lipsia, e di cui diremo nel
Paragrafo 5.7.

\subsection*{5.3. Le prime pubblicazioni}

\paragraph*{}
Torniamo agli articoli pubblicati. I primi, redatti tra il 1928 e il
1931, riguardano problemi di fisica atomica e molecolare: per lo
pi\'u questioni di spettroscopia atomica o di legame chimico (sempre,
s'intende, nell'ambito della meccanica quantistica). Come scrive E.
Amaldi, un esame approfondito di questi lavori lascia colpiti per la
loro alta classe: essi rivelano sia una profonda conoscenza dei dati
sperimentali anche nei pi\'u minuti dettagli, sia una disinvoltura
non comune, soprattutto a quell'epoca, nello sfruttare le propriet\`a
di simmetria degli ``stati quantistici'' per semplificare
qualitativamente i problemi e per scegliere la via pi\'u opportuna
per la risoluzione quantitativa.  Tra questi primi articoli ne
scegliamo un solo:

``Atomi orientati in campo magnetico variabile'' apparso sulla
rivista {\em Nuovo Cimento}, vol. {\bf 9} (1932) pp.43-50. \`E
l'articolo, famoso tra i fisici atomici, in cui viene introdotto
l'effetto ora noto come Effetto Majorana-Brossel. In esso Ettore
prevede e calcola la modificazione della forma delle righe spettrali
dovuta a un campo magnetico oscillante; e ci\`o in connessione a un
esperimento tentato a Firenze qualche anno prima (bench\'e senza
successo) da G. Bernardini ed E. Fermi. Questo lavoro \`e rimasto
anche un classico della trattazione dei processi di ribaltamento
``non adiabatico'' dello {\em spin} (o ``spin-flip''). I suoi
risultati ---una volta estesi, come suggerito dallo stesso Majorana,
da Rabi nel 1937 e quindi, nel 1945, da Bloch e Rabi (i quali,
entrambi premi Nobel [Rabi: 1944; Bloch: 1952], contribuirono a
diffondere  quanto trovato da Ettore tredici anni prima)--- hanno
costituito la base teorica del metodo sperimentale usato per
ribaltare anche lo {\em spin} dei neutroni con un campo a
radiofrequenza: metodo impiegato ancor oggi, ad esempio, in tutti gli
spettrometri a neutroni polarizzati.

In questo articolo viene introdotta anche la cosiddetta ``Sfera di
Majorana" (per rappresentare spinori mediante set di punti di una
superficie sferica), di cui ha parlato entusiasticamente ---per esempio---
Roger Penrose nei suoi ultimi libri semi-divulgativi (si vedano in
Bibliografia le citazioni di Penrose e Zimba \& Penrose, e quelle pi\'u
recenti di Corrado Leonardi et al.).

Gli ultimi tre articolo di Ettore sono tutti di tale importanza che
nessuno di essi pu\`o restare senza commento.

\subsection*{5.4. L'equazione a infinite componenti}

\paragraph*{}
L'articolo ``Teoria relativistica di particelle con momento
intrinseco arbitrario'' {\em Nuovo Cimento}, vol. {\bf 9} (1932) pp.
335-344) \`e il tipico esempio di lavoro che precorre talmente i
tempi da venire compreso e valutato a fondo solo molti anni dopo.

A quel tempo era opinione comune che si potessero scrivere equazioni
quantistiche compatibili con la Relativit\`a (cio\`e
``relativisticamente invarianti'') solo nel caso di particelle a spin
zero o un mezzo. Convinto del contrario, Ettore comincia a costruire
opportune equazioni quanto-relativistiche per i successivi valori
possibili per lo spin (uno, tre mezzi, ecc.); finch\'e scopre che si
pu\`o scrivere un'{\em unica} equazione rappresentante una serie
infinita di casi, cio\`e un'intera famiglia infinita di particelle a
spin qualsiasi (si ricordi che allora le particelle note ---che ora
sono centinaia--- si contavano sulle dita di una mano!). Tralascia
allora tutti i singoli casi studiati ---senza pi\'u pubblicarli--- e
si dedica solo a queste equazioni ``a infinite componenti'', senza
trascurare l'osservazione che esse possono descrivere non solo
particelle ordinarie ma anche tachioni.

Per realizzare questo programma inventa una tecnica per la
``rappresentazione di un gruppo'' vari anni prima della ``scoperta''
di queste tecniche  da parte di Eugene Wigner (premio Nobel 1963).
Pi\'u ancora, Majorana ricorre per la prima volta ---inventandole---
alle rappresentazioni unitarie del Gruppo di Lorentz {\em a infinite
dimensioni}: rappresentazioni riscoperte da Wigner in lavori del 1939
e 1948. Per comprendere l'importanza di quest'ultimo aspetto,
rifacciamoci a quanto Ettore stesso ---pur tanto schivo--- riferisce
a suo padre da Lipsia il 18 febbraio 1933:
\begin{quotation}
\footnotesize{
$<<$Nell'ultimo mio
articolo apparso sul ``Nuovo Cimento'' \`e contenuta una importante
scoperta matematica, come ho potuto accertarmi mediante un colloquio
col professor van der Waerden, olandese che insegna qui, una delle
maggiori autorit\`a in teoria dei gruppi$>>$.}
\end{quotation}

Questa teoria \`e stata reinventata da matematici sovietici (in
particolare Gelfand e collaboratori) in una serie di articoli del
1948-1958, e finalmente applicata dai fisici in anni ancora pi\'u
tardi. L'articolo iniziale di Ettore, anzi, rimarr\`a in ombra per
ben 34 anni, cio\`e fino a quando Amaldi lo traduce e segnala al
fisico americano D. Fradkin, il quale a sua volta strabilia i teorici
delle alte energie rendendo finalmente di pubblico dominio (nel
1966)\footnote{D. Fradkin: {\em American Journal of Physics} {\bf 34}
(1966) 314.} quanto compiuto da Majorana tanti anni prima. Dalla data
del 1966, la fama di Ettore comincia a crescere costantemente anche
tra i fisici delle particelle fondamentali.

\subsection*{5.5. Le forze di scambio}

\paragraph*{}
Non appena, al sorgere del 1932, giunge a Roma notizia degli
esperimenti dei Joliot-Curie [premi Nobel 1935 per la chimica],
Ettore comprende che essi avevano scoperto il ``protone neutro''
senza accorgersene. Prima ancora, quindi, che ci fosse l'annuncio
ufficiale della scoperta del {\em neutrone}, effettuata poco dopo da
Chadwick [premio Nobel 1935 per la fisica], Majorana \`e in grado
di spiegare la struttura e la stabilit\`a dei nuclei atomici
mediante protoni e neutroni. (I suoi manoscritti inediti ci dicono
che egli si era gi\`a cimentato su questo problema ricorrendo,
invano, a protoni ed elettroni: che erano le uniche particelle in
precedenza note). Ettore precorse cos\'{\i} anche il lavoro
pionieristico di D. Ivanenko. Ma non volle pubblicarne nulla, n\'e
permise a Fermi di parlarne a Parigi agli inizi di luglio: ci\`o \`e
narrato da Segr\'e e da Amaldi. I suoi colleghi ricordano che gi\`a
prima di Pasqua era giunto alle conclusioni pi\'u importanti della
sua teoria: che protoni e neutroni fossero legati da forze
quantistiche originate semplicemente dalla loro {\em
indistinguibilit\`a}; cio\`e da ``forze di {\em scambio}'' delle
rispettive posizioni spaziali (e non anche degli spin, come invece
far\`a Heisenberg), cos\'{\i} da ottenere la particella alfa (e non
il deutone) quale sistema saturato rispetto alla  energia di legame.

Solo dopo che Heisenberg pubblica il proprio articolo sullo stesso
argomento, Fermi riesce a indurre Majorana a recarsi a Lipsia
presso il grande collega. E, finalmente, Heisenberg sa convincere
Ettore a pubblicare (anche se tanto in ritardo) i propri risultati:
``\"{U}ber die Kerntheorie'', lavoro apparso il 3 marzo 1933 su {\em
Zeitschrift f\"{u}r Physik}, vol. {\bf 82} (1933) pp.137-145.

Le forze ``di scambio'' nucleari sono ora chiamate forze di
Heisenberg-Majorana. Ettore ne parla al padre, con grande modestia,
nella stessa lettera prima citata (del 18.2.1933):
\begin{quotation}
\footnotesize{$<<$Ho
scritto un articolo sulla struttura dei nuclei che a Heisenberg \`e
piaciuto molto bench\'e contenesse alcune correzioni a una sua
teoria$>>$. Sempre su questo lavoro scrive pochi giorni dopo, il 22
febbraio, alla madre: $<<$Nell'ultimo ``colloquio'', riunione
settimanale a cui partecipano un centinaio tra fisici, matematici,
chimici, etc., Heisenberg ha parlato della teoria dei nuclei e mi ha
fatto molta r\'eclame a proposito di un lavoro che ho scritto qui.
Siamo diventati abbastanza amici$\ldots >>$.}
\end{quotation}

Probabilmente la pubblicazione sulla stabilit\`a dei nuclei venne
subito riconosciuta dalla comunit\`a scientifica (in particolare dai
fisici nucleari) ---evento raro, come sappiamo, per gli scritti di
Ettore--- anche grazie a questa opportuna ``propaganda'' fattane da
Heisenberg, che proprio pochi mesi dopo ricever\`a il premio Nobel.

L'avversione a pubblicare le proprie scoperte, quando esse fossero
risultate, all'esame del suo senso ipercritico, o di carattere non
abbastanza generale o espresse in forma matematica non abbastanza
stringente ed elegante, divenne per Ettore anche motivo di vezzo.
Racconta Amaldi:
\begin{quotation}
\footnotesize{
$<<$Talvolta nel corso di una conversazione con qualche collega
diceva quasi incidentalmente di aver fatto durante la sera precedente
il calcolo o la teoria di un fenomeno non chiaro che era caduto sotto
l'attenzione sua o di qualcuno di noi in quei giorni. Nella
discussione che seguiva, sempre molto laconica da parte sua, Ettore a
un certo punto tirava fuori dalla tasca il pacchetto delle sigarette
Macedonia (era un fumatore accanito) sul quale erano scritte, in una
calligrafia minuta ma ordinata, le formule principali della sua
teoria o una tabella di risultati numerici. Copiava sulla lavagna
parte dei risultati, quel tanto che era necessario per chiarire il
problema, e poi, finita la discussione e fumata l'ultima sigaretta,
accartocciava il pacchetto nella mano e lo buttava nel cestino$>>$.}
\end{quotation}

Estremamente interessanti sono pure due altri passi di lettera. Il
14.2.1933, sempre da Lipsia, Majorana racconta alla madre:
\begin{quotation}
\footnotesize{
$<< \ldots$
L'ambiente dell'istituto fisico \`e molto simpatico. Sono in ottimi
rapporti con Heisenberg, con Hund e con tutti gli altri. {\em Sto
scrivendo alcuni articoli in tedesco. Il primo \`e gi\`a pronto}, e
spero di eliminare qualche confusione linguistica durante la
correzione delle bozze$>>$.}
\end{quotation}
Il lavoro ``gi\`a pronto'' \`e
naturalmente quello sulle forze nucleari di cui si sta parlando; il
quale, per\`o, rimase l'{\em unico} in lingua tedesca.

Ancora: nella lettera del 18 febbraio dichiara al padre
\begin{quotation}
\footnotesize{
$<<\ldots$
{\em pubblicher\`o in tedesco, estendendolo, anche l'ultimo mio
articolo apparso sul ``Nuovo Cimento''}$>>$.}
\end{quotation}

In realt\`a Ettore non pubblic\`o pi\'u nulla, n\'e in Germania, n\'e al
rientro in Italia, a parte  l'articolo (del 1937) di cui stiamo per dire.

Di notevole importanza \`e quindi sapere che Ettore stesse scrivendo
altri lavori: in particolare, che stesse {\em estendendo} il suo
articolo sulla equazione a infinite componenti. Nel brano alla madre,
\`e probabile si riferisca pure alla sua teoria simmetria di
elettrone e anti-elettrone, pubblicata solo quattro anni pi\'u tardi.

\subsection*{5.6. Il neutrino di Majorana}

\paragraph*{}
Dai manoscritti ritrovati pare, come si \`e detto, che Majorana
formulasse in quegli stessi anni (1932-33) le linee essenziali anche
della sua teoria  simmetrica per l'elettrone e l'anti-elettrone: che
le formulasse, cio\`e, non appena si diffuse la notizia della scoperta
dell'anti-elettrone, o ``positone''. Anche se Ettore pubblica tale
teoria solo molto pi\'u tardi, accingendosi a partecipare al Concorso
a cattedra di cui sappiamo: ``Teoria simmetrica dell'elettrone e del
positrone'', {\em Nuovo Cimento}, vol. {\bf 14} (1937) pp.171-184.
Questa pubblicazione viene inizialmente notata quasi esclusivamente per aver
introdotto la famosa {\em rappresentazione di Majorana} delle
``matrici di Dirac'' in forma reale\footnote{Si noti, per\`o, che
l'algebra $I\!\!R (4)\simeq I\!\!R_{3,1}$ cos\'{\i} introdotta da
Majorana \`e del tutto diversa dall'algebra $C\!\!\!\!I\ (4)\simeq I\!\!R_{4,1}$
introdotta da Dirac. Osserviamo, en passant, che l'algebra di
Majorana \`e una delle {\em due} algebre associabili naturalmente
allo spazio di Minkowski (la seconda essendo $I\!\!R_{1,3}\simeq
I\!\!H(2)$, ove $I\!\!H(2)$ \`e l'algebra delle matrici
quaternioniche $2\times 2$).}.  Conseguenza di tale teoria \`e  che un
``fermione'' neutro debba coincidere con la propria antiparticella:
ed Ettore suggerisce che i neutrini possano essere particelle di
questo tipo.

Ettore ci teneva molto a questa sua elaborazione teorica; ci\`o \`e
testimoniato da Carrelli, che ne discusse con Ettore durante il breve
periodo di lezioni a Napoli.

Come per altri scritti di Majorana, anche questo articolo ha
cominciato ad avere fortuna solo vent'anni dopo, a partire dal 1957.
Dopo di che ha goduto di fama via via crescente tra i fisici delle
particelle relativistiche e delle teorie di campi\footnote{Nel 1981,
ad esempio, una rivista giapponese di fisica ha ripubblicato in
lingua inglese (con traduzione a cura di Luciano Maiani) questo
articolo di circa quarantacinque anni prima.}. Ora sono di gran moda
espressioni come ``spinori di Majorana'', ``massa di
Majorana'', ``neutrini di Majorana''.

Le pubblicazioni di Majorana (ancora poco note, nonostante tutto) sono per
la Fisica una miniera. Recentemente, ad esempio, Carlo Becchi ha osservato
come nelle prime pagine di questo scritto si trovi una formulazione
estremamente chiara del principio d'azione quantistico, che in anni
successivi, attraverso i lavori di Schwinger e Symanzik, ha portato agli
sviluppi recenti pi\'u importanti di teoria dei campi quanto-relativistici.

\subsection*{5.7. Esistono altri manoscritti scientifici inediti?}

\paragraph*{}
Tornando alla lettera del 18 febbraio al padre, in essa troviamo una
notizia molto interessante: $<<${\em Pubblicher\`o in tedesco, estendendolo,
anche l'ultimo mio articolo apparso sul ``Nuovo Cimento"}$>>$. \
Questo progetto non verr\`a poi realizzato; ma \`e importante che Ettore
avesse in mente di generalizzare il lavoro in cui aveva introdotto la sua
equazione a infinite componenti. \ Anzi, la questione diviene del massimo rilievo
quando si leggano le lettere inviate in quel periodo al CNR (ritrovate presso
gli archivi del C.N.R., e a me pervenute attraverso la cortesia di
G.Fioravanti e soprattutto del collega M.De Maria).
Nella prima (21.1.33) Ettore specifica: $<<${\em Attendo attualmente
alla elaborazione di una teoria per la descrizione di particelle con momento
intrinseco arbitrario che ho iniziata in Italia, e di cui ho dato notizia
sommaria nel {\rm Nuovo Cimento} (in corso di stampa)...}$>>$. \ Nella seconda
(3.3.33) dichiara addirittura, riferendosi al medesimo lavoro: $<<${\em Ho
inviato alla {\rm Zeitschrift f\"ur Physik} un articolo sulla teoria dei
nuclei. Ho pronto il manoscritto di una nuova teoria delle particelle
elementari e lo invier\`o alla stessa rivista fra qualche giorno...}$>>$.
Se ricordiamo che l'articolo qui considerato
come ``notizia sommaria" di una nuova teoria era gi\`a di altissimo livello,
si comprende come sarebbe di enorme interesse scoprire una copia della teoria
completa: la quale nel marzo 1933 aveva gi\`a assunto la forma di un manoscritto
compiuto, forse gi\`a dattiloscritto in lingua tedesca.

Ma Ettore non ne fece pi\'u nulla; tanto che nella sua relazione finale
(14.9.33) al CNR non la menziona neppure pi\'u: era divenuta
tab\`u. \ Dopo avervi ricordato l'articolo sulla ``Teoria dei nuclei", infatti,
Majorana passa subito a parlare delle ricerche iniziate nel {\em secondo}
periodo di Lipsia: $<<${\em Nell'ultimo periodo della mia residenza a Lipsia ho
iniziato altri lavori che non ho potuto in seguito, per motivi di salute, n\'e
completare n\'e avvicinare alla conclusione. Credo inutile parlarne}$>>$. \
Perch\'e? Perch\'e Ettore, poi, non ne fece niente? Si potrebbe pensare
che {\em all'ultimo momento} abbia riscontrato qualche grave errore, che
inficiasse la sua nuova teoria. Ma, conoscendo Majorana, non lo riteniamo
probabile. Propendiamo, semmai, per un'altra possibile spiegazione: il
``referee" della rivista tedesca pu\`o avere respinto il suo manoscritto,
tanto pionieristico, non avendolo capito (purtroppo l'archivio di quegli anni
della {\em Zeitschrift f\"ur Physik} pare sia andato perduto durante la Seconda
guerra mondiale). Ed Ettore non era persona da mettersi a combattere con gli
sciocchi. Il colpo di grazia pu\`o essergli venuto da quei burocrati del CNR i
quali pretenderanno che gli articoli di Majorana, che avrebbero recato lustro
alla migliore rivista internazionale di fisica, uscissero sulle (allora
ancora provinciali) riviste di lingua italiana. Ettore
rispose a tono (il 9.5.33), ma poi potrebbe avere preso il
sopravvento in lui quella noia, quel  malessere per la stupidit\`a umana che
in un genio, in lui pur cos\'{\i} \ affettuoso col prossimo, doveva agire
ancora pi\'u prepotentemente che nei comuni mortali.

Non dimentichiamo per\`o che la citata lettera a Quirino del 16.1.1936 ci
ha rivelato che successivamente Ettore continu\`o a lavorare in fisica
teorica, occupandosi a fondo ---per lo meno--- di elettrodinamica quantistica.
 \ Dove sono finiti gli appunti, gli scritti, gli articoli relativi a tutta
questa attivit\`a?

In seguito ad una approfondita ricerca\footnote{B. Russo: ``Ettore
Majorana -- Un giorno di marzo",
programma televisivo trasmesso il 18.12.90 (Rai Tre -- Sicilia).}
effettuata ---in qualit\`{a} di
regista televisivo--- per conto della  Rai-3, Sede di Palermo,
Bruno Russo ha rintracciato e opportunamente intervistato, nel 1990,
gli studenti che seguirono le lezioni universitarie tenute da  Majorana a
Napoli nei primi mesi del 1938. Si \`{e} cos\'{\i} venuti a sapere che
Majorana, il giorno prima di salpare da Napoli (e successivamente sparire),
consegn\`{o} alla propria studentessa Sig.na Gilda Senatore [ora Prof.ssa
Senatore] una cartelletta di carte scientifiche.  Si ha ragione
di credere che tale cartelletta contenesse anche alcuni almeno dei risultati
del lavoro svolto da Majorana, in isolamento (e senza pubblicarne nulla:
eccezion fatta per il
materiale confluito nella ``tarda" pubblicazione n.9), tra il 1933 e il 1938.
Tali risultati sarebbero di {\em straordinaria} importanza, pi\'{u} ancora
che storica, per la stessa fisica teorica contemporanea.

\h Avvenne che la Sig.na Senatore mostr\`{o} i manoscritti di Majorana al Dottor
Cennamo, suo futuro marito, allora Assistente del Direttore Antonio Carrelli,
e questi ritenne opportuno consegnarli  ---in via burocratica e gerarchica---
al Professor Carrelli; \ e, per quanto a noi ora consta, essi si persero. \
Tale perdita, per la fisica teorica moderna, \`{e} davvero grave. \
Al riguardo ha dato nuova, ampia, interessante testimonianza la stessa
Prof.ssa Gilda Senatore, durante le celebrazioni organizzate nel 1998 dalla
memore Universit\`a di Napoli per i sessant'anni dalla scomparsa di Majorana.

\subsection*{5.8. Testimonianze di colleghi}

\paragraph*{}
Molte altre idee di Ettore, quando non restarono nella sua mente,
hanno lasciato traccia soltanto nelle sue carte inedite, o nella
memoria dei colleghi.

Una delle testimonianze pi\'u interessanti che abbiamo raccolto \`e
di GianCarlo Wick. Da Pisa il 16 Ott. 1978 scrive:

\begin{quotation}
\footnotesize{
$<<$Caro Prof. Recami: $\ldots$Il contatto scientifico {\em [tra me ed
Ettore]} di cui le accenn\`o Segr\'e non avvenne a Lipsia, ma a Roma
in occasione del Congresso Volta (dunque assai prima del soggiorno di
Majorana a Lipsia). La conversazione ebbe luogo in un ristorante, in
presenza di Heitler, e dunque senza lavagna n\'e formule scritte; ma
nonostante l'assenza di dettagli quello che Majorana descrisse a
parole era una ``teoria relativistica di particelle cariche di spin
zero basata sull'idea di quantizzazione dei campi'' (seconda
quantizzazione). Quando assai pi\'u tardi vidi il lavoro di
Pauli\footnote{Premio Nobel 1945.} e
Weisskopf rimasi assolutamente convinto che quello che Majorana aveva
descritto fosse la stessa cosa. Beninteso, Majorana non pubblic\`o
nulla e probabilmente non ne parl\`o a molti. Non ho nessunissima
ragione di pensare che Pauli e Weisskopf ne sapessero nulla$\ldots$
--- Cordialmente --- Suo G.C. Wick$>>$.}
\end{quotation}

E dal M.I.T. (Cambridge, Mass.), il 16 maggio 1984, Victor Weisskopf
ci scriver\`a:
\begin{quotation}
\footnotesize{
$<<$Dear Dr.Recami: $\ldots$ I am very glad that you
have found a letter in which Majorana says that he had good relations
with me$\ldots$ I have only a vague recollection that I did have a
discussion {\em [at Copenhagen, in 1933]}, with Majorana about the
newest developments in quantum electrodynamics$>>$.\footnote{$<<$Sono
molto contento che lei abbia rapporti con me$\ldots$ Io ricordo solo
vagamente che ebbi in effetti {\em [a Copenaghen, nel 1933]} a
discutere con Majorana intorno ai pi\'u recenti sviluppi
dell'elettrodinamica quantistica$>>$.}}
\end{quotation}

L'articolo di Pauli e Weisskopf a cui accenna GianCarlo Wick
usc\'{\i} nel 1934 [{\em Helvetica Physica Acta} {\bf 7} (1934) 709].
Continua Wick
\begin{quotation}
\footnotesize{
$<<\ldots$ Non ebbi mai occasione in seguito di parlare
a Heitler di questo episodio$\ldots$ Non ci sarebbe da stupirsi se se
ne fosse dimenticato, perch\'e Majorana aveva parlato della cosa con
quel tono distaccato e ironico che spesso usava anche a proposito
delle cose sue. Insomma, senza darsi importanza$\ldots >>$.}
\end{quotation}

Un'altra testimonianza ci giunge, anche se indirettamente, dalla
grande e tragica figura di Bruno Touschek. Il 29.10.76   da Rieti ci
scriveva infatti Eliano Pessa:
\begin{quotation}
\footnotesize{
 $<< \ldots$ Abbiamo discusso con
Touschek il tuo lavoro su Majorana\footnote{E. Recami: ``Nuovo
notizie sulla scomparsa del fisico E. Majorana', {\em Scientia} {\bf
110} (1975) 577-598.} in {\em Scientia} {\bf 110} (1975) 577; ha
avuto da dire per ci\`o che riguarda il tuo elenco delle opere
scientifiche di Majorana a pag. 585. Secondo lui si dovrebbe
aggiungere la teoria dell'``oscillatore di Majorana'', che \`e
implicitamente contenuta nella sua teoria del neutrino. L'oscillatore
di Majorana \`e descritto da un'equazione del tipo
$\ddot{q}+\omega^2q=\varepsilon .\delta (t)$, dove $\varepsilon$ \`e
una costante e $\delta$ \`e la funzione delta di Dirac. Secondo
Touschek le propriet\`a di questo oscillatore presentano un notevole
interesse, specie per ci\`o che riguarda lo spettro energetico. Non
vi \`e, comunque, una bibliografia in merito$\ldots >>$.}
\end{quotation}
Il problema,
in verit\`a, sembra essere non tanto quello di risolvere l'equazione
(ben nota), quanto di intendere cosa avesse in mente (quali
condizioni al contorno, ad esempio) Bruno Touschek.

\subsection*{5.9. Erice}

\paragraph*{}
Anche se a tutti noto, non vogliamo tralasciare di ricordare come al
nome di Ettore Majorana sia stato intitolato fin dal 1963 il
``Centro di cultura scientifica E. Majorana'' di Erice, presso
Trapani. Fondato e diretto da Antonino Zichichi, questo Centro ---cui
localmente prestano la loro operosa esperienza Miss Maria Za\'{\i}ni e
i dottori A. Gabriele
e P. Savalli--- \`e sede ogni anno di una sessantina di congressi.
Erice \`e una isolata, incantevole cittadina di origini
elimo-puniche, non lontana dai capolavori greci di Segesta e
Selinunte. L'atmosfera del Centro Majorana \`e stata paragonata da
qualcuno a quella mitica della Copenaghen degli anni venti. Con la
fama conquistatasi, ha contribuito a tenere vivo il nome di Ettore
tra  gli studiosi di un centinaio di Paesi.

\subsection*{5.10. Wataghin}

\paragraph*{}
Approfittiamo, infine, dei ricordi di Wataghin per ritornare ai
giorni di Lipsia. Gleb Wataghin, il noto fisico italiano di origine
ucraina recentemente scomparso, fondatore della fisica
brasiliana, ce ne ha lasciato una testimonianza nel 1975 presso
l'Universit\`a di Campinas (Stato di San Paolo del Brasile), in una
intervista raccolta in lingua portoghese presso l'Istituto di Fisica
che da lui prende il nome. Il linguaggio, ovviamente, \`e
colloquiale:
\begin{quotation}
\footnotesize{
$<<$A Lipsia, ove lavorava Heisenberg, incontrai
Jordan, Debye, Max Born che vi stava arrivando, ed anche Ettore
Majorana: giovane che {\em pareceu, como era realmente, um verdadeiro
genio$\ldots$} Il cameratismo, l'amicizia esistente tra gli
scienziati$\ldots$ si manifestava, per esempio, nel modo in cui si
svolgevano le discussioni scientifiche, cos\'{\i} come le
manifestazioni sportive. A Lipsia ci si riuniva, per un seminario di
due ore, dalle due alle quattro del pomeriggio. Di mattina i teorici
dormono.\footnote{Ricordiamo la ``definizione'' di fisica di Orear:
$<<$ La Fisica \`e quella cosa che fanno i fisici la sera tardi$>>$.}
Dopo si andava a giocare a ping-pong nella migliore biblioteca, su un
tavolo per gli studenti. Posso dire che il campione era Heisenberg.
Poi si andava a piedi in una birreria, e magari si giocava a scacchi.
Si  giocava a scacchi anche all'Istituto di fisica. Poich\'e
Heisenberg era uno dei direttori, nessuno protestava che si giocasse
a ping-pong o a scacchi in biblioteca: cosa impensabile,  a quel
tempo, in altri Istituti$\ldots$ Ai seminari giungeva gente di tutto
il mondo. Per esempio, ricordo che una volta il seminario fu tenuto
da Norzig e un suo collega: furono obbligati a una discussione molto
impegnativa {\em depois das perguntas que faziam o Heisenberg e o
Ettore Majorana} (a seguito delle domande che fecero H. ed E.M.)$>>$.}
\end{quotation}
Ancora, dichiara Wataghin nell'intervista:
\begin{quotation}
\footnotesize{
$<<$Vorrei ricordare in
particolare la figura di Majorana, che ---secondo il giudizio di
molti, e in particolare dello stesso Fermi--- era un genio
eccezionale$\ldots$ Ammalato, soffriva di ulcera, mangiava quasi
esclusivamente latte; non praticava sport o ginnastica; molte volte
faceva delle lunghe passeggiate da solo. Poco comunicativo. Ma lo
incontravamo ogni tanto, il sabato. Era molto critico: trovava che
{\em toda gente que ele encontrava era n\~ao preparada, ou
est\'upida, etc.} Si occupava molto di leggi statistiche applicate
alla materia nucleare$\ldots$ La simmetria di scambio tra protoni e
neutroni poteva essere completa, compresi carica e spin; o
riguardare solo la carica, o lo spin. Ci\`o non era stato proposto o
studiato da altri. E la simmetria per scambio delle sole posizioni di
protoni e neutroni (senza toccare lo spin) permetteva di comprendere
statisticamente perch\'e la materia nucleare dovesse avere una
densit\`a costante$\ldots$ Il che faceva s\'{\i} che la teoria di
Majorana avesse un grande vantaggio rispetto a quella proposta da Heisenberg$>>$.}
\end{quotation}

\section*{6. Interesse Scientifico e Storico delle Lettere}

\paragraph*{}
Nel marzo del 1972 abbiamo ritrovato, in collaborazione con la
sorella Maria Majorana, l'epistolario di Ettore, costituito
principalmente da due gruppi di lettere: quelle del 1933 (da Lipsia e
Copenaghen) e quelle del 1938 (da Napoli). Fissiamo l'attenzione,
qui, sulle lettere del '33. Ettore arriva a Lipsia il 19 gennaio
1933; e il 22 pu\`o scrivere:
\begin{quotation}
\footnotesize{
$<<$Cara mamma, $\ldots$ All'Istituto di Fisica mi hanno accolto
molto cordialmente. Ho avuto una lunga conversazione con Heisenberg
che \`e una persona straordinariamente cortese e simpatica. Sono in
ottimi rapporti con tutti, specie con l'americano Inglis che avevo
conosciuto a Roma e ora mi tiene frequentemente compagnia e mi fa da
guida. Il mio tedesco migliora a vista d'occhio.

Fra pochi giorni avr\`o la visita di Bernardini\footnote{Si
riferisce a queste parole la lettera indirizzataci da Gilberto
Bernardini (ved. Cap. 4) il 9.5.84.} che risiede a Berlino-Dalhem
e ritorna temporaneamente in Italia. Il clima \`e piacevole; un po'
pi\'u freddo che a Roma ma senza vento$\ldots$ Se arrivano degli
estratti del Nuovo Cimento\footnote{Si riferisce all'articolo
``Teoria relativistica di particelle con momento intrinseco
arbitrario'', {\em Nuovo Cimento} {\bf 9} (1932) 335.} ti prego di
farne inviare qui solo un piccolo numero, dieci al massimo, in busta
aperta raccomandata (stampati raccomandati)$\ldots$ Se a Turillo} [il
fratello Dr. Salvatore] {\em capita di andare al Ministero dopo il 27
gennaio pu\`o ritirare il mio decreto di libera docenza (da
Borsi)$\ldots >>$.}
\end{quotation}

Il 7 febbraio aggiunge:
\begin{quotation}
\footnotesize{
$<<$\`E arrivato a Roma Feenberg, un altro fisico americano con cui
avevo stretto amicizia. Ci intendiamo abbastanza bene in tedesco.
Domani comincia a Lipsia la cosiddetta ``settimana magnetica'' che
richiama qui i fisici di quasi tutta la Germania; rivedr\`o parecchie
vecchie conoscenze. Io conto di restare a Lipsia fino alla fine di
febbraio: in marzo e aprile qui fanno infatti vacanza. Ne approfitter\`o per
recarmi probabilmente a Zurigo da Pauli, uno dei pi\'u celebri
scienziati viventi. Verranno con me anche Bloch, uno svizzero che ho
conosciuto qui e che ha pure la virt\'u di parlare perfettamente
l'italiano, e Inglis$>>$.}
\end{quotation}

E al termine della ``settimana magnetica'', il 14 febbraio:
\begin{quotation}
\footnotesize{
$<<$Cara mamma, $\ldots$ Si \`e svolto a Lipsia un congresso
internazionale di Fisica tra grande animazione. Ho stretto relazioni
personali con vari illustri personaggi, particolarmente con
Ehrenfest\footnote{Paul Ehrenfest, uno dei fondatori della meccanica
quantistica, morto tragicamente (suicida) quello stesso anno 1933.}
che mi ha costretto a spiegarli minutamente alcuni miei lavori e mi
ha invitato a recarmi in Olanda.

Il primo marzo andr\`o a Copenaghen in luogo di Zurigo, perch\'e in
Svizzera come in Germania in marzo e aprile le scuole sono chiuse,
mentre in Danimarca si segue l'uso italiano. A Copenaghen trover\`o
Bohr {\em [premio Nobel 1922]} e altri che gi\`a conosco
personalmente. \`E con Lipsia il centro pi\'u importante in Europa
per la fisica teorica$\ldots >>$.}
\end{quotation}

Dopo pochi giorni, il 18, conferma il progetto di recarsi a
Copenaghen scrivendo al padre:
\begin{quotation}
\footnotesize{
$<<$Caro pap\`a, $\ldots$ Il $1^{\underline{0}}$ marzo mi recher\`o a
Copenaghen da Bohr, il maggiore ispiratore della fisica moderna, ora
un po' invecchiato e sensibilmente rimbambito$\ldots >>$.}
\end{quotation}

Non deve sorprendere che Ettore sia sarcastico con Bohr. Come un
rocciatore sestogradista, che ama le ``direttissime'', non degna di
molta considerazione chi conquista le cime seguendo i sentieri
naturali, o arrivandovi col paracadute, cos\'{\i} Ettore
---raffinatissimo, elegante e rigoroso arrampicatore dei picchi della
fisica teorica--- non guardava con simpatia ai colleghi, quando
questi preferivano invece farsi un'idea intuitiva del panorama
generale della fisica senza troppa attenzione al rigore. Ettore,
inoltre, parla di Bohr senza averlo ancora conosciuto di persona:
probabilmente influenzato da un collega pettegolo. In ogni caso, di
lui dice, giustamente, essere il maggior ispiratore della fisica
moderna: alla scuola di Bohr (il quale fin dal 1913 aveva edificato
il modello dell'atomo a tutti noto) si erano successivamente formati
quasi tutti i costruttori della meccanica quantistica.

Il 28 febbraio dichiara, per\`o, al padre:
\begin{quotation}
\footnotesize{
$<<$Io mi fermer\`o
probabilmente a Lipsia ancora due o tre giorni, perch\'e devo
chiacchierare con Heisenberg. La sua compagnia \`e insostituibile e
desidero approfittarne finch\'e egli rimane qui$\ldots$ Ho a
Copenaghen un vecchio amico, Placzek, che \`e stato un anno a Roma
l'anno scorso$>>$.}
\end{quotation}
 Sentimenti analoghi aveva gi\`a espresso alla
madre il giorno 22:
\begin{quotation}
\footnotesize{$<<$Mi dispiace molto di dover lasciare
Lipsia dove ho trovato un'accoglienza  molto cordiale, e vi
ritorner\`o volentieri fra due mesi$\ldots$ \ [Heisenberg  ed io]
siamo  diventati abbastanza amici in seguito a molte discussioni
e ad alcune partite a scacchi. Le occasioni per queste sono offerte
dai ricevimenti che egli offre tutti i marted\'{\i} sera ai
professori e studenti dell'istituto di fisica teorica$\ldots >>$.}
\end{quotation}

Ettore arriva a {\em K\o benhavn} il 4 marzo. Dal libro degli ospiti
dell'Istituto di fisica di Copenaghen risulta che Ettore frequenta
quello che ora si chiama ``Niels Bohr Institutet''\footnote{Situato,
come oggi, in 15-17 Blegdamsvej (pronunciare {\em Pl\`aidams-v\`ai}).}
dal 5 marzo al 12 aprile 1993. Dopo avere incontrato Bohr, gi\`a il 7
di marzo annota:
\begin{quotation}
\footnotesize{
$<<$Cara mamma,$\ldots$ Bohr \`e un bonaccione;
gli piace che io parli il tedesco peggio di lui e si \`e molto
preoccupato di trovarmi una pensione vicino all'istituto. Sono in
buoni rapporti con Mo$\!\!\!/$ller e Weisskopf. Placzek \`e invisibile
perch\'e da tempo immemorabile \`e occupato a scrivere l'Handbuch.
Mi ha telefonato pi\'u volte parlando ancora un buon
trasteverino. Questa sera sono a cena da lui\footnote{Di George
Placzek dice Segr\'e: $<<$Placzek aveva vasti interessi anche al
di fuori della fisica; conosceva bene molte lingue e letterature, la
musica, e aveva profonde conoscenze storiche e politiche. Era un uomo
pieno di saggezza, spiritoso, e di un'onest\`a e forza di carattere
difficilmente eguagliabili$>>$.} con Weisskopf$>>$.}
\end{quotation}
La lettera di
Weisskopf che abbiamo incontrato nel Paragrafo 5.7 si riferisce a
queste frasi di Ettore; naturalmente, pi\'u di cinquant'anni dopo,
Victor Weisskopf non ricordava molto.

 Il giorno 18 di marzo Ettore aggiunge:
\begin{quotation}
\footnotesize{
$<<$Cara mamma, $\ldots$ Bohr \`e partito per una decina di giorni.
\`E adesso in montagna con Heisenberg per riposarsi$\ldots$ A
Copenaghen \`e molto popolare. Il proprietario di una grande fabbrica
di birra gli ha costruito e offerto una graziosa villetta a cui si
accede passando attraverso montagne di botti. \`E un problema
notoriamente difficile quello di scoprirne l'ubicazione per chi vi si
reca per la prima volta. Ci sono andato una volta per un t\`e. Bohr
in persona ha guidato i miei passi poich\'e sono stato abbastanza
fortunato da incontrarlo mentre gironzolava in bicicletta per i
dintorni $\ldots >>$.}
\end{quotation}

E il 29 di marzo:
\begin{quotation}
\footnotesize{
$<<\ldots$\`E tornato Bohr. Un nuovo
acquisto: Rosenfeld, cio\`e tutta quanta la fisica teorica del
Belgio $\ldots >>$.}
\end{quotation}
Ancora una volta il tono di Ettore \`e scherzoso:
sappiamo quanto egli sia sedotto dall'ironia. E si tenga presente che
egli scrive non a colleghi, ma a familiari al di fuori del mondo
internazionale della fisica teorica. Il suo atteggiamento va dunque
considerato pi\'u irriverente e canzonatorio che sarcastico.

Dopo il 12 di aprile, Ettore torna brevemente a Roma. (Gi\`a il
23.2.33 aveva annunciato: $<<$Conto di partire da Copenaghen qualche
giorno prima di Pasqua direttamente per Roma$>>$; e il 29.3.33
scriver\`a: $<<$Non desidero i fascicoli del Nuovo Cimento che trover\`o
a Roma fra quindici giorni$\ldots$ Partir\`o intorno al 12 aprile$>>$.)

All'inizio di maggio riprende a scrivere a casa  da {\em Leipzig}. Il
23, ad esempio, scrive alla mamma:
\begin{quotation}
\footnotesize{
$<<$La situazione tedesca \`e
molto tranquilla. La mia in particolare anche di pi\'u. Sono in
rapporti cordiali con Heisenberg che ama le mie chiacchiere e mi
insegna pazientemente il tedesco. Di questo devo fare uso esclusivo
nella conversazione dopo la partenza di Bloch, buon conoscitore del
toscano. --- Si annuncia col pi\'u grande rumore la prossima partenza
dei 24 apparecchi di Balbo per l'America$\ldots >>$.}
\end{quotation}

\section*{7. Le Lettere del 1938}

\subsection*{7.1. L'annuncio delle lezioni}

\paragraph*{}
Le lettere del 1938, aventi rilevanza per le circostanze della sua
scomparsa, sono state da noi analizzate altrove\footnote{E. Recami:
{\em Il caso Majorana: Epistolario, testimonianze, documenti}, $2^{\rm a}$
edizione (Mondadori, serie ``Oscar"; Milano, 1991; esaurita); e $3^{\rm a}$
edizione (Di renzo; Roma, genn.2001)}. Accenniamo brevemente
solo a quelle che qui ci interessano, riferendosi al periodo in cui Majorana
svolge le sue lezioni universitarie.

Verso la fine del 1937, Ettore riceve a casa, in Roma, la
``partecipazione di nomina'' a firma del Ministro dell'Educazione
Nazionale, Giuseppe Bottai:
\begin{quotation}
\footnotesize{
$<<$Si comunica a V.S. che, in
applicazione dell'art. 8 del R.D.L. 20 giugno 1935-XIII, n. 1071, si
\`e disposta la nomina della S.V., indipendentemente dalla usuale
procedura del concorso, a ordinario di fisica teorica presso la
Facolt\`a di scienze della Regia Universit\`a di Napoli, per l'alta
fama di singolare perizia cui Ella \`e pervenuta nel campo degli
studi riguardanti la detta disciplina, con decorrenza dal 16 novembre
1937-XVI$>>$.}
\end{quotation}

Majorana si reca a Napoli dopo l'Epifania (verso il 10 gennaio 1938),
e il 12 ---dalla sua nuova sede universitaria--- risponde alla
Direzione Generale della Istruzione Superiore:
\begin{quotation}
\footnotesize{
$<<$ Ricevo
comunicazione diretta da S.E. il Ministro per la mia nomina a
ordinario di Fisica Teorica presso la Regia Universit\`a di Napoli,
in applicazione dell'art. 8 del R.D.L. 20 giugno 1935-XIII, n.
1071. Nel porgere rispettosamente a S.E. il ministro l'espressione
del mio grato animo per l'alta distinzione concessami, tengo ad
affermare che dar\`o ogni mia energia alla scuola e alla scienza
italiane, oggi in cos\'{\i} fortunata ascesa verso la riconquista
dell'antico e glorioso primato. --- Con osservanza --- Ettore
Majorana$>>$.}
\end{quotation}
Queste parole possono sembrare retoriche. Effettivamente
sono ``di circostanza''; ma sono vere: nel senso che dagli inizi
degli anni trenta, come sappiamo, la fisica italiana era davvero in
rapido sviluppo. Ed Ettore poteva ben permettersi di affermare il
proprio contributo a tale ascesa. L'eventuale retorica \`e temperata,
in ogni caso, dal noto spirito ironico di Ettore; proprio riguardo a
questa nomina il giorno prima aveva scritto  alla madre:
\begin{quotation}
\footnotesize{
$<<$Cara mamma, $\ldots$ oggi abbiamo comprato i mobili per il mio
studio, graziosamente offerti dalla Facolt\`a. Praticamente
l'Istituto si riduce alla persona di Carrelli, del vecchio aiuto
Maione e del giovane assistente Cennamo. Vi \`e anche un professore
di fisica terrestre difficile a scoprire. Ho trovato giacente da ben
due mesi una lettera del Rettore in cui mi annunciava la mia nomina
``per l'alta fama di singolare perizia''. Non avendolo trovato, gli
ho risposto con una lettera altrettanto elevata$\ldots >>$.}
\end{quotation}

Mentre in precedenza, il 21 novembre del 1937, cos\'{\i} si era espresso con
l'amico e collega Giovanni Gentile jr.: $<<$Caro Gentile, [...] Mi meraviglio
che per quanto mi riguarda tu
dubiti del mio buono stomaco, in senso metaforico. \ Pio XI \`e molto
vecchio ed io ho ricevuto un'ottima educazione cristiana; se al prossimo
conclave mi fanno papa per meriti eccezionali, accetto senz'altro [...] $>>$.

Desta comunque sorpresa l'impegno a prodigarsi per la scuola e la
scienza: impegno senza dubbio sincero, espresso due mesi soltanto
prima del marzo 1938.

Nella medesima lettera dell' 11.1.38 da Napoli alla madre, Ettore scrive:
\begin{quotation}
\footnotesize{
$<<$Cara mamma, --- Ho annunziato l'inizio del corso per
gioved\'{\i} 13 alle ore nove. Ma non \`e stato possibile verificare
se vi sono sovrapposizioni d'orario, cos\'{\i} che \`e possibile che
gli studenti non vengano e che si debba rimandare. Ho visto il
preside con cui ho concordato di evitare ogni carattere ufficiale
all'apertura del corso, e anche per questo non vi consiglierei di
venire. Carrelli \`e stato molto gentile$\ldots$ L'istituto \`e molto
pulito e in ordine, bench\'e poco attrezzato.

L'albergo Napoli \`e discreto, con prezzi ragionevoli; cos\'{\i} \`e
probabile che vi rimarr\`o per qualche tempo. Napoli, almeno nella
parte centrale, ha un aspetto molto decoroso, bench\'e sia strana la
scarsit\`a di veicoli. Vi scriver\`o gioved\'{\i} sulle vicende della
prima lezione.\footnote{La famiglia, invece, si present\`o puntuale
il gioved\'{\i} 13 gennaio 1938, alle ore nove, per assistere alla
prolusione di Ettore: come ricordava la sorella Maria.}

Saluti affettuosi --- Ettore$>>$.}
\end{quotation}

Trascriviamo qui di seguito gli appunti di Ettore per la sua lezione
inaugurale. Il manoscritto, ritrovato nel 1972, \`e stato da noi reso
noto nel 1982 ({\em Corriere della Sera} del 19.10.82) in occasione
del cinquantesimo anniversario di molte delle sue ricerche pi\'u
importanti. Trattandosi di note stese a scopo  personale, e non di
pubblicazione, esse sono state leggerissimamente ``editate''.

\subsection*{7.2. Gli appunti per la lezione inaugurale (13.1.38)}

$<<$In questa prima lezione di carattere introduttivo illustreremo
brevemente gli scopi della fisica moderna e il significato dei suoi
metodi, soprattutto in quanto essi hanno di pi\'u inaspettato e
originale rispetto alla fisica classica.

\vspace{2mm}
\noindent
$<<$La fisica atomica, di cui dovremo principalmente occuparci,
nonostante le sue numerose e importanti applicazioni  pratiche ---e
quelle di portata pi\'u vasta e forse rivoluzionaria che l'avvenire
potr\`a riservarci---, rimane anzitutto una scienza di enorme
interesse {\em speculativo}, per la profondit\`a della sua indagine
che va veramente fino all'ultima radice dei fatti naturali. Mi sia
perci\`o consentito di accennare in primo luogo, senza alcun
riferimento a speciali categorie di fatti sperimentali e senza l'aiuto
del formalismo matematico, ai caratteri generali della concezione
della natura che \`e accettata nella nuova fisica.

\vspace{1mm}
\begin{center}
---   ---  ---
\end{center}
\vspace{1mm}

\noindent
$<<$La {\em fisica classica} (di Galileo e Newton) all'inizio del
nostro secolo era interamente legata, come si sa, a quella concezione
{\em meccanicistica} della natura che dalla fisica \`e dilagata non
solo nelle scienze affini, ma anche nella biologia e perfino nelle
scienze sociali, informando di s\'e in tempi a noi abbastanza vicini
tutto il pensiero scientifico e buona parte di quello filosofico;
bench\'e, a dire il vero, l'utilit\`a del metodo matematico che ne
costituiva la sola valida giustificazione sia rimasta sempre
circoscritta esclusivamente alla fisica.

\vspace{2mm}
\noindent
$<<$Questa concezione della natura poggiava sostanzialmente su due
pilastri: l'esistenza oggettiva e indipendente della materia, e il
determinismo fisico. In entrambi i casi si tratta, come vedremo, di
nozioni derivate dall'esperienza comune e poi generalizzate e rese
universali e infallibili soprattutto per il fascino irresistibile che
anche sugli spiriti pi\'u profondi hanno in ogni tempo esercitato le
leggi esatte della fisica, considerate veramente come il segno di un
assoluto e la rivelazione dell'essenza dell'universo: i cui segreti,
come gi\`a affermava Galileo, sono scritti in caratteri matematici.

\vspace{2mm}
\noindent
$<<$L'{\em oggettivit\`a} della materia \`e, come dicevo, una
nozione dell'esperienza comune, poich\'e questa insegna che gli
oggetti materiali hanno un'esistenza a s\'e, indipendente dal fatto
che essi cadano o meno sotto la nostra osservazione. La fisica
matematica classica ha aggiunto a questa constatazione elementare la
precisazione o la pretesa che di questo mondo oggettivo \`e possibile
una rappresentazione mentale completamente adeguata alla sua
realt\`a, e che questa   rappresentazione mentale pu\`o consistere
nella conoscenza di un serie di grandezze numeriche sufficienti a
determinare in ogni punto dello spazio e in ogni istante lo stato
dell'universo fisico.

\vspace{2mm}
\noindent
$<<$Il {\em determinismo} \`e invece solo in parte una nozione
dell'esperienza comune. Questa d\`a infatti al riguardo delle
indicazioni contraddittorie. Accanto a fatti che si succedono
fatalmente, come la caduta di una pietra abbandonata nel vuoto, ve ne
sono altri ---e non solo nel mondo biologico--- in cui la successione
fatale \`e per lo meno poco evidente. Il determinismo in quanto
principio universale della scienza ha potuto perci\`o essere
formulato solo come generalizzazione delle leggi che reggono la
meccanica celeste. \`E ben noto che un {\em sistema} di punti
---quali, in rapporto alle loro enormi distanze, si possono considerare
i corpi del nostro sistema planetario--- si muove e si modifica
obbedendo alle leggi di Newton$\ldots$ {\em (omissis)}$\ldots$ Ne
segue che la configurazione futura del {\em sistema} pu\`o essere
prevista con il calcolo purch\'e se ne conosca lo stato iniziale
(cio\`e l'insieme delle posizioni e velocit\`a dei punti che lo
compongono). E tutti sanno con quale estremo rigore le osservazioni
astronomiche abbiano confermato l'esattezza della legge di Newton; e
come gli astronomi siano effettivamente in grado di prevedere con il
suo solo aiuto, e anche a grandi distanze di tempo, il minuto preciso
in cui avr\`a un'eclisse, o una congiunzione di pianeti o altri
avvenimenti celesti.

\vspace{1mm}
\begin{center}
---   ---  ---
\end{center}
\vspace{1mm}

\noindent
$<<$Per esporre la {\em meccanica quantistica} nel suo stato attuale
esistono due metodi pressoch\'e opposti. L'uno \`e il cosiddetto
metodo storico: ed  esso spiega in qual modo, per indicazioni
precise e quasi immediate dell'esperienza, sia sorta la prima idea
del nuovo formalismo; e come questo si sia successivamente sviluppato
in una maniera obbligata assai pi\'u dalla necessit\`a interna che
non dal tenere conto di nuovi decisivi fatti sperimentali. L'altro
metodo \`e quello matematico, secondo il quale il formalismo
quantistico viene presentato fin dall'inizio nella sua pi\'u generale
e perci\`o pi\'u chiara impostazione, e solo successivamente se ne
illustrano i criteri applicativi. Ciascuno di questi due metodi, se
usato in maniera esclusiva, presenta inconvenienti molto gravi.

\vspace{2mm}
\noindent
$<<$\`E un fatto che, quando sorse la meccanica quantistica, essa
incontr\`o per qualche tempo presso   molti fisici sorpresa,
scetticismo e perfino incomprensione assoluta, e ci\`o soprattutto
perch\'e la sua consistenza logica, coerenza e sufficienza appariva,
pi\'u che dubbia, inafferrabile.  Ci\`o venne anche, bench\'e del
tutto erroneamente, attribuito a una particolare oscurit\`a di
esposizione dei primi creatori della nuova meccanica; ma la verit\`a
\`e che essi erano dei fisici, e non dei matematici, e che per essi
l'evidenza e giustificazione della teoria consisteva sotrattutto
nell'immediata applicabilit\`a ai fatti sperimentali che l'avevano
suggerita. La formulazione generale, chiara e rigorosa \`e venuta
dopo, e in parte per opera di cervelli matematici. Se dunque noi
rifacessimo semplicemente l'esposizione della teoria secondo il modo
della sua apparizione storica, creeremo dapprima inutilmente uno
stato di disagio o di diffidenza, che ha avuto la sua ragione
d'essere ma che oggi non \`e pi\'u giustificato e pu\`o essere risparmiato.
Non solo, ma i fisici ---che sono giunti, non senza qualche pena,
alla chiarificazione dei metodi quantistici attraverso le esperienze
mentali imposte dal loro sviluppo storico--- hanno quasi sempre
sentito a un certo momento il bisogno di una maggiore coordinazione
logica, di una pi\'u perfetta formulazione dei princ\'{\i}pi, e non
hanno sdegnato per questo compito l'aiuto dei matematici.

\vspace{2mm}
\noindent
$<<$Il secondo metodo, quello puramente matematico, presenta
inconvenienti ancora maggiori. Esso non lascia in alcun modo
intendere la genesi del formalismo e in conseguenza il posto che la
meccanica quantistica ha nella storia della scienza. Ma soprattutto
esso delude nella maniera pi\'u completa il desiderio di intuirne in
qualche modo il significato fisico, spesso cos\'{\i} facilmente
soddisfatto dalle teorie classiche. Le applicazioni, poi, bench\'e
innumerevoli, appaiono rare, staccate, perfino modeste di fronte alla
sua soverchia e incomprensibile generalit\`a.

\vspace{2mm}
\noindent
$<<$Il solo mezzo di rendere meno disagevole il cammino a chi
intraprende oggi lo studio della fisica atomica, senza nulla
sacrificare della genesi storica delle idee e dello stesso linguaggio
che dominano attualmente, \`e di premettere un'esposizione il pi\'u
possibile ampia e chiara degli strumenti matematici essenziali della
meccanica quantistica, in modo che essi siano gi\`a pienamente
familiari quando verr\`a il momento di usarli e non spaventino allora
o sorprendano per la loro novit\`a: e si possa cos\'{\i} procedere
speditamente nella derivazione della teoria dai dati dell'esperienza.

\vspace{2mm}
\noindent
$<<$Questi strumenti matematici in gran parte preesistevano al sorgere
della nuova meccanica (come opera disinteressata di matematici che non
prevedevano un cos\'{\i} eccezionale campo di applicazione), ma la
meccanica quantistica li ha ``sforzati'' e ampliati per soddisfare
alle necessit\`a pratiche; cos\'{\i} essi non verranno da noi esposti
con criteri di matematici, ma di fisici. Cio\`e  senza preoccupazioni
di un eccessivo rigore formale, che non \`e sempre facile a
raggiungersi e spesso del tutto impossibile.

\vspace{2mm}
\noindent
$<<$La nostra sola ambizione sar\`a di esporre con tutta la chiarezza
possibile l'uso effettivo che di tali strumenti fanno i fisici da
oltre un decennio, nel quale uso ---che non ha mai condotto a
difficolt\`a o ambiguit\`a--- sta la fonte sostanziale della loro certezza$>>$.

\ \hfill{Ettore Majorana}

\subsection*{7.3. Le successive lezioni}

\paragraph*{}
Ettore faceva lezione la mattina dei ``giorni pari'' della settimana:
marted\'{\i}, gioved\'{\i} e sabato. Avendo iniziato il gioved\'{\i}
13, sabato 22 gennaio aveva gi\`a svolto la sua quinta lezione
universitaria. Nella stessa data, questa volta su carta intestata
dell'{\em Istituto di Fisica, R. Universit\`a di Napoli, Via A. Tari
3}, scrive infatti:

\begin{quotation}
\footnotesize{
$<<$22 gennaio 1938-XVI --- Cara mamma, --- Ho avuto la tua lettera e
il pacco della biancheria. Non sono raffreddato. Ho finito adesso la
quinta lezione. Sono ancora al {\em [l'albergo]} Terminus, ma andr\`o
prossimamente in una pensione. Carrelli \`e tuttora a Roma. Il tempo si
\`e rimesso al bello. Nel corso della prossima settimana sar\`o a
corto di denari; perci\`o potresti  pregare Luciano di
ritirare la mia parte del conto alla banca e magari di mandarmela
tutta, tenendo conto dei prelevamenti precedenti e dopo averti
restituito le mille lire che mi hai dato ultimamente. Ho buoni
indirizzi per pensioni fornitimi dall'infermiera. Credo che verr\`o
fra pochi giorni ma solo per poche ore perch\'e debbo ritirare un
libro da Treves e altri da casa.

Saluti affettuosi e arrivederci --- Ettore$>>$.}
\end{quotation}

Seguono tre lettere informative, brevi, ma che non fanno presagire (o
non lasciano trasparire) alcun travaglio interiore. Nella prima
Ettore dice, tra l'altro:
\begin{quotation}
\footnotesize{
$<<$Napoli, 23-2-1938-XVI --- Cara mamma ---
Sono all'albergo Bologna, via Depretis, che \`e abbastanza buono e
molto pulito. Personale quasi tutto bolognese. Ho una stanza
discreta; oggi me ne daranno una migliore $\ldots$ Verr\`o forse
dopo carnevale. Saluti affettuosi --- Ettore$>>$.}
\end{quotation}

La seconda, del 9 marzo, rispondendo alle preoccupazioni materne,
recita:
\begin{quotation}
\footnotesize{
$<<$Cara mamma --- Ho avuto la tua lettera. Prender\`o tutte
le precauzioni per la biancheria. Dei cartelli affermano con molta
enfasi che i servizi di stireria e lavanderia sono inappuntabili. ---
Qui c'\`e un tempo bellissimo, ideale per navigare nel golfo$\ldots$
--- Spero di venire in fine settimana. --- Saluti affettuosi$>>$.}
\end{quotation}

La terza \`e inviata al fratello Salvatore:
\begin{quotation}
\footnotesize{
$<<$Napoli,
19-3-1938-XVI --- Caro Turillo, --- Ho avuto la tua. Per ora non
vengo perch\'e luned\'{\i} ho alcune faccende da sbrigare
all'anagrafe e altrove. Vedr\`o se \`e possibile avere il libretto
per la mamma, ma non vedo come si possa affermare la convivenza
perch\'e io ho l'obbligo di prendere la residenza a Napoli, anzi l'ho
gi\`a presa provvisoriamente qui in albergo, alias via Depretis 72.
--- Vi mando un telegramma perch\'e non mi aspettiate stasera, ma
verr\`o certamente sabato prossimo$>>$.}
\end{quotation}

Il ``sabato prossimo'' sar\`a quello (26 di marzo) della sua
scomparsa. Della sua ultima lettera: da Palermo ---dopo il viaggio in
nave Napoli-Palermo--- a Carrelli.

Venerd\'{\i} 25, infatti, riprende in mano la penna per redigere la
sua ``prima lettera a Carrelli'' (da noi riportata in nota$^{(11)}$,
Cap. 4) e la missiva {\em ``Alla mia famiglia''}$\ldots$

Ma noi qui ci fermiamo. Le questioni propriamente biografiche, anche
se di straordinario interesse, non  trovano posto in questa sede, e
sono state da noi esposte altrove$^{(32)}$. Chiudiamo invece con
alcune note circa la calligrafia di Ettore.

\section*{8. Lo Stile. Epilogo}

\paragraph*{}
Anche l'esame della calligrafia di Ettore, aiutandoci forse a
comprendere il suo animo, pu\`o essere per noi illuminante. Un colto
amico, storico e saggista in incognito ---che nascostamente si
diletta anche di grafologia---, acconsent\'{\i} nel 1972 a studiare
uno scritto di Majorana. Ecco il responso del Dr. Gianni Sansoni:
\begin{quotation}
\footnotesize{
$<<$Carissimo, \ mai come in questo momento la grafologia si rivela una
cosa inutile ed anche irriguardosa. Io penso che a Majorana possono
benissimo adattarsi le raccomandazioni di Cesare Pavese: non fate
pettegolezzi. Ma quando mai c'\`e stato rispetto per il
solitari?$\ldots$ Dico questo perch\'e la conversazione di ieri sera
ha rafforzato un convincimento che mi ero fatto su Ettore Majorana, e
cio\`e che prima di tutto egli era (non \`e escluso che lo sia
ancora) di elevate qualit\`a morali e che molto deve avere
sofferto$\ldots$ Comunque, la prima impressione  \`e che si tratta di
un soggetto {\em irrequieto, mai soddisfatto di s\'e n\'e degli
altri}. Ci\`o che aggrava il quadro \`e la seconda scoperta, cio\`e
che il suo animo {\em non sembra ancorato a una divinit\`a o ad un
mito}. Anzi, rilevo, nell'uso di certe lettere, una {\em scarsa
propensione per vedere ``al di l\`a delle cose''}. Ti riconfermo che
il soggetto \`e fondamentalmente positivista, dotato di {\em logica
stringente, consequenziale}. Terza ``sgradita'' sorpresa: l'{\em
introversione}, che si  alimenta con un vasto fondo di {\em
scetticismo} e di {\em pessimismo}. Insomma, sin qui abbiamo un
quadro di individuo un po' gattopardiano, influenzato da una classe e
da una situazione secolare ben definite. Capisci? La logica naturale,
e la {\em forte volont\`a}, innestandosi sullo scetticismo, lo
costringono probabilmente a veri e propri dubbi e tormenti, cui non
vuole sottrarsi se non superandoli logicamente. Un tragico circolo
chiuso che solo una finalit\`a, uno scopo, un'autentica teleologia
potevano sciogliere. Permettimi alcuni accenni sull'aspetto generale
delle lettere. Ottimo stile, elegante, signorile, il che si attaglia
al suo aspetto fisico. Esse rivelano poi un amore di precisione, un
desiderio estremo di chiarezza, una logica ---ancora--- profonda e
ineliminabile. E la citazione della ``ragazza ibseniana'' \`e solo un
vezzo culturale o, sotto sotto, non svela il timore di identificarsi
nel desiderio di un rinnovamento? Abbiamo parlato del mare che lo ha
``rifiutato'' e dell'espressione ``vi ricorder\`o fino alle undici'':
altalena di sentimenti (le canzoni tango dell'epoca sono piene di lui
e di lei che vanno, vengono, scappano, tornano, $\ldots$), o non
piuttosto senso panico di una scadenza, di un dovere cui non ci si
pu\`o sottrarre? Come vedi non possiamo chinarci su di una persona
senza iniziare un discorso che non pu\`o che essere troppo lungo$\ldots$
Posso per\`o dire questo: che il Majorana doveva essere {\em persona mite
e buona, bisognosa d'affetto pi\'u che mai} e penso che miglior
elogio non gli si possa fare che avvicinandoci alle sue vicende con
rispetto e comprensione$>>$.}
\end{quotation}

\section*{9. Elenco delle Pubblicazioni di\\
\h Ettore Majorana}

\h Prima di passare al Capitolo II, in cui elenchiamo i manoscritti
scientifici inediti lasciatici dal Majorana (e forniamo in particolare il
catalogo dei ``Quaderni", inediti), ricordiamo qui gli scritti da
Majorana {\em pubblicati}: i quali pure, come gi\`a si diceva,
sono una miniera ancora parzialmente inesplorata di idee e di tecniche di
alta fisica teorica. \ Abbiamo gi\`a detto, per fare solo un esempio, che
nell'articolo n.6 (quello in cui viene scoperto l'effetto Majorana--Brossel)
Ettore introdusse anche
la ``Sfera di Majorana" per rappresentare spinori mediante punti su di una
superficie sferica. Tale invenzione \`e nota soltanto da quando R.Penrose,
accortosene in anni non lontani, ne ha fatto opportuna propaganda in
{\em 300 Years of Gravity}, ed. by S.W.Hawking \& W.Israel (Cambridge
Univ.Press.; 1987). Questa ``Sfera" viene attualmente studiata da un gruppo
di Palermo (C.Leonardi, F.Lillo, A.Vaglica e G.Vetri: ved. la Bibliografia).

\begin{description}
\item[1] --- ``Sullo sdoppiamento dei termini Roentgen ottici a causa
dell'elettrone rotante e sulla intensit\`a delle righe del Cesio'', in
collaborazione con Giovanni Gentile Jr.: {\em Rendiconti Accademia
Lincei}, vol. {\bf 8}, pp.229--233 (1928).
\item[2] --- ``Sulla formazione dello ione molecolare di He'': {\em
Nuovo Cimento}, vol. {\bf 8}, pp.22-28 (1931).
\item[3] --- ``I presunti termini anomali dell'Elio'': {\em Nuovo
Cimento}, vol. {\bf 8}, pp.78-83 (1931).
\item[4] --- ``Reazione pseudopolare fra atomi di Idrogeno'': {\em
Rendiconti Accademia Lincei}, vol. {\bf 13}, pp.58-61 (1931).
\item[5] - ``Teoria dei tripletti $P'$ incompleti'': {\em Nuovo
Cimento}, vol. {\bf 8}, pp.107-113 (1931).
\item[6] --- ``Atomi orientati in campo magnetico variabile'': {\em
Nuovo Cimento}, vol. {\bf 9}, pp.43-50 (1932).
\item[7] --- ``Teoria relativistica di particelle con momento
intrinseco arbitrario'': {\em Nuovo Cimento}, vol. {\bf 9} , pp.
335-344 (1932).
\item[8] --- ``\"{U}ber die Kerntheorie'': {\em Zeitschrift f\"{u}r
Physik}, vol. {\bf 82}, pp.137-145 (1933).
\item[8bis] --- ``Sulla teoria dei nuclei'': {\em La Ricerca
Scientifica}, vol. {\bf 4} (1), pp.559-565 (1933).
\item[9] --- ``Teoria simmetrica dell'elettrone e del positrone'':
{\em Nuovo Cimento}, vol. {\bf 14}, pp.171-184 (1937).
\item[10] --- ``Il valore   delle leggi statistiche nella fisica e
nelle scienze sociali'' (pubblicazione postuma, a cura di G. Gentile
Jr.): {\em Scientia}, vol. {\bf 36}, pp.55-66 (1942).
\end{description}

{\em Alcuni commenti:}\hfill\break

7) Questo \`e il famoso articolo con l'equazione quanto-relativistica a
   infinite componenti.

8) Questo \`e l'articolo con le ``forze di scambio" nucleari di
   Majorana-Heisenberg (che spiega ad es. come esse si saturino per la
   paerticella alfa).

9) Questo \`e il manoscritto che E.M. estrasse dal cassetto nel 1937 (era
   pronto dal 1932/33), e pubblic\`o, circa i neutrini di Majorana,
   la ``massa di Majorana", gli spinori di Majorana, etc.  \ All'inizio,
   esso fu notato quasi soltanto per la presenza della rappresentazione
   (di Majorana) delle matrici di Dirac.

10) Questo scritto, postumo, fu estratto da Giovannino Gentile (figlio di
    uno dei pi\'u famosi filosofi italiani della prima met\`a del secolo,
    cio\`e dell'ex-Ministro dell'Educazione Nazionale G.Gentile sr.) dalle
    carte lasciate da E.M. \ Si tratta di un articolo semi-divulgativo.
    Tra parentesi, G.Gentile jr. fu il primo a introdurre le parastatistiche
    (tanto che in America Latina vari autori chiamano ``gentilioni" le
    particelle che obbediscono a parastatistiche).\\

\section*{Capitolo II}

\section*{{\large {ELENCO DEI MANOSCRITTI SCIENTIFICI INEDITI DI
E. MAJORANA}} \footnote{A cura di M. Baldo, R. Mignani e E. Recami.}}

\section*{1. Introduzione}

\paragraph*{}
\h Si vuole qui dare breve notizia preliminare dei manoscritti {\em scientifici} inediti
lasciatici da Ettore Majorana\footnote{Si vedano ad es. E. Recami: ``Ettore Majorana: lo
scienziato e l'uomo", in Bibl.$^{(37)}$, pp.131--174; \ ed E. Recami: {\em Il
caso Majorana: Epistolario, Documenti, Testimonianze}, $2^{\rm a}$ ediz.
(Mondadori; Milano, 1991).} e a noi finora noti, e del relativo Catalogo. \
La maggior parte di tali manoscritti si trovano presso gli archivi della
``Domus Galilaeana" di Pisa.\footnote{Cfr. ad es. E. Amaldi: {\em La vita e
l'opera di E.Majorana} (Acc. dei Lincei; Roma, 1966).} \ Un catalogo pi\`u
preciso \`e contenuto nei successivi volumi, pi\`u tecnici, da noi co-edited
e riproducenti parte dei manoscritti scientifici lasciati inediti da Ettore
Majorana: usciti presso Kluwer (in inglese, 2003) e Zanichelli (in italiano, 2006),
e infine presso Springer (in inglese, 2009): si veda ad esempio l'e-print
arXiv:0709.1183[physics.hist-ph].

\h Oltre ai suoi appunti per le proprie lezioni universitarie tenute a Napoli
tra il Gennaio e il Marzo 1938 ---appunti recentemente
pubblicati\footnote{{\em Ettore Majorana -- Lezioni all'Universit\`{a} di
Napoli}, a cura di B.Preziosi (Bibliopolis; Napoli, 1987).}--- essi
comprendono essenzialmente: \ (a) la tesi di laurea; \ (b) dodici {\em
fascicoli} (riordinati da R. Liotta;\footnote{R. Liotta: in Bibl.$^{(36)}$,
pag.91.}
\ (c) cinque {\em Volumetti} manoscritti; e \ (d) diciotto {\em Quaderni}.

\h I ``Volumetti" sono stati redatti da Majorana tra il 1927 e il 1930,
tranne l'ultimo che \`{e} stato presumibilmente scritto nel 1932 (non prima,
perch\'{e} il Volumetto V$^{0}$ contiene a pag.8 la schematizzazione
dell'interazione  nucleare, mediante scattering da buca sferica a profilo
rettangolare, sotto il titolo ``Urto tra protoni e neutroni": e il nome
{\em neutrone} venne coniato nel 1932.\footnote{Ved. ad es.
P. Caldirola \& E. Recami: Voci ``Teorie
fondamentali" e ``Componenti fondamentali della materia", in {\em Scienza e
Tecnica del Novecento} (EST/Mondadori; Milano, 1977).}
e non dopo, perch\'{e} verso il termine vi si incontrano i prodromi del suo
articolo n.7, uscito nel 1932). \ Essi sono quaderni--libro, ordinatissimi,
divisi in capitoli, con pagine numerate e indice. I loro indici sono stati
gi\`{a} resi noti da Liotta.$^{(36)}$ \ Nei Volumetti ---scritti ciascuno
nel tempo di un anno circa--- Ettore sintetizza tutto ci\`{o} che ritiene
essenziale dei suoi {\em studi}, prima di studente e poi di ricercatore.
Come si \`{e} gi\`{a} detto altrove, tali Volumetti potrebbero essere
riprodotti fotograficamente, cos\'{\i} come sono, e costituirebbero un
ottimo testo {\em moderno} di consultazione in fisica teorica per gli studenti
universitari di oggi. \ Essi, tra parentesi, mettono in evidenza una delle
caratteristiche pi\'{u} geniali di Ettore: cio\`{e} la capacit\`{a} di
scernere fra tutto il materiale gli elementi matematici e fisici pi\'{u}
importanti per gli sviluppi futuri.

\section*{2. I ``Volumetti'': Cenno}

\paragraph*{}
A volte i ``Volumetti'' contengono anche spunti originali. Qui
segnaliamo, in breve, quanto segue.
\begin{description}
\item[Vol. II$^{\underline{0}}$]: nel capitolo 31, a pag. 78, Majorana cerca di
ricavare la relazione $e^2=\alpha \hbar c$;
\item[Vol. III$^{\underline{0}}$]: nel cap. 18, a pag. 105, sotto il titolo ``Matrici
di Dirac e Gruppo di Lorentz'' (scritto tra il 28.06.29 e il
23.04.30), tratta il problema delle rappresentazioni di un numero
generico ${p}$ di matrici di Dirac con un numero arbitrario
${n}$ di dimensioni: cio\`e il problema dell'equazione
d'onda relativistica di un oggetto con spin arbitrario in uno
spazio-tempo ${p}$-dimensionale;
\item[Vol. V$^{\underline{0}}$]: nel cap. 2, a pag. 8, tratta ---come
si \`e detto--- dell'urto fra il protone e l'appena scoperto neutrone
(prescindendo dallo spin del neutrone: ``se esiste'', dice); nel
cap. 8,  a pag. 36, comincia la trattazione delle rappresentazioni
unitarie a infinite dimensioni del gruppo di Lorentz, che sfocer\`a
nell'articolo n. 7 del 1932.
\end{description}

\section*{3. I Quaderni Scientifici}

\paragraph*{}
Il materiale che richiama la maggior attenzione \`{e} costituito dai
diciotto {\em Quaderni} scientifici, in cui Ettore stende le parti pi\'{u}
importanti delle sue ricerche a noi note (dopo i primi tentativi eseguiti,
insieme coi calcoli numerici, su fogli a parte: raccolti ora nei
{\em fascicoli}). Di questi Quaderni agli inizi degli anni Settanta  non
esisteva ancora alcun catalogo accettabile, dato che in Bibl.$^{(38)}$ erano
stati solo elencati i ``titoli" che Majorana stesso, saltuariamente e
casualmente, aveva voluto mettere all'inizio di qualche sua indagine teorica:
salvo poi, magari, interrompere tale indagine dopo mezza pagina per iniziarne
---senza alcun segnale--- una diversa, continuandola per parecchie pagine.
In tali anni, quindi, ci si accinse a redarne un Catalogo (ved.
bibliografia), che qua e l\`{a} presenta ancora qualche incertezza.

\h I Quaderni non recano date, e la loro numerazione (preesistente al nostro
intervento) non segue l'ordine cronologico: per esempio, Ettore compil\`{o}
il Quaderno IX$^{0}$ ancora da studente. \
 \ Osserviamo, tra parentesi, come l'esame dei manoscritti inediti suggerisca
che anche
il materiale per l'articolo n.9 (pubblicato solo nel 1937, alle soglie del
Concorso a cattedre universitarie) sia stato sostanzialmente preparato da
Ettore entro il 1933.

\h Naturalmente tra il materiale inedito (e non solo nei Quaderni) molti spunti
e molte idee hanno ancora
interesse scientifico {\em attuale}; noi abbiamo operato una selezione di tale
materiale: alcune centinaia di pagine [trasmesse in copia anche al Center for
History of Physics dell'A.I.P., New York, e relativa Niels Bohr Library]
possono essere ancora utili per la ricerca contemporanea. Una piccola
{\em parte} di esse sono state da noi studiate, interpretate e
pubblicate.\footnote{M.Baldo, R.Mignani \& E.Recami: ``About a Dirac--like
equation for the photon, according to Ettore Majorana", {\em Lett. Nuovo
Cimento} {\bf 11} (1974) 568; \ E.Recami: ``Possible physical meaning of the
photon wave--function according to E.Majorana", in {\em Hadronic Mechanics and
Non-Potential Interactions}, a cura di M.Mijatovich (Nova Sc. Pub.; New York,
1990), p.231.}\footnote{E.Giannetto: {\em Lett. Nuovo Cimento} {\bf 44} (1985)
140; \ {\bf 44} (1985) 145; \ E.Giannetto: in {\em Atti IX Congresso Naz.le
Storia della Fisica}, a cura di F.Bevilacqua (Milano, 1988); \ E.Giannetto:
``E.Majorana and the rise of Elementary particle theoretical physics", accettato
per la pubblicaz. su {\em Physis}; \ ``On Majorana's theory of arbitrary spin
particles", in corso di stampa sui {\em Proceedings of the School on the
Scientific Heritage of E.Majorana - Erice, 1989}; \ ``E.Majorana e il problema
degli stati ad energia negativa", in corso di stampa sugli {\em Atti del
Convegno sui Beni Culturali - Pavia, 1990}.}

\section*{Catalogo dei Quaderni Scientifici}

\subsection*{4.1. Quaderno 1}

\begin{description}
\item[p. 1:] Risoluzione dell'eq. di Schroedinger con {\em campo
coulombiano regolarizzato nell'origine} (ad es. per il caso de
scattering di elio su idrogeno): \\
(a) metodo perturbativo, con sostituzione di $\beta /{r}$
con $\beta /\sqrt{{r}^2+{a}^2}$; \\
(b) tentativo di risoluzione, con $\beta /{r}$ per
${r} > {R}$, e costante negativa per
${r} < {R}$; \\
(c) trattazione standard dello scattering da potenziale coulombiano.
\item[p. 14:] {\em Gruppo di Lorentz} ed equazioni relativistiche del
moto: lontana anticipazione dell'articolo n. 7 del 1932, con
introduzione degli operatori ${a}$ e ${b}$ ivi
contenuti. Ricordiamo che il manoscritto di tale articolo \`e
allegato al fascicolo n. 8, busta $II^{\underline{a}}$, e che
interessante vi \`e una pagina poi cancellata da Majorana.
\item[p. 26:] {\em Algebra degli spinori} di Dirac, in relazione
anche al suddetto articolo n. 7 (1932) e all'articolo n. 9 del 1937.
\item[p. 37:] Di nuovo Gruppo di Lorentz e Algebre spinoriali: {\em
equazioni relativistiche} (in relazione all'articolo n. 7). \\
Equazioni relativistiche al limite non-relativistico (mediante
decomposizione degli spinori a 4 componenti o di spinori pi\'u generali).
\item[p. 42:] {\em  Atomo di Idrogeno relativistico}.
\item[p. 48:] Appunti vari (eq. di Dirac; Gruppo di Lorentz).
\item[p. 50:] Appunti sulle regole di quantizzazione tipo Dirac.
\item[p. 51:] Da capo, {\em Atomo di H relativistico}: trattazione
standard, con tabulazione delle funzioni d'onda angolari.
\item[p. 64:] Onde sferiche relativistiche.
\item[p. 66:] Quantizzazione del campo elettromagnetico libero
(principio variazionale; {\em trasformazioni di Lorentz del campo
elettromagnetico}; gauge di Coulomb; quantizzazione). A questa pagina
erano inseriti dei fogli di carta da lettere listati a lutto (forse
del 1934, anno della scomparsa del padre).
\item[p. 76:] Seguono 25 pagine lasciate in bianco, apparentemente
per lavoro ancora da svolgere.
\item[p. 101:] {\em Teoria dell'elettrone}.     \\
Caso di due elettroni liberi. \\
Tentativo per il caso di due elettroni interagenti.
\item[p. 106:] Scattering di particelle da un potenziale (teoria
formale dello scattering): (a) metodo di Dirac; (b) metodo di Born;
(c) tentativo di calcolo al secondo ordine.
\item[p. 144:] Onda piana in coordinate paraboliche.
\item[p. 118:] Inizio di studio delle frequenze di oscillazione
(piccole oscillazioni) nell'NH$_3$.
\item[p. 121:] Passaggio di un atomo orientato nei pressi di un punto
di campo magnetico nullo (cfr. articolo n. 6 del 1932).
\item[p. 132:] Equazioni relativistiche del moto: Quantizzazione
della equazione di Dirac.
\item[p. 141:] Inizio di tabella sulle funzioni di Bessel.
\item[p. 145:] {\em Teoria di Dirac} (?): tentativo di introduzione
di insoliti operatori di traslazione spazio-temporale.
\item[p. 150:] Equazione di Dirac a massa nulla (equazione di Weyl).
\\
Inizio di teoria a due componenti del neutrino.
\item[p. 154:] Corpo rigido (ved. anche pag. 180).
\item[p. 161:] Orbitali interni del Calcio. Calcolo con potenziale
coulombiano pi\'u potenziale schermato (fenomenologico): risoluzione
approssimata, apparentemente {\em originale}. Caso $1s$.
\item[p. 180:] Rappresentazione del Gruppo delle rotazioni: cenno.
\item[p. 186:] Appunti di teoria degli stati instabili (cfr. la Tesi
di laurea). Cenno sulla correlazione di incertezza energia-tempo.
\end{description}

\subsection*{4.2. Quaderno 2}

\begin{description}
\item[p. 1:] Calcoli vari di elettromagnetismo classico.
\item[p. 3:] Problema dei due centri (ad es. per molecola H$_2$):
soluzione per vari casi {\em generali}, con calcolo della normalizzazione.
\item[p. 35:] Piccoli calcoli di Relativit\`a generale.
\item[p. 37:] Continuazione da pag. 112 (vedi).
\item[p. 38:] Campo di Dirac: calcoli vari. (``Versuchenweiser'').
\item[p. 46:] Equazione di Dirac: decomposizione in quattro equazioni
componenti, disaccoppiate al limite non relativistico. Idem, dedotta
da un principio variazionale; e nuova decomposizione. Problema della
definizione positiva della densit\`a di carica.
\item[p. 60:] Idem: calcolo degli stati stazionari.
\item[p. 69:] Carica deformabile ({\em problema risolto}).
\item[p. 75:] Trasformazioni di Lorentz ed Equazioni di Maxwell.
\item[p. 79:] ?
\item[p. 81:] Equazione relativistica per particella libera o in
campo elettromagnetico.
\item[p. 86:] Approccio preliminare al nucleo atomico come formato da
bosoni positivi e bosoni negativi. Campo scalare complesso per
particelle {\em cariche}. (Nel complesso, teoria non di facile interpretazione).
\item[p. 98:] Carica in moto relativistico.
\item[p. 101:] Elettrodinamica quantistica (quantizzazione del campo
elettromagnetico), fino a pag. 112.
\item[p. 101:] Alla pagina 101 sono {\em allegate sette pagine} (pag.
101/1 - 101/7) con lo studio delle analogie tra le equazioni di
Maxwell e l'equazione di Dirac (cfr. anche Quaderno 3, p. 20). Tale
allegato ha dato origine all'articolo ``About a Dirac--like equation
for the photon, according to E. Majorana'' (di M. Baldo, R. Mignani e
E. Recami): {\em Lett. Nuovo Cimento} {\bf 11} (1974) 568; e pi\'u
recentemente, agli articoli di E. Giannetto: {\em Lett. Nuovo
Cimento} {\bf 44} (1985) 140 e 145.
\item[p. 113:] Spinori di Dirac.
\item[p. 115:] Calcoli numerici.
\item[p. 121:] Ancora sul problema dei due centri (un elettrone e due
nuclei).
\item[p. 130:] Ancora sul problema coulombiano puro (nuovo metodo di risoluzione).
\item[p. 137:] Ancora sull'equazione di Dirac.
\item[p. 141:] Sovrapposizione di campi di Maxwell e Dirac. Calcoli
vari. Quantizzazione.
\item[p. 150:] Ancora sugli stati stazionari per l'equazione di
Dirac. Questione dell'elicit\`a (?).
\item[p. 157:] Calcoli perturbativi. Correlazione d'incertezza
energia-tempo. Calcoli vari. Approssimazioni varie. Equazioni algebriche.
\item[p. 170:] Calcolo approssimato di un integrale.
\item[p. 171:] Calcoli vari.
\item[p. 176:] Atomo di H in un campo elettrico.
\item[p. 178:] idem, pi\'u carica infinitesima.
\item[p. 182:] Equazioni di Maxwell.
\item
[p. 184:] Spinore di Dirac; {\em tetra-corrente di Dirac}.
Calcoli di non facile interpretazione.
\item[p. 195:] Matrici di Pauli.
\end{description}

\subsection*{4.3. Quaderno 3}

\begin{description}
\item[p. 1:] {\em Teoria di Dirac generalizzata a spin superiori}.\\
Trasformazioni infinitesime di Lorentz nella rappresentazione
ordinaria (in 4 dimensioni cartesiane).
\item[p. 2:] idem: teoria di Dirac a $2n(n+1)$ componenti.
\item[p. 8:] idem: casi $n=1$ (quattro componenti) e $n=2$ (dodici
componenti).
\item[p. 11:] Irraggiamento: equazioni di Maxwell con vettore
elettromagnetico complesso
$\underline{Z}=\underline{E}+i\underline{H}$.
\item[p. 16:] idem: caso del campo agente su una carica in moto
radiale. Introduzione di formalismo analogo a quello quantistico per
la descrizione del campo elettromagnetico (attraverso la posizione
$\psi_j =E_j-iH_j)$.
\item[p. 20:] Tentativo di scrittura delle equazioni di Maxwell in
maniera simile alla equazione di Dirac. (Questione della realt\`a
fisica della funzione d'onda del fotone?).
\item[p. 26:] elenco di alcuni argomenti.
\item[p. 28:] Tabella.
\item[p. 29:] Urto fra due elettroni (metodo di Mo$\!\!\!/$ller).
\item[p. 31:] Elettrone in campo elettromagnetico: calcoli per
trovare l'Hamiltoniana.
\item[p. 34:] L'operatore $\sqrt{1-\Delta_2}$ (operatore energia);
equazione di Klein-Gordon {\em non} quadratica?
\item[p. 35:] Riprende un precedente esercizio.
\item[p. 36:] Equazione di Dirac con campo centrale (atomo di H
relativistico). Correzioni relativistiche standard all'atomo di H.
\item[p. 38:] Scattering elastico coulombiano.
\item[p. 41:] Effetto Compton (teoria di Dirac): solo impostazione.
\item[p. 42:] Pagine lasciate in bianco, fino a pag. 60.
\item[p. 61:] Campo elettromagnetico in una scatola cubica.
\item[p. 63:] idem: quantizzazione in coordinate cartesiane. \\
Allegatevi due pagine (Z/1 e Z/2): ved. avanti.
\item[p. 67:] Sui campi elettromagnetico e di Dirac (elettrodinamica
quantistica: Hamiltoniana di un elettrone nel campo
elettromagnetico), apparentemente con campo elettromagnetico
quantizzato e campo di Dirac non quantizzato.
\item[p. 71:] Trasformazioni di Lorentz rappresentate mediante
matrici $2\times 2$. Rappresentazione spinoriale del Gruppo di Lorentz.
\item[p. 75:] pagine in bianco, fino a pag. 94.
\item[p. 95:] Irraggiamento dipolare: Calcoli perturbativi.
\item[p. 99:] Piccolo problema di elettrostatica.
\item[p. 100:] Effetto Auger: tentativo abortito.
\item[p. 101:] Calcoli sullo spettro continuo dell'energia di un
sistema.
\item[p. 102:] Calcoli elementari combinatoriali sul gruppo
simmetrico di permutazione (Tabella di Young).
\item[p. 103:] Teoria della diffusione coerente e incoerente (metodo
di Dirac)?
\item[p. 108:] Seguono cinque pagine in bianco.
\item[p. 113:] Questioni dal teste di Wittaker e Watson (Formule di
Darboux; Numeri e polinomi di Bernouilli; Equazioni differenziali del
$2^{\underline{0}}$ ordine).
\item[p. 119:] idem: funzione Gamma e applicazioni.
\item[p. 131:] idem: funzione $\zeta$ di Riemann.
\item[p. 135:] Ancora calcoli sul problema dell'atomo di H con carica
infinitesima vicina.
\item[p. 138:] Seguono cinque pagine bianche.
\item[p. 143:] Semplici calcoli.
\item[p. 144:] Continuazione da pag. 188 (vedi). Calcoli con segno di
richiamo uguale a quello presente alle pagine 37 e 112 del Quaderno 2.
\item[p. 155:] Irraggiamento in una cavit\`a. Quantizzazione.
\item[p. 160:] Ancora sul campo  elettromagnetico con vettore
elettromagnetico complesso: nuovo metodo per descrivere il campo
elettromagnetico   in analogia a quello di Dirac (apparentemente per
esplorare il significato della funzione d'onda quantistica).
\item[p. 162:] Campo elettromagnetico prodotto da un quadrupolo?
\item[p. 163:] Calcolo sulle {\em cariche magnetiche} (con pagina strappata).
\item[p. 165:] Calcoli vari.
\item[p. 166:] Piccoli calcoli sulla teoria perturbativa.
\item[p. 169:] Calcoli sul vettore di Poynting.
\item[p. 170:] Inizio di lavoro sulla equazione di Dirac, che
continua alle pagg. 180-188 e alle pagg. 144 e segg.
\item[p. 174:] Bianca
\item[p. 175:] Fino a pag. 179, parentesi di calcoli in Seconda
Quantizzazione sui problemi delle tre particelle (con correlazioni a
due particelle), dell'oscillatore armonico, e delle particelle
identiche. (?)
\item[p. 180:] Fino a pag. 188, riprende il lavoro di pag. 170 sulle
onde piane a frequenze positive e negative dell'equazione di Dirac.
Teoria di non facile interpretazione, probabilmente in preparazione
dell'articolo n. 9 (pubblicato poi nel 1937).
\item[p. 188:] Calcoli.
\end{description}

{\em {Allegati: A questo Quaderno 3 sono allegati vari fogli:}}

\begin{description}
\item[pp. A/1-1$\div$A/4-3] (15 pagg. numerate): Calcoli sulla
equazione di Dirac generalizzata a spin superiori: caso della teoria
a 12 componenti: [Ciascuno dei fogli $A/1\div A/4$ \`e costituito da
due pagine, ovvero da {\em quattro} facciate].
\item[pp. B/2-1$\div$B/2-4] (quattro pagg.): Calcoli sul momento
angolare per l'equazione di Dirac.
\item[pp. C/1-1$\div$C/1-4] (quattro pagg.): Calcoli
sull'equazione di Dirac con campo elettromagnetico.
\item[pp. C/11-1$\div$C/11-4] (quattro pagg.): idem.
\item[pp. Z/1$\div$Z/2]: Seconda Quantizzazione del campo di Dirac?
\end{description}

\subsection*{4.4. Quaderno 4}

\begin{description}
\item[p. 1:] Calcoli numerici.
\item[p. 7:] Atomo di H perturbato, e calcoli.
\item[p. 23:] Trasformazioni di Lorentz. Esercizi.
\item[p. 27:] Calcoli vari.
\item[p. 30:] Equazioni (di d'Alembert) delle onde: semplici calcoli.
\item[p. 32:] Trasformate di Fourier.
\item[p. 38:] Calcoli vari (Algebra gruppale; Funzioni di Eulero);
Relazioni di Eulero per un solido geometrico; Gruppo simmetrico).
\item[p. 46:] Corpo nero: semplici calcoli.
\item[p. 48:] Calcoli vari di Geometria sferica; sul Gruppo delle
rotazioni in 4 dimensioni; ecc.
\item[p. 54:] Calcoli di non facile interpretazione.
\item[p. 55:] Matrici del momento angolare (nello spazio ordinario)
per vari valori del momento.
\item[p. 60:] Equazione differenziale del $2^{\underline{0}}$ ordine
(una equazione agli autovalori, sembra generatrice di funzioni del
tipo delle ipergeometriche).
\item[p. 63:] Rotatore rigido; problema agli autovalori.
\item[p. 65:] Calcoli vari (perturbativi, ecc.). Perturbazioni
dipendenti linearmente dal tempo, {\em per tempi brevi}.
\item[p. 69:] Equivalenza energetica tra insieme canonico e microcanonico.
\item[p. 70:] Calcoli di un integrale.
\item[p. 71:] Semplici formule di termodinamica statistica.
\item[p. 74:] Equazione di Schroedinger per uno ione molecolare di
idrogeno: cenno.
\item[p. 74:] Alla pag. 74 sono {\em allegate tre pagine} ($74/1\div
74/3$), con calcoli vari.
\item[p. 77:] Semplici algebra astratta.
\item[p. 78:] Termodinamica standard (trasformazioni di variabili;
equazione di Claperyon; abbozzo di calcolo termodinamico per vapore
saturo in presenza del proprio liquido).
\item[p. 82:] Bianca.
\item[p. 83:] (Listino di borsa).
\item[p. 84:] Tre pagine in bianco.
\item[p. 87:] Ancora sulle equazioni tipo Dirac generalizzate: {\em
sviluppo esplicito} del caso a 12 componenti.
\item[p. 95:] Tre pagine lasciate in bianco.
\item[p. 98:] Moto in coordinate polari (elementi). Sviluppo di
un'onda piana in funzioni di Bessel. Equazione di Schroedinger in
coordinate polari; caso del potenziale coulombiano.
\item[p. 100:] Risoluzione (della parte radiale) dell'equazione di
Schroedinger con potenziale coulombiano, col metodo delle trasformate
di Laplace.
\item[p. 102:] Studio dei polinomi di Legendre.
\item[p. 106:] Studio della funzioni radiale per l'atomo di H
(Funzioni coulombiane).
\item[p. 106:] Vi sono {\em allegate tre pagine} ($106/1\div 106/3$),
sullo stesso argomento.
\item[p. 108:] Generatori delle rotazioni spaziali. Laplaciano in
quattro dimensioni: scritto {\em esplicitamente}. {\em Coordinate
polari in quattro dimensioni euclidee}. Gruppo delle rotazioni in
quattro dimensioni.
\item[p. 121:] Seguono 16 pagine bianche.
\item[p. 137:] Tentativo di introduzione di coordinate polari nello
spazio-tempo di Minkowski?
\item[p. 138:] Bianca.
\item[p. 139:] Equazione di Hamilton: semplici calcoli (cancellati).
Vi sono {\em allegate due pagine} ($139/1-139/2$), sul moto
relativistico di particella in campo elettromagnetico, e sulle
funzioni ipergeometriche.
\item[p. 143:] Equazione di Dirac per l'atomo di H (campo centrale):
pi\'u precisamente, correzioni {\em relativistiche} alle ``correzioni
di Rydberg per la struttura iperfina''.
\item[p. 149:] Ancora sull'equazione di Dirac per l'atomo di H:
struttura  iperfina.
\item[p. 154:] Equazione di Dirac a 4 componenti.
\item[p. 155:] Equazione di Dirac a sedici componenti (matrici
$16\times 16$).
\item[p. 158:] idem, a 6 componenti.
\item[p. 160:] idem, a 5 componenti (caso di Parastatistica?).
\item[p. 65:] Momenti magnetici e struttura iperfina: trattazione
standard. Calcolo sulla struttura fina (Formula di Land\'e).
\item[p. 169:] Calcolo relativistico del momento magnetico dell'atomo
di H.
\item[p. 171:] Metodo {\em originale} per il calcolo del momento
magnetico degli atomi (con la teoria di Dirac a sei componenti?).
\item[p. 174:] Equazione di Dirac a 4 componenti, con interessanti
{\em modifiche}.
\end{description}

\subsection*{4.5. Quaderno 5}

\begin{description}
\item[p. 1:] Elettrodinamica e Relativit\`a: equazione di Dirac per
elettrone e positone. [Il frontespizio di tale Quaderno reca la
parola ``Ghenos''].
\item[p. 3:] Equazione di Schroedinger (da un libro).
\item[p. 5:] Teoria quantistica dei campi, col formalismo
variazionale. Caso delle energie negative. Equivalenza tra
$I^{\underline{a}}$ e $II^{\underline{a}}$ quantizzazione.
\item[p. 8:] Equazione di Schroedinger per un sistema di N particelle.
\item[p. 11:] Oscillatore armonico unidimensionale quantistico, in
Seconda Quantizzazione.
\item[p. 14:] Separazione degli operatori di creazione e
annichilazione (non normali) in parte hermitiana e antihermitiana.
\item[p. 15:] Trasformazioni canoniche, con esempi.
\item[p. 17:] (?)
\item[p. 18:] idem: trasformazioni canoniche lineari.
\item[p. 20:] Semplici calcoli su trasformazioni canoniche.
\item[p. 23:] Moto piano di un punto in campo centrale.
\item[p. 24:] Limite non-relativistico dell'equazione di Dirac.
\item[p. 26:] Propriet\`a dell'operatore di traslazione in Meccanica Quantistica.
\item[p. 28:] Equazioni di Maxwell: principio variazionale.
\item[p. 31:] Evoluzione temporale di un insieme statistico, in
meccanica classica e quantistica.
\item[p. 32:] Introduzione di una famiglia
${a}_{pq}$ di operatori allo scopo di costruire
{\em in generale} l'operatore quantistico
${A}({p},{q})$
corrispondente ad una variabile dinamica classica  ${A}({p},{q})$:
Confronto tra meccanica classica e quantistica, anche ai fini
dell'interpretazione di quest'ultima. ({\em continua\/}).
\item[p. 44:] Esempio di trasformazione canonica infinitesima unidimensionale.
\item[p. 45:] Continuazione delle pagg. 32-43. ({\em continua ancora\/}).
\item[p. 51:] Struttura iperfina per spettri complessi (accoppiamento
R.S., ecc.).
\item[p. 65:] Varie equazioni d'onda (campi ritardati?). Continua a
pag. 76.
\item[p. 71:] Continuazione dalle pagg. 45-50. ({\em continua ancora\/}).
\item[p. 74:] Lagrangiane e Hamiltoniane varie.
\item[p. 75:] Trasformazioni di Lorentz.
\item[p. 76:] (in fondo) Introduzione di un ritardo $\tau$ in campi
ritardati (?): ved.  pagg. 65-66.
\item[p. 77:] Polinomi di Legendre e loro propriet\`a. Funzione
ipergeometrica; momenti angolari; ecc.
\item[p. 100:] Semplici calcoli.
\item[p. 102:] Termini atomici e loro propriet\`a (per l'azione di
operatori di momento angolare).
\item[p. 109:] Continuazione delle pagg. 71-73, con sviluppi in
$\hbar$. ({\em continua ancora\/}).
\item[p. 117:] Ancora sul principio variazionale per le equazioni di Maxwell.
\item[p. 119:] Continuazione delle pagg. 109-116, con sviluppi fino
ad $\hbar^3$.
\item[p. 124:] Seguono sei pagine pressoch\'e bianche.
\item[p. 130:] Semplici tabelle di integrali.
\item[p. 131:] Matrici di momenti angolari, forse in connessione con
la equazione di Dirac in 5 {\em dimensioni}.
\item[p. 137:] Seguono 10 pagine bianche.
\item[p. 147:] Calcoli sui momenti angolari.
\item[p. 150:] Calcoli vari.
\item[p. 156:] Atomo di Elio (ved. articolo n. 3, pubblicato nel 1931).
Continua alle pagg. 166-175.
\item[p. 164:] Calcoli.
\item[p. 166:] Continuazione delle pagg. 156-163.
\item[p. 176:] Tabella di polinomi di Legendre.
\item[p. 177:] Funzioni sferiche e atomo di He.
\item[p. 181:] Seguono 11 pagine bianche.
\item[p. 192:] Tabella di funzioni sferiche (per la molecola di idrogeno?).
\item[p. 194:] Calcoli di integrali. Trasformata di Fourier del
potenziale coulombiano.
\end{description}

\subsection*{4.6. Quaderno 6}

\begin{description}
\item[p. 1:] Conti numerici sullo ione molecolare di He (ved.
articolo n. 2, pubblicato nel 1931).
\item[p. 6:] Calcoli sulle varie rappresentazioni dell'operatore di
spin per l'equazione di Dirac.
\item[p. 8:] Altri calcoli sullo ione molecolare di He (problema di
tre fermioni).
\item[p. 15:] Buca di potenziale rettangolare unidimensionale: esercizio.
\item[p. 26:] Tabelle e calcoli sulle configurazioni elettroniche in
atomi leggeri.
\item[p. 29:] Calcoli sui termini anomali dell'He (cfr. articolo n. 3).
\item[p. 31:] Oscillazioni armoniche a tre gradi di libert\`a.
\item[p. 37:] Oscillazioni armoniche a 2 gradi di libert\`a, o con
masse uguali, o con masse diverse (oscillatori accoppiati).
\item[p. 41:] idem, formulazione generale.
\item[p. 42:] Coefficienti di Clebsh-Gordon.
\item[p. 44:] Configurazioni elettroniche nella molecola di H$_2$.
\item[p. 46:] Calcoli per l'articolo n. 3, del 1931.
\item[p. 51:] Calcoli sulla molecola di H$_2$O.
\item[p. 52:] Fino a pag. 98, ricerca delle soluzioni radiali
dell'equazione di Schroedinger con potenziale
${V}_0 = (-1/{x}+11/16) \ \exp[-11 {x}/8]$ o, pi\'u in generale, con
potenziale $V=-{c} \ \exp[-2 {a} {x}]$.
\item[p. 99:] Conti sullo ione molecolare di He: elenco dei simboli usati.
\item[p. 101:] Tabelle e conti numerici.
\item[p. 107:] Calcoli di serie, integrali; {\em calcoli soprattutto
per} l'He, e per la reazione pseudopolare tra atomi di H (cfr.
articolo n. 4, del 1931).
\item[p. 118:] Fino a pag. 193, grosse tabelle e molti calcoli per
l'articolo n. 4 ({\em reazione pseudopolare tra atomi di} H).
\item[p. 194:] Programma di lavori e/o articoli futuri, sulla base
anche dei conti gi\`a eseguiti in questo Quaderno 6:
(i) La formazione di He$_2^+$, (ii) Reazione pseudopolare tra atomi
di H; (iii) Serie ortogonale di operatori anticommutanti; (iv) Il
termine anomalo $2p\ 2p^3\ P$ dell'Elio; (v) Energia di atomi leggeri;
(vi) Termini anomali di Roentgen; (vii) Il doppietto $2p^2P$ del
Litio; (viii) Intensit\`a nei Raggi X; (ix) Gruppi $pp'$ incompleti.
\end{description}

\subsection*{4.7. Quaderno 7}

\begin{description}
\item[p. 1:] Polinomi di Legendre, ecc.
\item[p. 6:] Fino a pag. 60, calcoli relativi all'articolo sui
termini anomali dell'Elio (cfr. la settima pagina dell'articolo n. 3).
\item[p. 61:] Fino a pag. 116, calcoli relativi all'articolo sulla
teoria dei tripletti $P'$ incompleti (articolo n. 5, del 1931).
\item[p. 117:] Risonanza tra un elettrone con $\ell =1$ e un
elettrone con $\ell$ generico, in un atomo.
\item[p. 123:] Tabelle e calcoli numerici sul ``potenziale
statistico'' di Thomas-Fermi: in preparazione dell'articolo n. 1, del
1928, uscito in coll. con G. Gentile Jr.
\item[p. 138:] idem, altri calcoli fino a pag. 161.
\item[p. 161:] Risoluzione dell'equazione di Dirac (con o senza campo
elettromagnetico), nelle rappresentazioni standard e spinoriale.
\item[p. 172:] Equazione di Dirac con campo centrale.
\item[p. 175:] Caso particolare: campo coulombiano.
\item[p. 178:] Equazione di Schroedinger; formulazione variazionale.
\item[p. 180:] Equazione di Pauli, e suo confronto con quella di Dirac.
\end{description}

\subsection*{4.8. Quaderno 8}

\begin{description}
\item[p. 1:] Calcolo combinatorio.
\item[p. 12:] Due pagine bianche.
\item[p. 14:] Meccanica statistica per la teoria del ferromagnetismo.
\item[p. 30:] Calcoli di teoria delle perturbazioni.
\item[p. 36:] Tre oscillatori accoppiati.
\item[p. 40:] Due pagine bianche.
\item[p. 42:] Sistemi di equazioni, lineari e non.
\item[p. 46:] Altri conti statistici, forse in relazione al
precedente problema del ferromagnetismo, ma ora quantistici. Da pag.
66  a pag. 74, per\`o, calcoli di equazioni integrali (ad es. di Fredholm).
\item[p. 86:] Fino a pag. 111, altri calcoli quantistici di non
facile interpretazione (Funzioni di Bessel; Risoluzione di equazioni
differenziali di $II^{\underline{0}}$ grado; Equazione di Bessel sferica).
\item[p. 112:] Calcoli di meccanica analitica, collegati con quanto
precede, e in particolare con quanto alle pagg. 75-78.
\item[p. 118:] Funzioni tipo Bessel  o Neumann, con grafici e
tabulazioni. Relazioni di ricorrenza.
\item[p. 125:] Risoluzione dell'equazione di d'Alembert in coordinate
sferiche.
\item[p. 132:] Polinomi di Legendre, ecc.; meccanica analitica.
\item[p. 144:] Ottica geometrica elementare; principio di Huygens; oscillazioni.
\item[p. 157:] Trascrizioni analitiche dei principi di Fermat e di
Huygens. Radiazione di onde elettromagnetiche.
\end{description}

\subsection*{4.9. Quaderno 9 \footnote{Quaderno da studente, di antica data.}}

\begin{description}
\item[p. 1:] Esercizi vari (ottica geometrica e fisica, ecc.).
\item[p. 28:] Scarica nei gas; esperienze varie; ioni (studio ed
esercizi); elettrostatica (esercizi); cariche in campi elettromagnetici.
\item[p. 41:] Studio di esperienze varie, specie per determinare la
carica dell'elettrone.
\item[p. 53:] Studio degli oscillografi; di vari Effetti; di valvole termoioniche.
\item[p. 66:] {\em Equazione delle onde classiche:} Ottica delle lenti.
\item[p. 76:] Esercizi vari (equazione del moto ellittico piano; trigonometria).
\item[p. 84:] Ottica fisica e geometrica.
\item[p. 106:] Integrazioni su iperboloidi; elementi di volume (in
relazione con l'Ottica). Integrali. Coniche; quadriche.
\item[p. 120:] Esercizi vari; integrali; ecc.
\item[p. 151:] {\em Equazione delle onde quantistiche} (di
Schroedinger). Esercizi. Momento angolare.
\item[p. 164:] Altri esercizi (calcolo di funzioni matematiche;
serie; meccanica analitica; equazione delle onde in coordinate sferiche).
\item[p. 188:] Esercizi di Meccanica Quantistica.
\end{description}

\subsection*{4.10. Quaderno 10 \footnote{Quest'altro Quaderno n. 10
contiene essenzialmente il materiale preparato per la Tesi di laurea.}}
%%%sembra contenere i capitoli di un libro, o di una pubblicazione
%%%universitaria (tipo ``Dispense''), o di un lavoro semi-divulgativo.}}

\begin{description}
\item[p. 1:] {\em $I^{\underline{0}}$ Capitolo} (pagg. 1-17): Ionizzazione
spontanea di un atomo di H posto in una regione a potenziale negativo.
\item[p. 18:] {\em $II^{\underline{0}}$ Capitolo} (pagg. 18-26):
Legge fondamentale dei fenomeni radioattivi.
\item[p. 29:] Seguono tre pagine bianche.
\item[p. 30:]   {\em $III^{\underline{0}}$ Capitolo} (pagg.
30-39): Urto di una particella alfa contro un nucleo radioattivo.
\item[p. 40:] Seguono quattro pagine bianche.
\item[p. 44:]  {\em $IV^{\underline{0}}$ Capitolo} (pagg.
44-52): Calcoli di Gamow e Huntermans (per la Tesi di laurea).
\item[p. 53:] Seguono tre pagine bianche.
\item[p. 56:] {\em Introduzione} (pagg. 56-65).
\item[p. 66:] Bianca.
\item[p. 67:] Integrali, ecc.; esercizi di Ottica.
\item[p. 98:] Soluzione di equazione differenziale col metodo delle
funzioni di Green. Ottica varia. Teoremi di Green e di Stokes.
\item[p. 190:] Permutazioni: esercizi.
\end{description}

\subsection*{4.11. Quaderno 11 \footnote{{\em Studi vari}, da testi universitari.}}

\begin{description}
\item[p 1:] Teoria dei gruppi: calcoli.
\item[p. 6:] Cinque pagine bianche.
\item[p. 12:] Calcoli numerici.
\item[p. 13:] Calcoli sull'Elio.
\item[p. 29:] Fino   a pag. 64, metodo di Hartree per atomi con due
elettroni (calcoli approssimati).
\item[p. 65:] Sette pagine bianche.
\item[p. 72:] Ancora calcoli per l'Elio.
\item[p. 86:] Polarizzazione dell'elio (metodo di Hartree).
\item[p. 88:] Studio di operatori differenziali lineari. Sistemi di
equazioni lineari. Matrici. Parentesi di Poisson. Sistemi completi di
operatori.
\item[p. 94:] Equazioni simboliche del parallelismo. Simboli di
Christoffel. Geometria riemanniana. Geodetiche.
\item[p. 107:] Equazione di Pauli, e passaggio all'equazione di
Dirac, in due rappresentazioni. Hamiltoniana relativistica di un
elettrone in campo elettromagnetico.
\item[p. 113:] Ancora geometria riemanniana. Derivazione covariante.
Geometria differenziale. Calcolo tensoriale in spazi non euclidei.
\item[p. 160:] Equazioni di Schroedinger e di Dirac; ecc.
\item[p. 166:] Calcolo tensoriale in spazi di Riemann; ecc.
\item[p. 172:] Semplice problema agli autovalori.
\item[p. 174:] Permutazioni.
\item[p. 180:] Esercizi vari di Meccanica Quantistica. Trasformazioni
di Fourier tridimensionali. Onde piane sviluppate in onde parziali.
Polinomi di Legendre; ecc.
\end{description}

\subsection*{4.12. Quaderno 12}

\begin{description}
\item[p. 1:] Elaborazione teorica di non facile interpretazione.
Meccanica Quantistica (polinomi vari, per calcoli di valori medi).
\item[p. 16:] Serie. Equazioni integrali.
\item[p. 21:] Meccanica Quantistica: esercizi; regole di
commutazione, ecc.; equazione di Schroedinger; esercizi vari.
\item[p. 32:] Equazione di Dirac, e suo limite relativistico. Idem,
con campo centrale e campo elettromagnetico.
\item[p. 45:] Meccanica analitica (?).
\item[p. 48:] Sviluppi sui polinomi di Legendre. Integrali vari.
\item[p. 51:] Esperienza di Townsend.
\item[p. 53:] Ancora sul limite non relativistico dell'equazione di
Dirac (elettrone rotante in campo centrale: cenno).
\item[p. 54:] Onde superficiali in un liquido.
\item[p. 58:] Nucleo di carica ${Z}{e}$ con due
elettroni: energia dello stato fondamentale. Calcoli col metodo
perturbativo, e col metodo del minimo (principio variazionale); e
studi vari su quest'ultimo metodo.
\item[p. 70:] Rappresentazioni integrali delle funzioni di Bessel.
\item[p. 76:] Oscillazioni forzate di un elettrone in campo elettrico
alternato.
\item[p. 79:] Ancora sulle funzioni di Bessel.
\item[p. 82:] Moto anarmonico classico (moto ``dispersivo'' di un elettrone).
\item[p. 88:] Meccanica analitica.
\item[p. 90:] Integrali nel piano complesso.
\item[p. 92:] Ancora funzioni di Bessel.
\item[p. 96:] Funzioni sferiche di Legendre. Integrali. Funzioni di
Legendre di $2^{\underline{a}}$
specie.
\item[p. 101:] Spazi vettoriali ad ${n}$ dimensioni.
Calcoli con matrici. Spazi duali. Teoria degli spazi vettoriali a
dimensione finita.
\item[p. 112:] Tabella di Mendeleev.
\item[p. 130:] Ancora spazi vettoriali ${n}$-dimensionali.
Trasformazioni hermitiane e unitarie in ${n}$ dimensioni.
Diagonalizzazione. Trasformazioni unitarie infinitesime.
\item[p. 142:] Passaggio agli spazi di Hilbert (spazi a dimensione infinita).
\item[p. 145:] Integrali: calcoli per l'articolo (n .3) sull'Elio.
\item[p. 151:] Bianca.
\item[p. 152:] Spazi di Hilbert (secondo il libro di Weyl).
\item[p. 154:] Integrali, per l'Elio.
\item[p. 155:] Equazione di Schroedinger.
\item[p. 156:] Cenno sul diamagnetismo.
\item[p. 157:] Fino a pag. 188, integrali per l'articolo sull'Elio.
\end{description}

\subsection*{4.13. Quaderno 13}

\begin{description}
\item[p. 1:] Calcoli numerici.
\item[p. 2:] Meccanica analitica: equazioni canoniche.
\item[p. 3:] Fino a pag. 13, rappresentazione di Majorana delle
matrici di Dirac (rappresentazione reale). Calcoli per l'articolo n.
9 (teoria simmetrica dell'elettrone e del positone).
\end{description}

\subsection*{4.14. Quaderno 14}

\begin{description}
\item[p. 1:] Fino a pag. 8, geometria negli spazi di Riemann.
\end{description}

\subsection*{4.15. Quaderno 15}

\begin{description}
\item[p. 1:] Equazione
${n}_1+2{n}_2+3{n}_3+\ldots = {N}$.
\item[p. 6:] Equazione di Schroedinger in coordinate sferiche.
\item[p. 8:] Due pagine cancellate.
\item[p. 10:] Calcoli sull'equazione di Dirac.
\item[p. 16:] Equazioni di Maxwell.
\item[p. 18:] Calcoli con trasformazioni di Lorentz. (?)
\item[p. 22:] Gruppo di Lorentz (ved. anche Quaderno 1).
Rappresentazioni a infinite componenti (per l'articolo n. 7, del 1932).
\item[p. 26:] Equazione di Dirac.
\end{description}

\subsection*{4.16. Quaderno 16 \footnote{Gli {\em studi} presenti in
questo Quaderno sembrano basati sul testo del Weyl}.}

\begin{description}
\item[p. 1:] Molecola di Elio.
\item[p. 23:] Tre pagine bianche.
\item[p. 26:] Ancora sull'Elio.
\item[p. 31:] Spazi ${n}$-dimensionali: Algebre e teoria
dei gruppi. Permutazioni. Algebra invariante a sinistra.
Rappresentazioni (equivalenti e non; irriducibili; ecc.) del gruppo
delle permutazioni di ${f}$ particelle.
\item[p. 50:] Equazione tipo Schroedinger per due particelle.
\item[p. 56:] Bianca.
\item[p. 57:] Sistemi di  ${f}$ particelle. Caratteri del
gruppo delle permutazioni. Teoria dei gruppi.
\item[p. 76:] Tre pagine bianche.
\item[p. 79:] Autofunzioni del Litio.
\item[p. 83:] Scattering di Thomson.
\item[p. 84:] Ancora sul Litio.
\item[p. 98:] Energia del Litio (termine fondamentale del Litio).
\item[p. 100:] Campo autoconsistente in atomo con due elettroni.
\item[p. 103:] Calcoli numerici.
\item[p. 112:] Tabella (per il Litio?).
\item[p. 114:] Calcoli numerici.
\item[p. 118:] Di nuovo Tabella.
\item[p. 120:] Calcoli numerici e Tabella.
\item[p. 134:] Calcoli algebrici.
\item[p. 141:] Semplici calcoli.
\item[p. 157:] Stato fondamentale di atomi pesanti (per
${Z}$ tendente all'infinito) con tre elettroni.
\item[p. 158:] Andamento asintotico per i termini ${s}$ dei
metalli alcalini.
\item[p. 162:] Termine fondamentale del Litio? Integrali. Serie.
\item[p. 174:] Algebre e gruppo delle permutazioni in ${n}$
dimensioni.
\item[p. 175:] Calcoli numerici.
\item[p. 185:] Quattro pagine bianche.
\item[p. 189:] Equazione agli autovalori in spazi a dimensione finita.
\item[p. 190:] Pagina cancellata.
\end{description}

\subsection*{4.17. - Quaderno 17}

\begin{description}
\item[p. 1:] Rappresentazione del Gruppo di Lorentz. Equazione di
Schroedinger. Studi vari. Zeri delle funzioni di Bessel.
\item[p. 8:] Teoria dei nuclei (per l'articolo n. 8, del 1933).
\item[p. 31:] Nuclei semplici: studio di due forme di interazione.
\item[p. 35:] Meccanica statistica di due  particelle.
\item[p. 36:] Momento magnetico e suscettivit\`a magnetica di un
atomo con ${m}$ elettroni (trattazione relativistica).
\item[p. 39:] Trasformazioni generali di matrici.
\item[p. 40:] Teoria simmetrica (in preparazione dell'articolo n. 9).
Continua alle pagg. 74-81.
\item[p. 43:] Trasformazione di matrici.
\item[p. 45:] Equazione di Dirac ``reale'' (senza campo, e con campo
non relativistico).
\item[p. 69:] Analogia Dirac-Maxwell (continuazione da pag. 160).
\item[p. 72:] Calcoli numerici. Momento angolare in coordinate
sferiche tetradimensionali; ecc.
\item[p. 74:] Altri calcoli per la teoria simmetrica elettrone/positone
(dopo $1^{a}$ pagina 81, proseguono alle pagg. 40-42).
\item[p. 82:] Polinomio (calcoli).
\item[p. 83:] Trasformazioni di spinori (studio della analogie tra
equazione di Dirac e equazioni di Maxwell).
\item[p. 88:] Bianca.
\item[p. 89:] Equazione di Dirac.
\item[p. 92:] Teoria perturbativa.
\item[p. 94:] Tre  pagine bianche.
\item[p. 97:] Gas degeneri.
\item[p. 98:] Calcoli in coordinate polari; equazione di Schroedinger.
\item[p. 104:] Studio delle relazioni tra prodotto quantistico e
prodotto classico.
\item[p. 127:] Due pagine bianche.
\item[p. 129:] Equazione d'onda del neutrone; e scattering di fotoni
su neutrone.
\item[p. 146:] Ancora teoria simmetrica di elettrone e positone.
\item[p. 150:] Fino a pag. 151{\em bis}, equazione d'onda del
neutrone; ecc.
\item[p. 151{\em ter}:] Fino a pag. 153, calcoli geometrici.
\item[p. 154:] Autofunzioni atomiche.
\item[p. 156:] Gruppo di Lorentz.
\item[p. 159:] Ancora sulle analogie Maxwell-Dirac.
\item[p. 161:] Autofunzioni atomiche: termini 2{\em s} del Litio.
\item[p. 167:] Calcoli di Meccanica Quantica. Integrali nel piano
complesso. Equazione radiale di Schroedinger; ecc. Altri integrali.
\item[p. 176:] Equazione di Dirac.
\item[p. 177:] Autofunzioni atomiche.
\item[p. 179:] Calcoli di Meccanica Quantica: commutatori, ecc.
\item[p. 183:] Fino a pag. 190,  Formulario.
\end{description}

\subsection*{4.18. Quaderno 18}

\begin{description}
\item[p. 1:] Equazioni di Maxwell: calcoli variazionali multidimensionali.
\item[p. 8:] Calcoli vari: integrali; rappresentazione integrale di
funzioni di Bessel; funzioni di Hankel; soluzioni di scattering della
equaz. di Schroedinger; funzione di Green generalizzata, e metodo
della fz. di Green; integrali vari; equazioni di Hamilton; equazione
di Schroedinger e sua risoluzione per serie; derivate; formule
trigonometriche; equazioni differenziali; calcoli di Meccanica Quantica.
\item[p. 34:] Equazione di Schroedinger per due particelle: metodo di
Ritz. Equazioni integrali. Integrali vari. Equazione differenziale
$y''=xy$. Regioni d'integrazione.
\item[p. 54:] Meccanica analitica. Calcoli numerici e algebrici.
Equazione di van der Waals. Termodinamica.
\item[p. 61:] Meccanica Quantica: semplici calcoli; teoria delle
perturbazioni; funzioni d'onda di molti-corpi.
\item[p. 69:] Seconda Quantizzazione.
\item[p. 74:] Calcoli vari; calcolo combinatorio; ecc.
\item[p. 89:] Termini anomali dell'Elio (per l'articolo n. 3);
Tabelle relative.
\item[p. 106:] Integrali; calcoli numerici; calcoli geometrici; altri
calcoli per l'articolo sull'Elio; e altro.
\item[p. 128:] Calcoli per l'articolo n. 4, del 1931 (reazione
pseudopolare tra atomi di Idrogeno?).
\item[p. 134:] Tabelle di integrali. Di nuovo calcoli sulla reazione
pseudopolare tra atomi di H (?); e {\em sull'Elio}.
\item[p. 156:] Equazioni differenziali; equazione
$y''_j-ky_j=0$; teoria di Pauli del paramagnetismo; orto- e para-elio.
\item[p. 158:] Ancora sui termini anomali dell'He.
\end{description}

\

\section*{Ringraziamenti e Avvisi}

Per la fattiva collaborazione ai fini della realizzazione di
questo lavoro, l'autore \`e molto grato a Marcello Baldo, Franco
Bassani, Francisco Caruso, Carlo Castagnoli, Francesco del Franco,
Susanna De Maron, Francisca V.Fortaleza-Gomes, M\'ario Giambiagi,
J. Leite Lopes, Ettore Majorana Jr., Alwyn van der Merwe, Roberto
Mignani, Paolino Papali, Pio Picchi, Bruno Preziosi, Renato A.
Ricci, M\'{\i}riam Segre Giambiagi, Am\'os Troper, Pasquale Tucci
e Carmen Vasini, oltre che alla famiglia Majorana di Roma e
Catania, e alla ``Domus Galilaeana'' di Pisa (Prof. Derenzini,
Prof. Maccagni, Prof.C.A.Segnini, Dr.ssa A.Colotto, Dr.D.Ronco,
Dr. Tricarichi, Sig.na Puccianti, e Sig. Guerri). Ringrazia
inoltre, per la generosa cooperazione, Dharam Ahluwalia, Edoardo
Amaldi, Carlo Becchi, Gilberto Bernardini, Nicola Cabibbo,
Giuseppe Cocconi, Aldo Covello, Mimmo De Maria, Antonino Drago,
Donatello e Fosco Dubini, Salvatore Esposito, Myron Evans, Alberto
Gabriele, Enrico Giannetto, Fran\c{c}oise Gueret, Philippe Gueret,
Antonio Insolia, Francesco Izzo, Corrado Leonardi, Fabrizio Lillo,
Annamaria Papa, Franco Rasetti, Umberto Recami, Tina Roberto,
Bruno Russo, Laura R. Sansoni, Gianni Sansoni, Edvige Schettino,
Leonardo Sciascia, Emilio Segr\'e, Gilda Senatore, Paolo Strolin,
Franco Strumia, Alexander Tenenbaum, Ettore G.Vaccaro, Victor
Weisskopf, Giancarlo Wick e Daniel Wisniveski. \ Tutto il
materiale di questo articolo \`e tratto dal nostro libro ``Il Caso
Majorana: Epistolario, Documenti, Testimonianze" (Mondadori,
Milano, 1987,1991; Di Renzo, Roma, 2000-2008), e si rimandano a
tale libro i lettori interessati a maggiori e pi\`u profonde
informazioni; nonch\'e, per questioni pi\`u tecniche, ai
successivi volumi riproducenti ad esempio parte dei manoscritti
scientifici lasciati inediti da Ettore Majorana: si veda, p.es.,
l'e-print arXiv:0709.1183v1[physics.hist-ph]. \ [Proprietà
letteraria riservata. \ Nessuna parte di questo articolo pu\`o
essere riprodotta o trasmessa in qualsiasi forma o con qualsiasi
mezzo elettro-nico, meccanico o altro senza l'autorizzazione
scritta dei proprietari dei diritti (Erasmo Recami e Maria
Majorana)]. {\em {\bf P.S.:} The present material is mainly  taken
from our book ``Il Caso Majorana: Epistolario, Documenti,
Testimonianze" (Mondadori, Milan, 1987,1991; Di Renzo, Rome,
2000-2008). That book presents practically all the serious
documents Existing on Majorana's life \& work [indeed, almost all
the biographical documents have been discovered or collected,
during a few decades, by the present author, who was the first to
publish them]. We address to such a book  (c/o www.direnzo.it,
``Arcobaleno" series) all the readers interested in more and
deeper information; as well as, for more technical topics, to the
subsequent volumes reproducing e.g. part of the scientific
manuscripts left unpublished by Ettore Majorana: see, for
instance, the e-print {\rm arXiv:0709.1183v1[physics.hist-ph]}}.\\

\

\section*{Bibliografia}

\paragraph*{}
AA.VV.: {\em Scienziati e tecnologi contemporanei: Enciclopedia
Biografica}, 3 voll., a cura di E.Macorini (Milano, 1974).\\

E. Amaldi: {\em La Vita e l'Opera di E.Majorana} (Accademia dei Lincei; Roma,
1966).\\

E. Amaldi: ``Ettore Majorana: Man and Scientist", in {\em Strong and Weak
Interactions}, a cura di A.Zichichi (New York, 1966).\\

E. Amaldi: ``Ricordo di Ettore Majorana", in {\em Giornale di Fisica} 9
(Bologna, 1968) p.300.\\

E. Amaldi: ``From the discovery of the neutron to the discovery of
nuclear fission", in {\em Physics Reports} 111 (1984) pp.1--322.\\

E. Amaldi: in {\em Il Nuovo Saggiatore} 4 (Bologna, 1988) p.13.\\

M. Baldo, R. Mignani e E. Recami: ``Catalogo dei manoscritti scientifici
inediti di E.Majorana", in {\em E.Majorana -- Lezioni all'Universit\`a di
Napoli} (Bibliopolis; Napoli, 1987), p.175.\\

M. Bunge: {\em La Causalit\`a} (Torino, 1970).\\

F.L. Cavazza e S.R. Granbard: {\em Il Caso Italiano: Italia Anni '70}
(Milano, 1974).\\

{\em Conferenze e Discorsi di Orso Mario Corbino} (Roma, 1939).\\

D.De Masi (a cura di): {\em L'Emozione e la Regola: I Gruppi Creativi in
Europa dal 1850 al 1950} (G.Laterza; Bari, 1989).\\

F. e D. Dubini: ``La scomparsa di Ettore Majorana", programma televisivo
trasmesso nel 1987 (TV svizzera).\\

G. Enriques: {\em Via D'Azeglio 57} (Zanichelli; Bologna, 1971).\\

S. Esposito: ``Covariant Majorana formulation of electrodynamics", in
{\em Found. of Phys.} 28 (1998) 231--244.\\

S.Esposito, E.Majorana jr., A. van der Merwe e E.Recami (editors):
{\em Ettore Majorana - Notes on Theoretical Physics}
(Kluwer; Dordrecht and N.Y., Nov. 2003); 512 pages.\\

S.Esposito e E.Recami (editors):
{\em Ettore Majorana - Appunti Inediti di Fisica Teorica}
(Zanichelli, Bologna, 2006); 552 pages.\\

S.Esposito, E.Recami, A. van der Merwe e R.Battiston (editors):
{\em E.Majorana --
Unpublished Research Notes on Theoretical Physics},
(Springer; Berlin, 2009); 487 pages.\\

G. Fraser: in {Cern Courier} 38, issues no.5 and 6 (Summer and Sept.,
1998).\\

M. Farinella: in {\em L'Ora} (Palermo), 22 e 23 luglio 1975.\\

E. Fermi: ``Un maestro: O.M.Corbino", in {\em Nuova Antologia} 72 (1937)
p.313.\\

L. Fermi: {\em Atomi in Famiglia} (Milano, 1954).\\

C. Fontanelli: ``Il caso Ettore Majorana: Aspetti storici e filosofici", tesi
di laurea: Relatore A. Pagnini (Fac. Lett. Filos.; Univ. di Firenze, 1999).\\

B. Gentile: ``Lettere inedite di E.Majorana a G.Gentile jr.", in
{\em Giornale critico della filosofia italiana} (Firenze, 1988) p.145.\\

E. Giannetto: ``Su alcuni manoscritti inediti di E.Majorana", in {\em
Atti IX Congresso Naz.le di Storia della Fisica}, a cura di F.Bevilacqua
(Milano, 1988) p.173.\\

G.C. Graziosi: ``Le lettere del mistero Majorana", in {\em Domenica del
Corriere} (Milano), 28 novembre 1972.\\

G. Holton: {\em The Scientific Information: Case Studies} (Cambridge,
1978).\\

C. Leonardi, F. Lillo, A. Vaglica e G. Vetri: `` Quantum visibility,
phase-difference operators, and the Majorana Sphere", preprint (Phys.Dept.,
Univ. of Palermo, Italy; 1998), to appear; \ ``Majorana and Fano
alternatives to the Hilbert space", in {\em Mysteries, Puzzles, and Paradoxes
in Quantum Mechanics}, ed. by R.Bonifacio (A.I.P.; Woodbury, N.Y., 1999),
pp.312-315. \ ved. anche F.Lillo: ``Aspetti Fondamentali nell'Interferometria
a Uno e Due Fotoni", Tesi di Dottorato  (relatore C.Leonardi), Dip.to di
Fisica, Universit\`a di Palermo, 1998.\\

A. Majorana: ``La questione degli spostati e la riforma dell'Istruzione
Pubblica in Italia", discorso alla Camera dell' 11 marzo 1899
(Roma, 1899).\\

{\em Ettore Majorana -- Lezioni all'Universit\`a di Napoli}, ed. by
B.Preziosi (Bibliopolis; Napoli, 1987).\\

G., A. e D. Majorana: {\em Della Vita e delle Opere di Salvatore
Majorana Calatabiano} (Catania, 1911).\\

R. Mignani, E. Recami e M. Baldo: ``About a Dirac--like equation for the
photon, according to E.Majorana", {\em Lett. Nuovo Cimento} 11 (1974) p.568.\\

R. Penrose: {Ombre della Mente (Shadows of the Mind)} (Rizzoli; 1996),
pp.338--343 e 371--375.\\

R. Penrose: ``Newton, quantum theory and reality", in {\em 300 Years of
Gravity}, ed. by S.W.Hawking \& W.Israel (Cambridge Univ.Press; 1987).\\

B. Pontecorvo: {\em Fermi e la Fisica Moderna} (Roma, 1972).\\

B. Pontecorvo: contributo al {\em Congresso sulla storia della fisica
delle particelle (Parigi, 1982)}.\\

S. Ponz de Leon: ``Speciale News: Majorana", trasmesso il 30.9.1987
(Canale Cinque).\\

E. Recami: {\em Il caso Majorana: Epistolario, Documenti, Testimonianze}, $2^{\rm a}$
edizione, nella serie ``Oscar" (Mondadori; Milano, 1991), pp.1--230: esaurita;
e $5^{\rm a}$ edizione (Di Renzo: Roma, 2008), pp.1-273. \ [Della seconda edizione di
questo volume esiste un'ottima traduzione in francese ad opera di F. \& Ph. Gueret (inedita)].\\

E. Recami: ``I nuovi documenti sulla scomparsa di E.Majorana", in
{\em Scientia} 110 (1975) p.577.\\

E. Recami: in {\em La Stampa} (Torino), 1 giugno e 29 giugno 1975.\\

E. Recami: in {\em Corriere della Sera} (Milano), 19 ottobre 1982 e
13 dicembre 1983.\\

E. Recami: ``E.Majorana: lo scienziato e l'uomo", in {\em E.Majorana --
Lezioni all'Universit\`a di Napoli} (Bibliopolis; Napoli, 1987), p.131; \
e ``A cinquant'anni dalla scomparsa di E.Majorana", in {\em Mondotre}
(Siracusa, 1988) p.119.\\

E. Recami: ``Ricordo di Ettore Majorana a sessant'anni dalla sua scomparsa:
L'opera scientifica edita e inedita", in {\em Quaderni di Storia della
Fisica (S.I.F.)}, 5 (1999), pp.19-68; \ e in {\em Ci\^encia \& Sociedade:
PERFIS}, a cura di F.Caruso e A.Troper (C.B.P.F.; Rio de Janiero, 1997),
pp.107--172.\\

V. Reforgiato: {\em Cenni Biografici e Critici su Angelo Majorana}
(Catania, 1895).\\

V. Reforgiato: {\em Raccolta di Recensioni e Giudizi sulle Opere del
Prof. Avv. Giuseppe Majorana} (Catania, s.d.).\\

A. Rocca: {\em Il Liberty a Catania} (Catania, 1984).\\

B. Russo: ``Ettore Majorana -- Un giorno di marzo", programma televisivo
trasmesso il 18.12.90 (Rai Tre -- Sicilia). Vedere anche il libro col
medesimo titolo (Flaccomio; Palermo, 1997).

G. Scavonetti: {\em La Vita e l'Opera di Angelo Majorana} (Firenze,
1910).\\

E. Schroedinger: {\em Scienza e Umanesimo} (Firenze, 1970).\\

L. Sciacca: {\em I Catanesi Com'Erano} (Catania, 1975).\\

L. Sciascia: {\em La Scomparsa di Majorana} (Torino, 1975).\\

E. Segr\'e: {\em Enrico Fermi, Fisico} (Bologna, 1971).\\

E. Segr\'e: {\em Autobiografia di un Fisico} (Il Mulino; 1995).\\

E. Segr\'e: ``Una lettera inedita di E.Majorana", in {Storia
contemporanea} 19 (1988) p.107.\\

C. Tarsitani: ``O.M.Corbino", in {\em Sapere} 49 (Roma, 1983), n.5.\\

S. Timpanaro: {\em Pagine di scienza: Leonardo} (Milano, 1926).\\

V. Tonini: ``Il Taccuino Incompiuto" (Armando; Roma, 1984) \ [pregevole
divagazione, che parte da una tipica finzione letteraria per indagare
liberamente sulla possibile ``vita segreta" di E.Majorana].\\

G. Wataghin: in {\em Bolet\'{\i} m \  Informativo, Instituto de F\'{\i} sica
Gleb Wataghin, Universidade Estadual de Campinas} (Unicamp; Campinas, S.P.),
6 e 13 settembre 1982.\\

J. Zimba e R. Penrose: {\em Stud. Hist. Phil. Sci.} 24 (1993) 697.\\

\end{document}